\documentclass[preprint,12pt]{elsarticle}

\usepackage{amssymb}
\usepackage{amsmath}
\usepackage{upgreek}    
\usepackage{amsthm}     
\usepackage{enumitem}   
\usepackage{mathtools}  

\usepackage{booktabs}
\usepackage{pbox} 
\newcommand{\whl}{\noalign{\hrule height 1pt}}  
\newcommand{\ra}[1]{\renewcommand{\arraystretch}{#1}}

\usepackage{xcolor}
\definecolor{orange}{RGB}{217,83,25}
\definecolor{blue}{RGB}{0,114,189}
\definecolor{magenta}{RGB}{200,0,200}
\definecolor{green}{RGB}{0,146,69}

\newcommand{\red}{\textcolor{red}}
\newcommand{\orange}{\textcolor{orange}}
\newcommand{\blue}{\textcolor{blue}}

\newcommand{\green}{\textcolor{green}}

\newtheorem{lemmaT}{Lemma} 
\newtheorem*{genericT}{\genericTname}
\newenvironment{lemma}{\lemmaT}{\endlemmaT\xdef\laststatement{\thelemmaT}}
\newenvironment{corollary}[1][]
{\def\genericTname{Corollary \laststatement#1}\genericT}
{\endgenericT}

\def\ba#1\ea{\begin{align}#1\end{align}} 
\def\bas#1\eas{\begin{align*}#1\end{align*}} 
\def\bat#1\eat{\begin{alignat}{3}#1\end{alignat}} 
\def\bats#1\eats{\begin{alignat*}{3}#1\end{alignat*}} 

\def\Xint#1{\mathchoice
   {\XXint\displaystyle\textstyle{#1}}%
   {\XXint\textstyle\scriptstyle{#1}}%
   {\XXint\scriptstyle\scriptscriptstyle{#1}}%
   {\XXint\scriptscriptstyle\scriptscriptstyle{#1}}%
   \!\int}
\def\XXint#1#2#3{{\setbox0=\hbox{$#1{#2#3}{\int}$}
     \vcenter{\hbox{$#2#3$}}\kern-.57\wd0}}

\def\dashint{\Xint-}

\newcommand{\av}{\mathbf a}
\newcommand{\bv}{\mathbf b}
\newcommand{\cv}{\mathbf c}
\newcommand{\Ev}{\mathbf E}
\newcommand{\Hv}{\mathbf H}
\newcommand{\Bv}{\mathbf B}
\newcommand{\Dv}{\mathbf D}
\newcommand{\Pv}{\mathbf P}
\newcommand{\xv}{\mathbf x}
\newcommand{\Qv}{\mathbf Q}
\newcommand{\Psiv}{\mathbf\Psi}
\newcommand{\Phiv}{\mathbf\Phi}

\DeclareMathOperator*{\argmin}{arg\,min}
\newcommand{\Exp}{{\operatorname{e}}}
\newcommand{\curl}{\nabla\times}
\newcommand{\veps}{\varepsilon}

\DeclarePairedDelimiter\floor{\lfloor}{\rfloor}

\newcommand{\CP}{\text{c}}

\let\conjugatet\overline

\renewcommand{\d}[1]{\ensuremath{\operatorname{d}\!{#1}}}
\newcommand{\dd}{\mathrm d}

\newcommand{\tomega}{\widetilde{\omega}}

\newcommand{\wrc}{\theta}

\newcommand{\Real}{{\mathbb R}}

\biboptions{sort&compress}

\newcounter{bla}

\journal{Computer Physics Communications}

\begin{document}

\begin{frontmatter}


\title{Gaussian dispersion analysis in the time domain: efficient conversion with Pad\'e approximants}


\author[a]{Ludmila J. Prokopeva\corref{author}}
\author[a]{Samuel Peana}
\author[a]{Alexander V. Kildishev}

\cortext[author] {Corresponding author.\\\textit{E-mail address:} lprokop@purdue.edu}
\address[a]{Birck Nanotechnology Center, Purdue University, IN 47906, USA}

\begin{abstract}

We present an approach for adapting the Gaussian dispersion analysis (GDA) of optical materials to time-domain simulations. Within a GDA model, the imaginary part of a measured dielectric function is presented as a sum of Gaussian absorption terms. Such a simple model is valid for materials where inhomogeneous broadening is substantially larger than the homogeneous linewidth.  The GDA model is the essential broadband approximation for the dielectric function of many glasses, polymers, and other natural and artificial materials with disorder. However, efficient implementation of this model in time-domain full-wave electromagnetic solvers has never been fully achieved.
We start with a causal form of an isolated oscillator with Gaussian-type absorption --- \emph{Causal Dawson-Gauss oscillator}. Then, we derive explicit analytical formulas to implement the Gaussian oscillator in a finite-difference time-domain (FDTD) solver with minimal use of memory and floating point operations. The derivation and FDTD implementation employ our generalized dispersive material (GDM) model --- a universal, modular approach to describing optical dispersion with Pad\'e approximants. We share the FDTD prototype codes that include automated generation of the approximants and a universal FDTD dispersion implementation that employs various second-order accurate numerical schemes. The codes can be used with non-commercial solvers and commercial software for time-domain simulations of light propagation in dispersive media, which are experimentally characterized with GDA models.

\end{abstract}


\end{frontmatter}



{\bf PROGRAM SUMMARY}

\begin{small} 
\noindent
{\em Program Title:} \textbf{MADIS}  

\noindent
{\em CPC Library link to program files:}


\noindent
{\em Code Ocean capsule:} (to be added by Technical Editor)

\noindent
{\em Licensing provisions(choose one):}
GPLv3

\noindent
{\em Programming language: MATLAB}                                  


\noindent
{\em Nature of problem:} 
 The problem of efficient time-domain simulation of the Gaussian absorption is essential for wideband modeling of the optical response from materials with inhomogeneous spectral broadening, such as glasses, polymers, and other natural and artificial composite materials with structural or phase disorder.

\noindent
{\em Solution method:} 
    A Coupled Oscillator (CO) approximation to the Gaussian absorption in both the frequency and time domains is derived to solve this problem. The time-domain CO approximation is coupled to the finite-difference time-domain (FDTD) solver for the Maxwell equations in the MADIS (MAterial DIspersion Simulator) package. 
Verification of the Maxwell solvers' accuracy and stability is performed with the FDTD solver coupled to a compact universal implementation of the CO model, employing second-order schemes (either ADE or RC). Code prototypes of these efficient CO schemes can be ported to other methods and platforms for implementing the Gaussian absorption in open-source codes or commercial software.

{\em Additional comments including restrictions and unusual features:}
  Patent pending. Restrict any commercial use, including for profit reproduction.

\end{small}


\section{Introduction} \label{sec:Introduction}

\noindent
Realistic numerical modeling of light-matter interaction requires accurate broadband approximations of the dispersive dielectric functions. The dispersion models used in the frequency domain for characterizing optical materials are very diverse. Models that substitute the tabulated experiment-based descriptions of the dielectric function are commonly used in variable angle spectroscopic ellipsometry (VASE) and frequency-domain (FD) simulations. However, only a limited subset of these frequency-domain models are directly applicable to time-domain simulations. To be efficient in the time domain (TD), the dielectric function must be computed recursively in time through a numerical approximation of either the auxiliary differential equations or a convolution integral. This requirement significantly limits the classes of available approximants. This is why standard methods of computational electromagnetics in the time domain (e.g. \cite{taflove2005computational}) are utilized only for classical models, such as the Lorentz damped oscillator, Debye relaxation, and Drude plasma models.

The classical models are most appropriate for crystalline solids, where all oscillators of a given class are immersed in the same local environment. In contrast to crystalline solids, the local environment in amorphous materials is not uniform; its variations create inhomogeneous broadening of the resonant line-shape. A convolution model of such broadening was initially proposed by Efimov et al. \cite{efimov1979analytical,efimov1985dispersion} and later by Brendel and Bormann \cite{brendel1992infrared}. This model treats absorption bands as a set of Gaussian distributions of homogeneously broadened Lorentz oscillators. To satisfy causality the approach was initially amended by De Sousa Meneses \cite{meneses2005causal}. Recently, a new, fully-causal and asymptotically consistent fix of the Efimov convolution has been proposed by Orosco et al. and used to approximate the dielectric functions of metals \cite{orosco2018causal, orosco2018optical}. 

If the inhomogeneous broadening is much larger than the homogeneous linewidth, then the absorption bands can be approximated by isolated Gaussian distribution peaks, providing a simple, experiment-based approximation to the convolution models. This approach --- Gaussian dispersion analysis (GDA), proposed by McDonald et al. \cite{macdonald2000dispersion} and improved by Kefee \cite{keefe2001curvefitting} --- is one of the most application-critical approximations commonly used to characterize the broadband dielectric function of polycrystalline and amorphous materials with disorder including oxides \cite{may2007optical,uprety2017spectroscopic,schoche2017optical}, polymers \cite{pallapapavlu2011characterization, rauch2012temperature, naqavi2018optical, hilfiker2018dielectric, patel2020diphenylsiloxane}, metals \cite{lonvcaric2011optical,synowicki2017optical}, and as an additional term to fit the complicated dispersive behavior of other disordered solids, such as, for example, phase change materials \cite{orava2008optical,abdel2018optical,ramirez2018thermal}. With this approach, the imaginary part of a dispersive dielectric function, $\Im \left[ \veps(\omega) \right] = \veps''(\omega)$, is presented as a sum of Gaussian absorption terms $\chi''_{\rm{G}}(\omega,i)$ that may correspond to separate oscillations in the phonon density of states and disordered light-matter interactions in the ultra-violet bands
\begin{equation}
    \veps''_\text{G}(\omega) =\sum_i{ \chi''_{\rm{G}}(\omega,i)}.
    \label{eq:GDA_Intro}
\end{equation}

\noindent
The important problem in employing the Gaussian-based dielectric functions in time-domain Maxwell solvers has not been addressed in the literature. In this paper, we adapt the dielectric function described with causal Dawson-Gauss oscillators for use with time-domain numerical solvers. To enable computationally efficient and accurate conversion from Gaussian absorption terms we use Pad\'e approximants of argument 
$s = -\iota\omega$. Standard numerical fitting approaches done individually for each set of dispersion parameters are inefficient and important asymptotic properties and symmetries could be lost. Hence, we derive explicit formulas for the approximants, preserving their analytical properties and connection to parameters of the original dispersion formula. 

The paper includes all the code prototypes for (i) analytically-derived Pad\'e approximants of different orders, and (ii) coupling of the converted model with a generic TD solver written in a high-level interpreted language (MATLAB) that could be ported to other languages if needed. For simplicity, we use the classical Yee’s Finite Difference Time Domain (FDTD) scheme \cite{yee1966numerical}, however the technique can be used with other explicit or implicit TD schemes, including Discontinuous Galerkin Time Domain (DGTD) \cite{ren2018continuous}, Finite Volume Time Domain (FVTD) \cite{prokopeva2011fvtd}, Finite Element Time Domain (FETD) \cite{abraham2018convolution}, or higher-order FDTD schemes \cite{angel2019high}.
This paper and codes are intended to allow for seamless integration of Gaussian dispersion analysis derived dielectric functions into commercial software as well as non-commercial Computational Electromagnetics (CEM) and multiphysics codes. For example, the proposed approach was successfully tested with commercial time-domain solvers, ANSYS Lumerical (FDTD) and COMSOL Multiphysics (DGTD).
\section{Mathematical Model}
\noindent
In this section, we describe a mathematical model of light propagation in a dispersive medium characterized by a finite set of Gaussian absorption profiles.
First in Section~\ref{sec:Gauss}, we introduce the dielectric function for an isolated Dawson-Gaussian oscillator in causal analytical form. Then we introduce a generalized dispersive material (GDM) formulation \cite{prokopeva2011optical,prokopeva2020time} in terms of the Pad\'e approximants in Section \ref{sec:GDM}. The GDM formulation is used later in Section~\ref{sec:GDMapprox} to adapt the Gaussian-based models for TD solvers. In the final Section \ref{sec:Maxwell},  the universal  TD-compatible GDM dielectric function is coupled with Maxwell equations to model light propagation in a dispersive medium.

\subsection{Gaussian Absorption Model}\label{sec:Gauss} 
\noindent
In the Gaussian model, the material absorption is characterized by a Gaussian distribution
$A\Exp^{-(\omega-\Omega)^2/\sigma^2}$ with a given amplitude $A$, center $\Omega$, and width $\sigma$. The causal, parity-consistent formulation of Gaussian absorption \cite{keefe2001curvefitting, cataldo2016submillimeter} requires \emph{adding a second term}, since $\chi'$ and $\chi''$ shall be even and odd functions of frequency respectively, $\chi'(-\omega)=\chi'(\omega)$ and $\chi''(-\omega)=-\chi''(\omega)$ \cite{nussenzveig1972causality}. Thus, we arrive at the causal Dawson-Gaussian oscillator model $\chi''_{\rm{G}}(\omega)$ that is broadly used in Gaussian dispersion analysis,
\begin{equation}
    \chi''_\text{G}(\omega) = A\left[ \orange{\Exp^{-(\omega - \Omega)^2 / \sigma^2}} 
    \blue{ - \Exp^{-(\omega + \Omega)^2 / \sigma^2}} \right].
    \label{eq:CDG_im}
\end{equation}
Here we use the orange and blue colors to keep track of the formulas symmetry. The blue term is needed for causality and is mandatory for the correct Fourier transform to the time domain. However, in frequency domain solvers, the blue term can be neglected if its contribution is small. For example, if $\Omega>3\sigma$, we have
$\blue{\Exp^{-(\omega + \Omega)^2 / \sigma^2}} < 2\cdot 10^{-4}$ for all $\omega>0$.
If the Gaussian width $\sigma$ is larger than the offset of Gaussian center $\Omega$, then the positive (red) and negative (blue) peak of the Gaussian absorption are not isolated from each other, the merged absorption peak becomes asymmetric and does not correspond to an isolated normal distribution. This $\sigma/\Omega$-ratio dependent behavior should be taken into account when permittivity data is fitted with a GDA model and the physical consistency of the fit is assessed. 

The real part of the causal Dawson-Gaussian model can be found from the imaginary part \eqref{eq:CDG_im} using the Kramers-Kronig relation
\begin{equation}
    \chi'_\text{G}(\omega)
    = \frac{1}{\pi}\dashint\limits_{-\infty}^\infty\! \frac{\chi''_\text{G}(\tomega)}{\tomega-\omega} \, \mathrm{d}\tomega
    = -\mathcal{H}\{\chi''_\text{G}\}
    = \frac{2A}{\sqrt{\pi}} \left[ \blue{F\left(\frac{\omega+\Omega}{\sigma} \right)} 
    \orange{ - F\left(\frac{\omega-\Omega}{\sigma} \right)}    \right] ,
    \label{eq:CDG_re}
\end{equation}
where the crossed integral sign is the Cauchy principal value integral and $F(y)$ is the Dawson function, connected to the Gaussian distribution through the Hilbert transform, 
$\mathcal{H}\{\Exp^{-x^2}\}(y) = 2\pi^{-1/2}F(y)$ 
\cite{weideman1995computing}. Note that while the Dawson function is an odd function $F(-x) = -F(x)$, the difference of the Dawson pair in \eqref{eq:CDG_re} is even, in compliance with the analytical properties of the dielectric function. The resulting causal dielectric function term is $\chi_\text{G}(\omega) = \chi'_\text{G}(\omega) + \iota\chi''_\text{G}(\omega)$. Throughout the paper {\it Gaussian (absorption) model} refers to the causal Dawson-Gaussian dispersion model \eqref{eq:CDG_im}-\eqref{eq:CDG_re}.

Another important relation is the connection of the Gaussian dielectric function to the Faddeeva function $w(z)$ \cite{faddeeva1961tables}, 
\begin{equation}
    \chi_\text{G}(\omega) = 
    \iota A 
    \left[
    \orange{w\left(\frac{\omega-\Omega}{\sigma}\right)} 
    \blue{-w\left(\frac{\omega+\Omega}{\sigma}\right)}
    \right].
    \label{eq:Faddeeva}
\end{equation}
The derivation can be found in the Appendix. The Faddeeva function is an exponentially scaled complex complementary error function 
$w(z) = \Exp^{-z^2}\mathrm{erfc}(\iota z)$ and is related to the Dawson function as $w(z) = \Exp^{-z^2} + 2\iota\pi^{-1/2}F(z)$. The Faddeeva function is closely related to another known special function --- a plasma dispersion function, $Z(\zeta) = \iota\sqrt{\pi}w(\zeta)$ --- commonly used in the theory of linearized waves in hot plasma, with or without a magnetic bias, whenever the velocity distribution is taken to be Gaussian \cite{fried2015TPDF}. 

After substitution, $z=\iota s$, function $w_s(s) = w(\iota s)$ becomes a \textit{Positive Real Function} (PRF) of the complex argument $s$. This follows from the fact that $w_s(s)$ is entire and non-negative real for pure imaginary argument $\Re[w_s(\pm\iota\omega)]=\Exp^{-\omega^2}\ge0$, see for example \cite{chen2009note}. By the definition, PRF is a function $f(s)$ of a complex variable $s$, which is (i) real-valued for real $s$ and (ii) has a non-negative real part whenever the real part of $s$ is positive, $\Re[s]>0 \Rightarrow \Re[f(s)]\ge0$, \cite{brune1931synthesis}. This property is important to guarantee that positive loss $\Im(\omega) = \Re[s]>0$ results in non-negative absorption $\chi''(\omega)\ge0$ in~\eqref{eq:Faddeeva}, which is important for physical consistency of the dielectric function and stability of time-domain numerical schemes. 

If a dielectric function is approximated for time domain modeling with a rational polynomial of order $n$, 
$\veps(\omega)\approx\veps_n(\omega)$, then the approximation $\veps_n(\omega)$ ideally should also maintain the non-negative absorption property for physical consistency. Testing for sufficient positive-reality (PR) conditions of rational approximation $[-\iota\veps_n(\iota s)]$ is not always straightforward, but a number of necessary PR conditions are easy to check, including constraints on the polynomial roots and coefficients that are mentioned in the next section~\ref{sec:GDM}. 

\textbf{Gaussian model in the time domain}. 
The inverse Fourier transform of the Gaussian model \eqref{eq:CDG_im}-\eqref{eq:CDG_re} may be done by introducing Lorentzian broadening $\Gamma>0$ in a convolution integral and taking the infinitely narrow band limit $\Gamma\rightarrow +0$,
\begin{equation}
    \chi_\text{LG}(\omega,\Gamma\rightarrow +0)
    = 
    \left.
    \frac{2A}{\pi}\int\limits_{-\infty}^{\infty}
    \frac{x\Exp^{-(x-\Omega)^2/\sigma^2}}
    {x^2-(\omega+\iota\Gamma)^2}
 \mathrm{d}x 
 \right|
 _{\Gamma\rightarrow +0}
= \quad \chi_\text{G}(\omega).
    \label{eq:chi_LG}
\end{equation}
Performing the Inverse Fourier Transform (IFT) 
$\chi_\text{LG}(\omega,\Gamma) \;\xrightarrow{\;{\rm IFT}\;}\;\chi_\text{LG}(t,\Gamma)$ and taking the limit $\Gamma\rightarrow +0$ yields the time-domain susceptibility for the Gaussian model
\begin{equation}
\chi_\text{LG}(t,\Gamma\rightarrow +0) 
    = \underbrace{\frac{2A\sigma}{\sqrt{\pi}}}_{a}
    \Exp^{-t^2\sigma^2/4}
    \sin(\Omega t) \theta(t)
    =\chi_\text{G}(t),
    \label{eq:chi_LGt}
\end{equation}
here we use the Heaviside step function $\theta(t)$, the Fourier pair for a Lorentz oscillator \eqref{eq:chi_L}, and the Gaussian integral 
$\int_{\mathbb{R}} \Exp^{-c(x+b)^2} \mathrm{d}x = \sqrt{\pi/c}$, $(c>0)$. 

Initially we defined the Gaussian amplitude parameter $A$ as the maximum absorption in ~\eqref{eq:CDG_im}. We can also define the time-domain amplitude parameter $a$ and substitute $A = a\sqrt{\pi}/(2\sigma)$ in the dielectric function formulas~(\ref{eq:CDG_im}-\ref{eq:CDG_re}). Formulation with amplitude $a$ should be used, when taking the zero-width limit $\sigma\rightarrow +0$ to obtain a physically consistent result.

\begin{figure}[!ht]
    \center
    \includegraphics[width=0.75\textwidth]{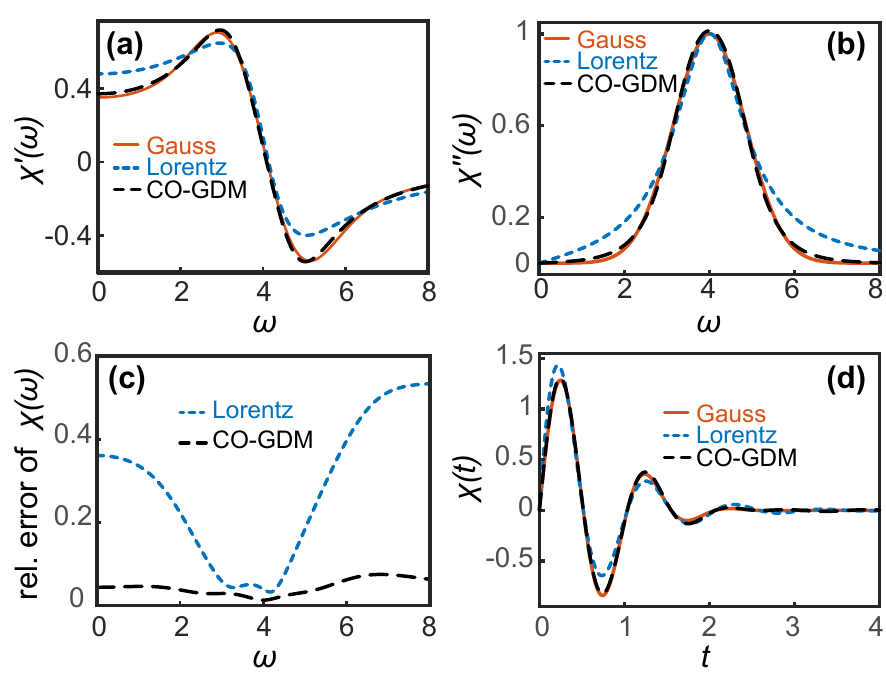}
    \caption{Approximation of the Gaussian oscillator (\orange{solid orange line, Gauss})  for $\Omega=4$, $\rm{FWHM}=1$ with a Lorentzian (\blue{dashed blue line, Lorentz}) and Coupled Oscillators model (this paper) that uses GDM formalism (\textbf{black dashed line, CO-GDM}): 
    (a,b)~the real and imaginary parts of susceptibility in the frequency domain,  
    (c)~relative error of the susceptibility approximated with a Lorentz oscillator and Coupled Oscillators, (d)~susceptibility in the time domain. 
    The lowest-order CO-GDM approximation ($n=2$) is used in all plots.}
    \label{fig:01_Gauss_Osc}
\end{figure}

Since the classical Lorentz oscillator has straightforward implementations in time-domain solvers, it is commonly used as a rough approximation to the Gaussian model. Figure~\ref{fig:01_Gauss_Osc} compares a Gaussian to its Lorentzian approximation with the same center $\Omega = 4$ and broadening
($\rm{FWHM} = 2\sigma\sqrt{\log(2)}=1$) that correspond to Lorentz parameters $\sqrt{b_0} = 4.031$ and $b_1 = 1.004$ obtained from ~\eqref{eq:chi_L}. As can be seen in Fig.~\ref{fig:01_Gauss_Osc}(abc) the Lorentz approximation fails to accurately reproduce the highly confined Gaussian absorption profile in both --- the frequency and time domains. By contrast, the model derived in this paper is based on the GDM decomposition principle (discussed in Section~\ref{sec:GDM}); it gives an accurate and TD-compatible representation in terms of {\it Coupled Oscillators} (CO) even for the lowest order ($n=2$), see Section \ref{sec:GDMapprox}.

It is important that the time-domain representation \eqref{eq:chi_LGt} shows that the Gaussian model represents a \textit{single} harmonic oscillator at frequency $\Omega$. In contrast to the Lorentz damped oscillator with constant homogeneous scattering $\Gamma$, the Gaussian oscillator has a time dependent scattering function $\Gamma_{\text G}(t) = t\sigma^2/4$ or \textit{inhomogeneous broadening}. Fitting Gaussian absorption with multiple Lorentz terms in frequency domain would inevitably produce multiple non-physical oscillations. However, with the analytically constrained coupled oscillator model derived in this paper, we will see that individual phase-relaxed oscillators effectively form a single oscillator model with a slow-varying harmonic modulation that approximates inhomogeneous broadening (Lemma 2).

\subsection{The Generalized Dispersive Material Model} 
\label{sec:GDM}
\noindent
Due to the diversity of dispersion laws in the optical range, modular approaches for the numerical implementation of dispersion in the time domain are more efficient compared with individual implementations for each dispersion model. In modular approaches, any dielectric function can be decomposed into a sum of generic terms with a known recursive TD implementation. For example, in the TD CEM literature, dielectric functions are often approximated with multi-pole Debye relaxation models, or  with a combination of Lorentz oscillator terms.

A known danger of such modular approaches is overfitting the experimentally measured dielectric function with too many parameters, producing nonphysical oscillations or relaxations. An interesting approach for evaluating model quality based on information criteria is proposed in \cite{likhachev2017model}.
In the method outlined in this paper, to avoid overfitting and resulting nonphysical terms, we use our prior approach \cite{prokopeva2011optical} of building a \textit{physically-constrained} rational approximation of argument $s=-\iota\omega$ based on general analytic properties of the dielectric function. It will be shown that the Gaussian dispersion is approximated best by employing a number of \textit{coupled oscillator pairs} with non-zero phase and possibly \textit{one} Lorentzian term (not multiple Lorentzian terms as a simple model-fitting approach might suggest).

In time-domain modeling, the dielectric function is assumed to be causal ($\veps(t)=0$ for $t<0$), and thus  $\veps(\omega)$ is analytic in the complex upper half-plane ($\Im[\omega]>0$), with the real and imaginary parts connected via Kramers-Kronig relations (Hilbert transform). By substituting $\omega = \iota s$, $s\ge0$, the Fourier transform of a causal dielectric function, $\veps(\omega) = \int_\mathbb{R}\veps(t)\Exp^{\iota\omega t}\d t$, becomes a real integral transform of a real-valued function which converges for a wide class of non-growing functions $\veps(t)$. Thus, the dielectric function $\veps(\omega)$  can always be represented as a rational approximation with real coefficients $p_i$, $q_i$ on the positive imaginary axis $\omega = \iota s$, $s\ge0$,
\begin{equation}
    \veps(\omega) = \veps_n(\omega) \approx \frac{P_n(s)}{Q_n(s)} = 
    \frac{p_0+p_1s+...+p_ns^n}{q_0+q_1s+...+s^n}, 
    \qquad s=-\iota\omega, \ p_i\in \mathbb{R},\  q_i\in \mathbb{R}^+,
    \label{eq:GDM_pq}
\end{equation}
and the domain of this definition can be analytically continued to the entire upper half-plane $\Im [\omega] \ge 0$, as long as all the poles are located in the lower half-plane.
Here $n$ is the order of denominator (or number of poles). The order of the numerator is smaller or equal to $n$ due to the finite permittivity limit at infinity. 

Basic properties of the linear causal dielectric function $\veps(\omega)$ that should be inherited by the approximation $\veps_n(\omega)$ and thus constrain coefficients $p_i,q_i$ include: 
\begin{enumerate}[label=(\roman*)]

\item {\textbf{Poles property:}}
$\veps_n(\omega)$ is analytic in the upper half-plane (causality principle). Therefore, $Q_n(s)$ is a Hurwitz polynomial, and thus its coefficients must be nonnegative, $q_i\ge0$, $i \in\overline{0,n-1}$.
\item {\textbf{Zeros property:}}
$\veps_n(\omega)$ should not have zeros in the upper half-plane.
Therefore $P_n(s)$ is a Hurwitz polynomial and hence its coefficients must be nonnegative,
$p_i\ge0$, $i \in\overline{0,n}$. (i.e. $\veps^{-1}_n(\omega)$ is analytic in the upper half-plane).
\item {\textbf{Non-negative absorption:}} 
$\veps_n(\omega)$ should have non-negative total absorption in the upper half-plane, i.e. $\veps_n''(\omega)\ge0$ for $\Im(\omega)>0$.
\end{enumerate}
Statements (i)-(ii) are corollaries of property (iii), which is in turn equivalent to the function $-\iota\varepsilon(\iota s)$ being a PRF of argument $s$, e.g. see criterion in \cite{kim1971modern}.

Non-negativity of absorption (iii) for the classical Lorentzian with $\Gamma\ge0$ is satisfied naturally. For non-Lorentzian dispersions, it may not be possible to achieve condition (iii) with a rational approximation, but the asymptotic version of (iii) will always follow from the approximation convergence: 
$\min\limits_\omega(\veps''_n(\omega)) \rightarrow 0$ as $n \rightarrow \infty$, which can be made exponentially fast with a proper approximation technique. 

For cases when (iii) is not satisfied exactly, testing for properties (i)-(ii) is necessary for stable time-domain modeling. Testing for `positive realness' is not always straightforward, a recent review of the PRF criteria, sufficient and necessary conditions, is presented in \cite{chen2009note}. In general, PRF properties are long and well studied in the context of network synthesis~\cite{brune1931synthesis,van1960introduction}, and can be utilized for physical consistency and stability analysis of approximated dielectric functions in time-domain CEM.

The poles property (i) should always be tested, since it does not depend on the presence of other dispersion terms in the dielectric function (unless poles cancel during the summation of terms); therefore, the constraint on $q_i$ is mandatory. However, the zeros property (ii) and absorption property (iii) depend on other dispersion terms in the sum: selected terms and partial sums could have zeros in the upper half-plane and negative absorption that are not present in the total sum. 

For example, the well known physically consistent Drude model with loss $\Gamma$ and plasma frequency $\Omega$, can be decomposed into two terms as
\begin{equation}
    \veps_\text{Drude}(\omega) = 
    \frac{\Omega^2}{s^2+s\Gamma}=
        \frac{\Omega^2}{s\Gamma} - \frac{\Omega^2}{\Gamma^2+s\Gamma}, \quad s = -\iota\omega.
    \label{eq:ex_Debye}
\end{equation}
Here, condition (ii) holds for the combined fraction since $p_0=\Omega^2\ge0$, but fails for the second term $p_0^{(2)} = -\Omega^2\Gamma^{-1}\le0$, after the conductivity pole is detached. In other words, after such decomposition, we obtain a Debye term with negative absorption, although the total Drude model does not have gain.

Similarly, when a sum of oscillators approximates the Gaussian dispersion, individual terms can have significant negative absorption, as it can be seen in Fig.~\ref{fig:GDMapprox_n}that cancels in the total sum. It can be shown that for a coupled oscillator pair approximating Gaussian dispersion, the properties (i)-(ii) hold as long as $\Omega > \sigma$, which is in agreement with the initial assumptions of the Gaussian model in Section~\ref{sec:Gauss}.


From these examples we observe that constraints (ii)-(iii) should be applied only to each \textit{physically independent} and \textit{spectrally isolated} dispersion term in its complete rational representation~\eqref{eq:GDM_pq}. 

For the time-domain implementation, we perform partial fraction decomposition of ~\eqref{eq:GDM_pq} assuming simple poles only (with multiplicity 1), where  properties (ii)-(iii) for each individual term in the sum are relaxed for the reasons explained above. The dielectric function is represented as a unique combination of nonzero real poles $\chi_i^{\rm D}(\omega)$ and conjugate pole pairs  $\chi_j^{\rm{L}_\varphi}(\omega)$,
\begin{equation}
    \veps(\omega) = \veps_{\infty}
    -\frac{\sigma}{\iota\omega\veps_0}
    +\sum\limits_{i=1}^{M}\chi_i^{\rm D}(\omega)
    +\sum\limits_{i=M+1}^{N}\chi_i^{\rm{L}_\varphi}(\omega).
    \label{eq:GDM}
\end{equation}
This universal TD-compatible description of material dispersion \eqref{eq:GDM} is called {\it the generalized dispersive material} (GDM) model \cite{prokopeva2011optical}. Constants $\veps_{\infty}$ and $\sigma$ are the high-frequency relative permittivity and conductivity parameters, respectively, and $\veps_0$  is the permittivity of vacuum. Every $i^{\rm{th}}$ real pole is {\it a Debye relaxation}, and its IFT in the time domain is
\begin{equation}
    \chi_{(i)}^{\rm D}(\omega) = \frac{a_1}{b_1-\iota\omega}
    \quad \xrightarrow{\quad{\rm IFT}\quad} \quad
    a_1\Exp^{-b_1t}\theta(t)=\chi_{(i)}^{\rm D}(t),
    \label{eq:chi_D}
\end{equation}
where $a_1 \in \mathbb{R}$ and $b_1>0$ are amplitude and decay parameters of the $i^{\rm{th}}$ Debye term, and $\theta(t)$ is the Heaviside step function. Note, for simplicity of notation we omit the sum indices $i$ when discussing individual terms. Finally, every $i^{\rm{th}}$ pole pair in the second sum is {\it a phase-relaxed Lorentz oscillator},
\begin{equation}
    \chi_{(i)}^{\rm{L}_\varphi}(\omega) =
    \frac{a_0 -\iota\omega a_1}{b_0-\iota\omega b_1-\omega^2}
    \quad \xrightarrow{\quad{\rm IFT}\quad} \quad
    a \Exp^{-\Gamma t} \sin({\Omega t - \varphi}) \theta(t) = \chi_{(i)}^{\rm{L}_\varphi}(t)
    \label{eq:chi_Lp}
\end{equation}
with real polynomial coefficients $a_0, a_1 \in \mathbb{R}$ and $b_0, b_1 \ge 0$. The time domain parameters of the phase-relaxed oscillator (amplitude $a$, broadening $\Gamma$, carrier frequency $\Omega$, and phase $\varphi$) may be derived from the FD coefficients $a_{0,1},b_{0,1}$ as
\begin{equation}
\begin{array}{rcl} 
    a      &=& \dfrac{\sqrt{a_0^2-a_0a_1b_1+a_1^2b_0}}{\Omega}
            = \dfrac{\sqrt{\left(a_0- a_1\Gamma-\iota a_1\Omega\right)
                          \left(a_0- a_1\Gamma+\iota a_1\Omega\right)}}{\Omega}, \\ [0.3cm]
    \Gamma &=& b_1/2, \\ [0.15cm]
    \Omega &=& \sqrt{b_0-\Gamma^2}, \\ [0.15cm]
    \varphi &=& \left\{
            \begin{array}{l}
            {\rm atan2} \left( -a_1\Omega, a_0-a_1\Gamma \right), \quad \text{if} \ b_1^2/4 < b_0 \\ [0.1cm]
            -\iota\log\dfrac{a_0-a_1(\Gamma+\iota\Omega)}{a\Omega}, \quad \text{otherwise} 
            \end{array}
        \right.
\end{array}        
\label{eq:GDM_FD2TD}
\end{equation}
Note if $ b_0< b_1^2/4$, then parameters $\Omega$, $a$, $\varphi$ are purely imaginary and the time-domain response $\chi^{\rm{L}_\varphi}(t)$ consists of two decaying exponential relaxations~\eqref{eq:chi_D}:  
$\pm 0.5\,  a\iota \, \mathrm{exp}\{(-\Gamma \mp \iota\Omega)t \pm \iota\varphi\}$. This case is usually referred to as {\it an overdamped oscillator}. Conversely, if $b_0>b_1^2/4$, then all the TD parameters are real and nonnegative, $a, \Gamma, \Omega\geq0$, and the angle $\varphi \in [0,2\pi)$ can be uniquely defined in the code with the two-argument function \texttt{atan2(y,x)} or alternatively with complex argument function \texttt{angle(x+1i*y)}. The backward conversion from TD parameters \eqref{eq:GDM_FD2TD} to FD coefficients is straightforward
\begin{equation}
\begin{array}{rcl} 
    a_0 &=& a\Omega\cos\varphi - a\Gamma\sin\varphi, \\
    a_1 &=& -a\sin\varphi, \\
    b_0 &=& \Omega^2 + \Gamma^2, \\
    b_1 &=& 2\Gamma.
\end{array}        
\label{eq:GDM_TD2FD}
\end{equation}
If phase $\varphi$ is zero, then $a_0=a\Omega$ and $a_1=0$, and thus the oscillator degenerates into a classical Lorentz oscillator
\begin{equation}
    \chi_{(j)}^{\rm{L}}(\omega) =
    \frac{a_0}{b_0-\iota\omega b_1-\omega^2}
    \quad \xrightarrow{\quad{\rm IFT}\quad} \quad
    a \Exp^{-\Gamma t} \sin({\Omega t}) \theta(t) = \chi_{(j)}^{\rm{L}}(t).
    \label{eq:chi_L}
\end{equation}
The phase-relaxed Lorentz model \eqref{eq:chi_Lp} can be written using TD parameters
\begin{equation}
\begin{aligned}
\chi_{(j)}^{\rm{L}_\varphi}(\omega) &= 
\frac{a_0 -\iota\omega a_1}{b_0-\iota\omega b_1-\omega^2} = 
a\frac{\Omega\cos\varphi + \iota(\omega+\iota\Gamma)\sin\varphi}{\Omega^2-(\omega+\iota\Gamma)^2},
\end{aligned}
\label{eq:chi_CP}
\end{equation}
and can be decomposed into a pair of conjugate poles
\begin{equation}
\begin{aligned}
    \chi_{(j)}^{\rm{L}_\varphi}(\omega) &= 
    \frac{a}{2} 
    \left[ \frac{\Exp^{-\iota\varphi}}{\omega+\iota\Gamma+\Omega}
          -\frac{\Exp^{ \iota\varphi}}{\omega+\iota\Gamma-\Omega}     \right].
\end{aligned}
\label{eq:chi_CP2}
\end{equation}
This form, also referred to as
{\it critical points (CP) model} of order $\nu = -1$, was initially used to approximate the dielectric functions of semiconductors \cite{kim1992modeling, kim1993modeling, aspnes1980handbook}. Beyond semiconductors,  \eqref{eq:chi_CP2} has been successfully employed to approximate the optical properties of metals \cite{etchegoin2006analytic,etchegoin2007erratum,little2011analysis} and gain materials \cite{campoy-quiles2004ellipsometric}. The asymptotic behavior of the CP model was improved by Leng et al. \cite{leng1998analytic}. 

The algebraic decomposition \eqref{eq:GDM} indicates that with a modular GDM approach, a non-zero phase should be added to the Lorentz oscillator, making the absorption peak of the oscillator asymmetric. Using one (as in Fig.~\ref{fig:01_Gauss_Osc}) or more symmetric Lorentz terms (\ref{eq:chi_L}) to approximate Gaussian absorption results in poor accuracy and generates non-physical behavior. However, as we show in Section \ref{sec:GDMapprox}, a combination of just two coupled CP terms \eqref{eq:chi_Lp} with non-zero phase, approximates a Gaussian oscillator with a few percent of relative error. This accuracy level is sufficient for approximating most experimental data.

Summarizing Sections \ref{sec:Gauss}--\ref{sec:GDM}, we conclude that a combination of Gaussian absorption terms enables the compact, physics-driven, and accurate FD characterization of optical materials with disorder. By utilizing a GDM decomposition approach FD Gaussian models can be adapted for TD numerical schemes with acceptable accuracy. In particular this approach is physically-constrained, and computationally efficient in TD as outlined in detail in Section~\ref{sec:Numerical}.

\subsection{Dispersive Maxwell Equations}\label{sec:Maxwell}

\noindent
To model light propagation in dispersive media, we couple the previously discussed dispersive dielectric function $\chi(t)$ with the time domain Maxwell equations \eqref{eq:Maxwell_TD}. These equations connect the inducing fields --- the electric and magnetic field vectors,  $\Ev(\xv,t)$ and $\Hv(\xv,t)$, with the induced fields --- the magnetic flux density $\Bv(\xv,t)$ and electric displacement $\Dv(\xv,t)$, where all quantities are functions of $\xv$ and time $t$,
\begin{equation}
\left\{
\begin{array}{lcrl}
    \partial_t\Dv &=&  \curl\Hv, &\quad\Dv = \veps_0 \, \veps *\Ev, \medskip\\
    \partial_t\Bv &=& -\curl\Ev, &\quad\Bv = \mu_0\Hv,
\end{array}
\right.
\qquad  \quad t \ge 0 .
\label{eq:Maxwell_TD}
\end{equation}
Here $\veps_0$ and $\mu_0$ are the electric permittivity and magnetic permeability in vacuum respectively. They are connected to the speed of light in vacuum via $c_0 = 1/\sqrt{\mu_0\veps_0}$.
The expression $\veps *\Ev$ in \eqref{eq:Maxwell_TD} denotes a convolution in time, i.e.
\begin{equation}
    \Dv(\xv,t) = \veps_0 \, \veps * \Ev  = \veps_0 \,
    \int\limits_{-\infty}^{\infty}\veps(\tau) \, \Ev(\xv,t-\tau) \, \rm{d}\tau,
\label{eq:Dconvolution}
\end{equation}
where $\veps(t)$ is a causal dielectric function, requiring that $\veps(t) = 0$ for all $t<0$.
For simplicity, we assume isotropic non-magnetic optical materials ($\mu = 1$), however the exact same GDM model~\eqref{eq:GDMfinal} is applicable for arbitrary dispersive magnetic permeability, including magnetic Gaussian terms. GDM can also be generalized to the anisotropic and bi-anisotropic cases.

In the frequency domain, the system of equations in~\eqref{eq:Maxwell_TD} becomes
\begin{equation}
\left\{
\begin{array}{lcrl}
    -\iota\omega\Dv(\omega) &=& \;\;\;\curl\Hv(\omega),    &\quad\Dv(\omega) = \veps_0\, \veps(\omega)\, \Ev(\omega)  ,\medskip\\
    -\iota\omega\Bv(\omega) &=& - \curl\Ev(\omega),        &\quad\Bv(\omega) = \mu_0\, \Hv(\omega) ,
\end{array}
\right.
\label{eq:Maxwell_FD}
\end{equation}
where the frequency domain functions are obtained through the Fourier transform of their corresponding quantities in the TD.
For example, $\Ev(\xv,\omega)$ corresponds to $\Ev(\xv,t)$ via the transform pair
\begin{equation}
\begin{array}{ll}
   \Ev(\xv,\omega) & = \int\limits_{-\infty}^{\infty} \Exp^{\iota\omega t} \Ev(\xv,t)\, {\rm d} t, \qquad
   \Ev(\xv,t) = \frac{1}{2\pi} \int\limits_{-\infty}^{\infty} \,  \Exp^{-\iota\omega t} \Ev(\xv,\omega)\, {\rm d} \omega.
\end{array}
\end{equation}

The dielectric function is usually given in the frequency domain, $\veps = \veps(\omega)$,
and it defines the dispersion characteristics of the material. In this paper we assume that the dispersion properties of the material are given in GDM form \eqref{eq:GDM}, i.e.
\begin{equation}
   \veps(\omega) = \veps_{\infty} - \frac{\sigma}{\iota\omega\veps_0} 
   + \sum_{i=1}^N
   \underbrace{\frac{a_{0,i} -\iota\omega a_{1,i}}{b_{0,i} -\iota\omega b_{1,i} - \omega^2}}_{\chi_i(\omega)}. 
\label{eq:GDMfinal}
\end{equation}
where $a_{0,i},a_{1,i}\in\Real$ and $b_{0,i},b_{1,i}\ge0$. The first-order GDM terms (i.e. Debye terms) are accounted for in the sum as a special case of $\chi_i(\omega)$, where $a_{0,i}=b_{0,i}=0$. The conductivity term is added separately, so the case of zero pole $a_{0,i}=b_{0,i}=b_{1,i}=0$ is forbidden in the sum. In the codes, conductivity and first-order GDM terms are implemented separately from the second-order GDM terms for optimized performance. The conductivity term can be implemented in Yee's scheme without any additional storage. The first-order terms require only one additional vector storage per term, while each second-order term requires two. 

The corresponding constitutive relations for electromagnetic fields in the time and frequency domains are
\begin{equation}
    \Dv(t) = \veps_0\, \veps_{\infty} \Ev(t) + \sigma\int\limits_0^t \Ev(\tau) \,{\rm d}\tau  + \sum_{i=1}^N\Pv_i(t),
\label{eq:Dsum_TD}
\end{equation}
\begin{equation}
    \Dv(\omega) = \veps_0\, \veps_{\infty} \Ev(\omega) - \frac{\sigma}{\iota\omega}\Ev(\omega)  + \sum_{i=1}^N\Pv_i(\omega),
\end{equation}
where the partial polarization terms are $\Pv_i(\omega) = \veps_0\chi_i(\omega) \Ev(\omega)$.

The FDTD-GDM method described in the next Section~\ref{sec:Numerical} offers a universal second-order accurate numerical solution to the TD Maxwell equations (\ref{eq:Maxwell_TD},\ref{eq:GDMfinal}) for light propagation in dispersive media. This FDTD-GDM method works for the dispersion models introduced through GDM decomposition (or approximation), including the GDM approximation of the Gaussian absorption model derived in this paper. In the FDTD-GDM method, different second-order ADE and RC numerical schemes are implemented in a universal and compact form~\cite{prokopeva2011optical,prokopeva2020time}, including the conventional ADE, bilinear ADE, TRC, quasi-TRC, PCRC, PLRC methods.
\section{Numerical implementation}\label{sec:Numerical}

\noindent
The numerical implementation section contains two subsections. In Section~\ref{sec:GDMapprox} we derive the analytical GDM approximation of Gaussian absorption with controlled accuracy. In Section~\ref{sec:FDTD_GDM} we implement the GDM model into Yee's FDTD scheme using a compact universal formalism applicable to various second-order-accurate ADE and RC schemes.

\subsection{GDM approximation of the Gaussian absorption model}\label{sec:GDMapprox} 
\noindent
    The goal of this section is to present a GDM approximation for the Gaussian absorption model to enable accurate time-domain simulations. One possible approach for generating a GDM approximation is to use optimization algorithms and fit the Gaussian data with a GDM sum. This would require solving an optimization problem for each and every combination of parameters and given materials. Also, important analytic properties and symmetries of the original Gaussian absorption may be lost. Instead of curve fitting, we \textit{analytically} derive an explicit GDM formula with coefficients that are explicit functions of the Gaussian parameters $[A,\Omega,\sigma]$. This derivation consists of two steps. (1) finding a constrained minimax rational approximation of the real Dawson function $F(x)$ with real argument $x$, see Section~\ref{sec:DawsonApprox}. (2) analytical derivation of the GDM approximation for the complex Gaussian susceptibility as a function of 4 variables $\chi_{\text G}(\omega; A,\Omega,\sigma)$ and its Fourier transform in TD, see Section~\ref{sec:GDMformula}. 
    
    In step 1, we can use known rational approximations for 
    (i)~the Dawson function $F(x)$, and the relation~\eqref{eq:CDG_re} to $\chi_{\text G}'(\omega)$; 
    (ii)~the Gaussian function $G(x) = \Exp^{-x^2}$, and the relation~\eqref{eq:CDG_im} to $\chi_{\text G}''(\omega)$; or 
    (iii)~the Faddeeva function $w(x)$ (or a closely related plasma dispersion function, $Z(x) = \iota\sqrt{\pi}w(x)$), along with the susceptibility formula~\eqref{eq:Faddeeva}. Since the imaginary and real parts of the permittivity are connected through the Hilbert transform (Kramers-Kronig relations), it is sufficient to approximate any of the functions (i)-(iii) and then derive the complex multi-parametric susceptibility formula analytically. However, known approximations are often built for high-precision computation of special functions with guaranteed double-precision accuracy ($10^{-15}-10^{-13}$). Furthermore, almost all known computational algorithms use different formulas for separate segments of the complex plane where their convergence is fastest. 
    The maximum error across the entire half-plane of a specific low-order approximation is usually unacceptably large for most known algorithms, unless a \textit{minimax optimization} method is used across the entire domain.
    
    In this paper, we focus on the lowest-order polynomials that provide a relative error at practically acceptable levels (starting from few percent). Every increase in polynomial order causes substantial increases in storage and floating point operations in TD Maxwell solvers. Worse, using higher-order polynomials can result in over-fitting and nonphysical oscillations in the time domain. The "optimal" rational approximant for a given  number of poles $n$ can be found by solving a \textit{constrained minimax optimization problem}. This method gives a solution with uniform and controlled exponentially converging error across the entire spectral domain, while maintaining analytical properties (such as asymptotic behaviour at infinity). The error of the $n$-th order minimax approximation (Fig.~\ref{fig:DawsonConverge}(a)) has a typical profile with uniform $(2n-1)$ alternating peaks, according to the Chebyshev alternation theorem and the Remez algorithm. 
    
\begin{figure}[!ht]
    \center
    \includegraphics[width=0.75\textwidth]{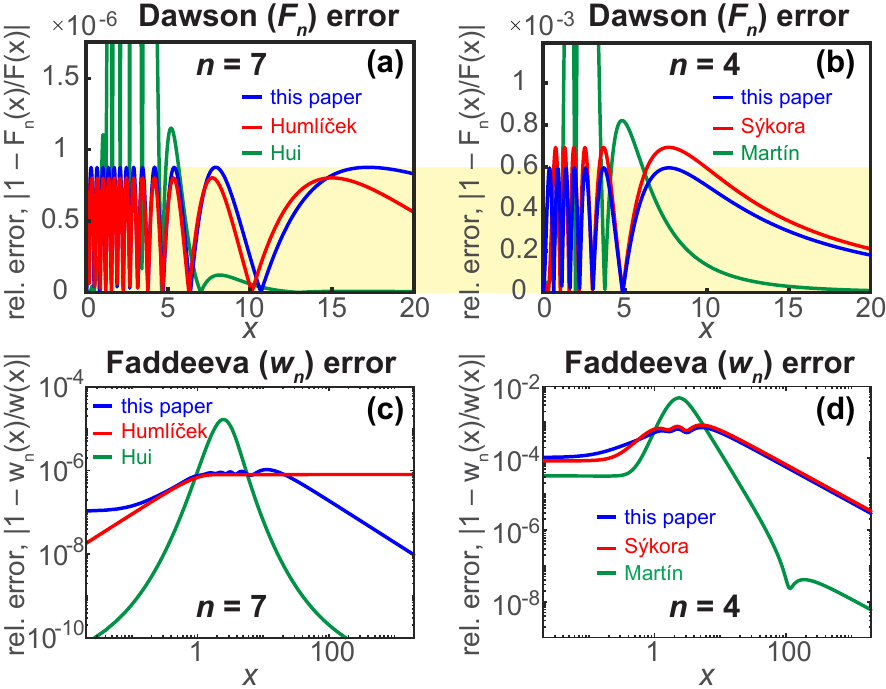}
    \caption{Comparison of the Dawson/Faddeeva rational approximation errors using our constrained minimax optimization   (Table~\ref{tab:Approx_Dawson}-\ref{tab:Approx_Faddeeva}) (\blue{solid blue line, ``this paper''}) versus known approximations to Dawson/Gauss/Faddeeva functions for $n = 7$ 
    (Left column: 
    \red{solid red line, Huml{\'\i}{\v{c}}ek} \cite{humlivcek1982optimized} and \green{solid green line, Hui} \cite{hui1978rapid}), 
    and $n = 4$, 
    (Right column: \red{solid red line, S{\`y}kora} \cite{sykora4dawson} and \green{ solid green line, Mart{\'\i}n} \cite{martin1980modified}). Conversion between the approximation coefficients for Dawson/Faddeeva functions is done using Eqs.~(\ref{eq:Dawson_poles}-\ref{eq:Faddeeva_poles}). The orders $n = \{7, 4\}$ are chosen to match the published data. }
    \label{fig:Appr_comparison}
    \end{figure}

    To show the benefit of minimax versus non-minimax rational approximations in the literature, we compare their relative errors, for both the Dawson and Faddeeva functions. In Figure~\ref{fig:Appr_comparison}(b,d) the green line shows the typical error behaviour of a non-minimax method, in this case, an asymptotic 2-point Pad\'e approximation method \cite{martin1980modified}. In this method, the coefficients for the plasma dispersion function are found by matching the asymptotic expansions at 0 and infinity. The convergence at infinity is better than the minimax method by many orders of magnitude, but the maximum error is unacceptably large. Similar behaviour is observed for other non-minimax methods: for example, the widely used McCabe's continued fractions method \cite{mccabe1974continued} and Abrarov's sampling methodology \cite{abrarov2018rational} require 24-25 poles to reach the same maximum error as the minimax solution with 8-9 poles. Generally, non-minimax rational approximations demonstrate exceptional performance in limited ranges, making them a good fit for segmented computational algorithms.
    
    We also reviewed known minimax rational approximations. The earliest approximations by Hastings~\cite{hastings1955approximations} and Cody~\cite{cody1970chebyshev} do not have the required analytical properties and/or have a limited range.
    Hui~\cite{hui1978rapid} built a minimax approximation for the Faddeeva function on the imaginary axis $w(\iota x)$, but on the real axis its maximum error is too large, as shown in Fig.~\ref{fig:Appr_comparison}(a,c). Humlicek~\cite{humlivcek1982optimized} built an \textit{unconstrained} minimax rational approximation for the Dawson function. The maximum error compared to our approximation is only slightly better, achieved at the cost of constant non-converging error at the infinity, see Fig.~\ref{fig:Appr_comparison}(c). This is not desirable for time-domain simulations. Finally, both Lether~\cite{lether1997constrained} and Sykora~\cite{sykora4dawson} solved the same constrained minimax optimization problem for the Dawson function for $2 \le n\le 6$. Lether finds a \textit{near}-minimax constrained solution through the modified Remez algorithm. His approximations have a slight deviation from the true constrained minimax solution presented in this paper, see Fig.~\ref{fig:AppendixFig}(a). Sykora attempts to find the constrained minimax optimum numerically, but doesn't reach the global minimum for orders $4 \le n\le 6$, see Fig.~\ref{fig:Appr_comparison}(b,d).

    As we can see from the comparison of these different approximation approaches, finding a \textit{constrained minimax rational approximation} to the Dawson function gives the lowest-order approximation for a given maximum error to the Faddeeva function on the real axis with the correct asymptotic behaviour. This approach is used in the next Section~\ref{sec:DawsonApprox}.

\subsubsection{Rational approximation of the Dawson function}\label{sec:DawsonApprox}
\noindent
We seek a rational approximation $F_n(x)$ of the Dawson function $F(x)$ that satisfies the following analytical properties of $F(x)$:
    \begin{enumerate}[label=(\roman*)]
        \item odd parity $F(-x) = -F(x)$;
        \item at $x=0$ all even derivatives vanish, $F^{(2k)}(0) = 0$ for all $k>0$;
        \item at $x=0$ the first derivative is equal to one, $F'(0)=1$;
        \item asymptotic behavior $F(x) \approx 0.5x^{-1}$ as $x\rightarrow\infty$.
    \end{enumerate}
    Such an approximation thus should have the form
    \begin{equation}
        F(x) \approx F_n(x) = x\frac{\overbrace{1}^{p_0}+p_1y + ... +p_{n-1} y^{n-1}}
        {\underbrace{1}_{q_0} + q_1y + ... + q_{n-1} y^{n-1} + (\underbrace{2p_{n-1}}_{q_n})y^n},
        \ y = x^2,\  n\geq 2,
    \label{eq:Dawson_n}
    \end{equation}
    where the coefficients are real positive numbers, $p_i,q_i>0$, and the properties (iii) and (iv) enforce constraints $p_0/q_0=1$ and $q_n = 2p_{n-1}$, respectively. The approximation order $n$ defines the order of the denominator and thus the number of poles. Since $p_i,q_i>0$, $F_n(x)$ does not have real positive zeros nor does it have real positive poles (in the general case, the number of zeros/poles satisfies Descartes' rule of signs). 
    
    The lowest order we consider is $n=2$. We skip the first-order polynomial $F_1 = x/(1+2x^2)$, since it gives a 40\% error when the coefficients are constrained by (iii)-(iv), and 25\% unconstrained. The rational polynomial $F_1(x)$ corresponds to approximating the dielectric function with one Lorentz oscillator. As we have seen earlier in Fig.~\ref{fig:01_Gauss_Osc} it can not reproduce the Gaussian lineshape accurately.
    
\begin{figure}[!ht]\center
        \includegraphics[width=0.75\textwidth]{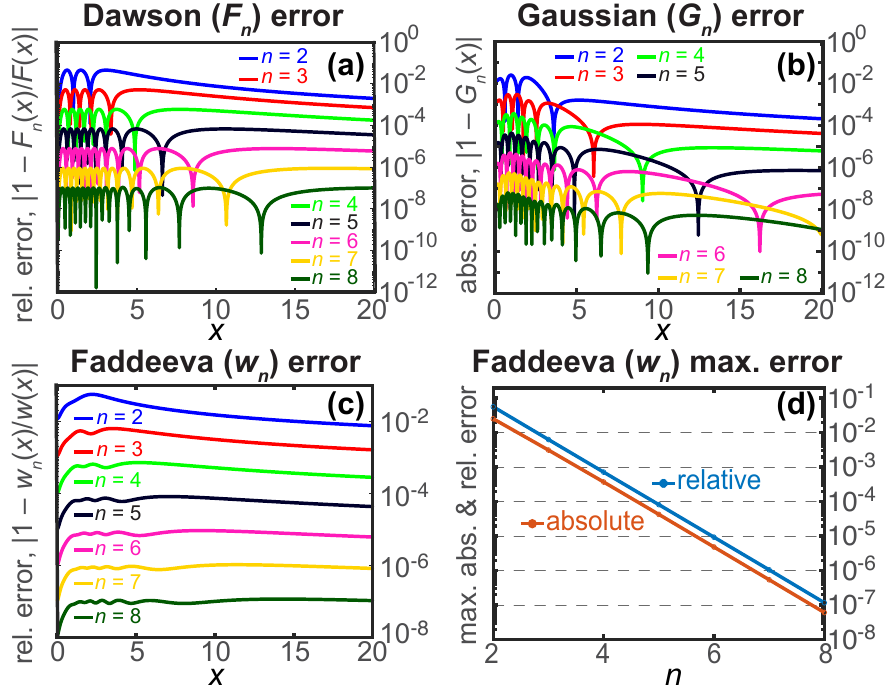}
        \caption{ (a-c) Approximation errors of the Dawson ($F_n(x)$), Gaussian ($G_n(x)$), and Faddeeva ($w_n(x)$) functions, their coefficients can be found in Tables~\ref{tab:Approx_Dawson}-\ref{tab:Approx_Faddeeva} for orders $2 \le n \le 8$; (d) convergence rate of the maximum absolute and relative approximation errors of $w_n$ vs the approximation order $n$. }
        \label{fig:DawsonConverge}
    \end{figure}

    Note that the initial problem of approximating a multi-parametric complex dielectric function $\chi(\omega;A,\Omega,\sigma)$ in~(\ref{eq:CDG_im}-\ref{eq:CDG_re})  is reduced to approximating a real function of one real variable \eqref{eq:Dawson_n}. As discussed earlier, instead of the minimizing root mean square deviation ($l_2$-norm), standard for curve fitting, we formulate the minimax optimization problem to minimize the maximum relative approximation error ($l_\infty$-norm) to find coefficients $p_i,q_i$,
    \begin{equation}
        [p_1,..,p_{n-1},q_1,..,q_{n-1}] = \argmin\limits_{p_1,..,p_{n-1},q_1,..,q_{n-1}>0}
        \max\limits_{x\ge0}
        |1-F_n(x)/F(x)|.
    \label{eq:minimax}
    \end{equation}
    This general constrained optimization problem is then solved numerically, using the MATLAB function \texttt{fminimax} over a finite range of arguments \texttt{x=linspace(0,20,1000)}. The function is based on a goal attainment algorithm \cite{gembicki1974vector}. The obtained approximations (Table~\ref{tab:Approx_Dawson}) for $2\le n\le8$ give an ideal uniform minimax profile for the error 
    (Fig.~\ref{fig:DawsonConverge}(a)) that converges exponentially for both the Dawson and Faddeeva functions (Fig.~\ref{fig:DawsonConverge}(d)).  The approximation error goes to zero for large $x$ (Fig.~\ref{fig:Appr_comparison}(c,d)) because of the constraint $q_n = 2p_{n-1}$ which matches the Dawson asymptote. The solutions found are a slight improvement over Lether's published data, $n\le6$ \cite{lether1997constrained} with his \textit{constrained near-minimax formulation} (see Fig.~\ref{fig:AppendixFig})(a)) and are a significant improvement over Sykora's published data for $n\le6$ \cite{sykora4dawson} (see Fig.~\ref{fig:Appr_comparison})(b)). Also the solutions are found for an extended range of orders $2 \le n \le 8$ (vs. $2 \le n \le 6$ in \cite{lether1997constrained, sykora4dawson}).

    The first two approximations for $n=2,3$ are the most important for its application in TD Maxwell solvers, and so we write them here for reference, 
    \begin{equation}
        \begin{array}{rcl}\label{eq:Dawson_23_pq}
            F_2(x) &=& x \, \dfrac{1 + 0.45823x^2}{1 + 0.80414x^2 + 0.91646x^4}, \\ [11pt]
            F_3(x) &=& x \, \dfrac{1 + 0.13298x^2 + 0.09960x^4 }{1 + 0.85449x^2 + 0.22599x^4 + 0.19920x^6}.
        \end{array}
    \end{equation}

    After coefficients $p_i,q_i$ are found numerically with the correct constraints, we perform partial fraction decomposition of $F_n(x)$ to $n$ individual poles, and obtain Eq.~\eqref{eq:Dawson_poles} (complex coefficients $a_i,b_i$ are listed in Table~\ref{tab:Approx_Dawson}). This pole expansion is used to restore the complex function via the Hilbert transform. However, using this form directly in the optimization problem~\eqref{eq:minimax} would make the constraints complicated.
    \begin{equation}
    F(x) \approx F_n(x) = x\sum_{i=1}^{n} \dfrac{b_i}{x^2 - a_i}
        = x\left(
        \dfrac{b_0}{x^2 - a_0} +
        2\Re\left[\sum_{i=1}^{\floor*{n/2}} 
                \dfrac{b_i}{x^2 - a_i}
            \right]
        \right)
    \label{eq:Dawson_poles}
    \end{equation}
    If the order of $n$ is \textit{odd} then the decomposition consists of complex conjugate pole pairs ($\Re[a_i],\Im[a_i],\Re[b_i],\Im[b_i]\neq 0$) with
    $a_{i+\floor*{n/2}} = \conjugatet{a_i}$ and 
    $b_{i+\floor*{n/2}} = \conjugatet{b_i}$, for $i=1,...,(n-1)/2$,  and one real pole $b_n=b_0>0$, $a_n=a_0<0$. If $n$ is \textit{even} then the real pole vanishes 
    ($b_0=0$) and the sum consists of $n/2$ conjugate pole pairs only. Poles $a_i$, as well as zeros, can be located anywhere in the complex plane except for the positive axis $x>0$ because of the constraints $p_i,q_i>0$. The forward and backward conversions between polynomial coefficients $[p_i,q_i]$ and pole expansion coefficients $[a_i,b_i]$ is done numerically with MATLAB functions \texttt{[b,a] = residue(p,q)} and \texttt{[p,q]=residue(b,a,0)}, respectively. 
    
    The corresponding approximation of the Gauss function is obtained by performing the Hilbert transform for every pole in the decomposition~\eqref{eq:Dawson_poles}
    \begin{equation}
    \Exp^{-x^2} \approx G_n(x) = -\frac{2}{\sqrt{\pi}}\mathcal{H}\left\{ F_n(x) \right\}
    = \frac{2}{\sqrt{\pi}}\sum_{i=1}^{n} \dfrac{b_i \sqrt{-a_i}}{x^2 - a_i}.
    \label{eq:Gauss_poles}
    \end{equation}
    We used the Hilbert transform pair $x/(x^2+c_i^2) \, \rightarrow \, -c_i/(x^2+c_i^2)$, where $c_i = \sqrt{-a_i}$ is a \textit{principal branch} square root. The transform condition $\Re[c_i]>0$ is satisfied since the roots of the minimax Dawson solution $a_i$ are either real and negative or complex with a nonzero imaginary part (see Table~\ref{tab:Approx_Dawson}). The algebraic structure of the Gaussian approximation~\eqref{eq:Gauss_poles} is the same as the Dawson approximation~\eqref{eq:Dawson_poles}. If $n$ is odd then the sum has exactly one real negative pole $a_n<0$ with a positive residue $b_n\sqrt{-a_n}>0$ and the rest of the poles are complex conjugate pairs. If $n$ is even then the sum consists of conjugate pole pairs only (see Table~\ref{tab:Approx_Gauss}). The poles $a_i$ are the same and can not be real positive numbers. In contrast to the Dawson approximation, the zeros of the Gaussian approximation can not be avoided on the positive real axis in this approach. However, the minimum negative value of the Gaussian approximation converges to zero exponentially with order $n$. To illustrate this, we plot the logarithm of the additive inverse of the approximation $\log[-G_n(x)]$, so that only the ``criminal'' negative values of $G_n(x)$ show up for each order $n$ in the plot (while positive values go to NaN and are not displayed), see Fig.~\ref{fig:AppendixFig}(b). We can see that even approximation orders $n$ (without the Lorentzian center) result in infinitely long negative tails, while odd orders $n$ (with the Lorentzian center) produce finite ranges of negative values.
    
    Summation of \eqref{eq:Dawson_poles} and \eqref{eq:Gauss_poles} gives the corresponding rational approximation of the Faddeeva function
    \begin{equation}
    w(x) \approx w_n(x) = G_n(x) + \frac{2\iota}{\sqrt{\pi}}F_n(x) 
    = \frac{2}{\sqrt{\pi}}\sum_{i=1}^{n} \dfrac{b_i}{(-\iota x) + \sqrt{-a_i}}.
    \label{eq:Faddeeva_poles}
    \end{equation}
    During this summation, the opposite pole $[(-\iota x) - \sqrt{-a_i}]$ cancels out. As a result, all the poles and zeros of the approximation $w_n(x)$ are located in the left half-plane (check Table~\ref{tab:Approx_Faddeeva}). In other words, if we combine the poles back into a single rational polynomial $w_n(\iota s) = P_n(s)/Q_n(s)$, then the resulting polynomials $P_n(s), Q_n(s)$ are Hurwitz polynomials. This property is important for the stability of FDTD schemes (and other TD solvers). Even though the necessary positive-realness condition is met, the function $w_n(\iota s)$ is not a PRF. For real arguments $s$, the function $w_n(\iota s)$ is real. But for complex arguments $s$ with a positive real part $\Re[s]>0$, the real part of the function $w_n(\iota s)$ can achieve small negative values. It means that the dielectric function $\varepsilon_n(\omega)$ defined by Eq.~\eqref{eq:Faddeeva} can also achieve small negative values. However, the minimum negative value is proportional to the approximation error and goes to zero exponentially as the approximation order $n$ increases, see Fig.~\ref{fig:AppendixFig}(c). In that sense, the derived approximation $w_n(\iota s)$ can be called \emph{asymptotically positive real}. Real materials usually have non-zero Lorentzian broadening ($\Gamma>0$) in addition to the inhomogeneous broadening ($\sigma>0$) which provides a shift $\omega \rightarrow \omega+\iota\Gamma$ in the complex plane and compensates for the small negative absorption values. Also, one can add numerical dissipation in the time-domain scheme, if needed, to cancel out these negligible negative absorption values. 
    
    Alternatively, approximations with strictly positive absorption can be built by posing a minimax optimization problem directly on Gaussian function. For example, the first two approximations, $G_2(x)\approx \exp^{-x^2}$ and $G_3(x)\approx \exp^{-x^2}$, in this alternative approach read
    \begin{equation}
        \begin{array}{rcl}\label{eq:Gauss_23_pq}
            G_2(x) &=& \dfrac{1}{1 + 0.68488x^2 + 1.227526x^4}, \\ [11pt]
            G_3(x) &=& \dfrac{1}{1 + 1.121306x^2 + 0.03068x^4 + 0.546042x^6}.
        \end{array}
    \end{equation}
    These approximations, however, do not converge exponentially and give significantly larger approximation errors than the Dawson constraint minimax approach used in this paper.
    
    The resulting pole expansions for Dawson/Gauss/Faddeeva functions and all final coefficients for TD simulations are summarized in Tables~\ref{tab:Approx_Dawson}-\ref{tab:Approx_Faddeeva} for $2\le n\le8$. These coefficients can be used more broadly for calculating the Dawson/Faddeeva functions with a guaranteed accuracy of $10^{-7}-10^{-3}$ depending on the order $n$, as indicated in (Fig.~\ref{fig:DawsonConverge}(d)). The polynomial order $n$ should be chosen depending on the required accuracy and computational complexity. The approximation $n=2$ gives a relative error on the order of few percent, which in practice is comparable to experimental error much of the time. Higher precision, $<$1\%,
    is achievable with three poles ($n=3$). The maximum number of poles calculated was $n=8$, yielding an approximation with single machine precision, 1e-7.
    
    Note that the constrained minimax solutions were found \textit{numerically} for relatively low orders $2\le n \le 8$ using standard global optimization techniques within MATLAB. The results demonstrated an ideal minimax error profile with uniform peaks (Fig.~\ref{fig:DawsonConverge}(a)). An ideal exponential convergence rate of the maximum error with the polynomial order $n$ is verified in Fig.~\ref{fig:DawsonConverge}(d) and confirms that the mimimax solutions were found correctly.
        
    We speculate that finding exponentially converging constrained minimax solutions (either numerically and/or using a modified Remez algorithm) is possible for higher orders ($n > 8$). According to the convergence figure~\ref{fig:DawsonConverge}(a), double precision accuracy ($10^{-15}$) is expected for $n=16$ poles, but extended precision arithmetics and advanced fraction decomposition algorithms may be required. For example, in the derivation of the Faddeeva function approximation, we used numerical partial fraction decomposition (with MATLAB function \texttt{residue}). Numerical partial fraction decomposition is a known ill-posed problem and likely to become inaccurate as the error approaches computer round-off for larger orders and thus special care is needed for larger orders $n$.

    \subsubsection{GDM formula for Gaussian absorption}\label{sec:GDMformula} 
    \noindent
    Before deriving the GDM formula, we summarize the previous nomenclature.
    First, we have the Gaussian parameters [$A$, $\Omega$, $\sigma$] --- the absorption maximum, center, and width --- that define the Gaussian susceptibility function $\chi_\text{G}(\omega)$ according to Eq.~(\ref{eq:CDG_im}-\ref{eq:CDG_re}). From the time-domain formula~\eqref{eq:chi_LGt}, we see that the absorption amplitude parameter $A$ can be swapped with the time-domain amplitude parameter $a$ using a linear relation $a = 2A\sigma/\sqrt{\pi}$,
    \begin{equation}
        \chi_\text{G}(t) = 
            \underbrace{\frac{2A\sigma}{\sqrt{\pi}}}_{a} \Exp^{-t^2\sigma^2/4}
            \sin(\Omega t) \theta(t).
    \label{eq:GaussParam_lemma}
    \end{equation}
    Second, we have fixed approximation constants $a_i,b_i$ (see Table~\ref{tab:Approx_Dawson}) that define a rational polynomial approximation of the Dawson function for a given number of poles $n$, 
    \begin{equation}
        F(x) \approx F_n(x) = x\sum\limits_{i=1}^n \frac{b_i}{x^2-a_i}.
    \label{eq:DawsonApprox_lemma}
    \end{equation}
    Substituting the Dawson function with its rational approximation \eqref{eq:DawsonApprox_lemma} in the CDG model~(\ref{eq:CDG_im}-\ref{eq:CDG_re}), gives the following causal approximation of the susceptibility function suitable for stable FDTD simulations (Lemma 1).
    
    \newcommand{\subG}{{\normalfont\text G}}
    \newcommand{\subc}{{\normalfont\text c}}
    \begin{lemma}[General oscillator formula for Gaussian absorption]\label{lemma1}
    The Gaussian absorption model can be approximated with a sum of phase-relaxed damped oscillators in the time domain
    \begin{align}\label{eq:lemma1A}
    \begin{split}
        \chi_\subG(t; a,\Omega,\sigma) = a\Exp^{-t^2\sigma^2/4} \sin(\Omega t) \theta(t) & \\
        \approx 
            \sum\limits_{i=1}^n a_\subc^i\Exp^{-\Gamma_\subc^i t} \sin(\Omega^i_\subc t - \varphi_\subc^i) \theta(t) &
        = \chi_\subG^n(t; a,\Omega,\sigma).
    \end{split}
    \end{align}
    The latter can be equivalently written in the frequency domain
    \begin{align}\label{eq:lemma1B}
    \begin{split}
        \chi_{\normalfont\text G}(\omega; a,\Omega,\sigma) =  
\frac{a}{\sigma} \left[ F\left(\frac{\omega+\Omega}{\sigma} \right) - F\left(\frac{\omega-\Omega}{\sigma} \right) \right] + \\
\frac{\iota a\sqrt{\pi}}{2\sigma}\left[ \Exp^{-(\omega - \Omega)^2 / \sigma^2} - \Exp^{-(\omega + \Omega)^2 / \sigma^2} \right] \approx \\
            \sum\limits_{i=1}^n a_\subc^i\frac{\Omega_\subc^i\cos\varphi_\subc^i + \iota(\omega+\iota\Gamma_\subc^i)\sin\varphi_\subc^i}{(\Omega_\subc^i)^2-(\omega+\iota\Gamma_\subc^i)^2} =\\
            \sum\limits_{i=1}^n \frac{a_\subc^i}{2} \left[ \frac{\Exp^{-\iota\varphi_\subc^i}}{\omega+\iota\Gamma_\subc^i+\Omega_\subc^i}
            -\frac{\Exp^{ \iota\varphi_\subc^i}}{\omega+\iota\Gamma_\subc^i-\Omega_\subc^i}     \right]
        = \chi_{\text G}^n(\omega; a,\Omega,\sigma).
    \end{split}
    \end{align}
    The oscillator parameters 
    [$a_\subc^i$, $\Omega_\subc^i$, $\Gamma_\subc^i$, $\varphi_\subc^i$] 
    are explicit functions of Gaussian parameters 
    [$\color{red}a$, $\color{red}\Omega$, $\color{red}\sigma$] and approximation constants [$a_i$,$b_i$] from Table~\ref{tab:Approx_Dawson}:
    \begin{equation}
    \begin{array}{rcll} 
        a_\subc^i       &=& 2 |b_i|{\color{red}a}, & \quad\quad[amplitude] \\ [0.15cm]
        \Gamma_\subc^i  &=& \Re[\sqrt{-a_i}]{\color{red}\sigma}, & \quad\quad[broadening] \\ [0.15cm]
        \Omega_\subc^i  &=& \Im[\sqrt{-a_i}]{\color{red}\sigma} + {\color{red}\Omega}, & \quad\quad[center] \\ [0.15cm]
        \varphi_\subc^i &=& \arg[b_i]. & \quad\quad[phase] \\ [0.15cm]
    \end{array}        
    \label{eq:lemma1C}
    \end{equation}
    \end{lemma}
    \noindent
    Here $\arg[z]$ is an argument function of a complex number $z$ (can be computed using \texttt{angle(z)} or \texttt{atan2(imag(z),real(z))} in MATLAB). The square root denotes the principal branch ($\Re(\sqrt{z})\ge0$). One way of proving Lemma 1 is to substitute the Faddeeva function with its approximation~\eqref{eq:Faddeeva_poles} into the frequency domain susceptibility formula~\eqref{eq:Faddeeva}, and then use formulas~\eqref{eq:GDM_FD2TD} to find the time-domain oscillator parameters.

\begin{figure}[!ht]
    \center
    \includegraphics[width=0.99\textwidth]{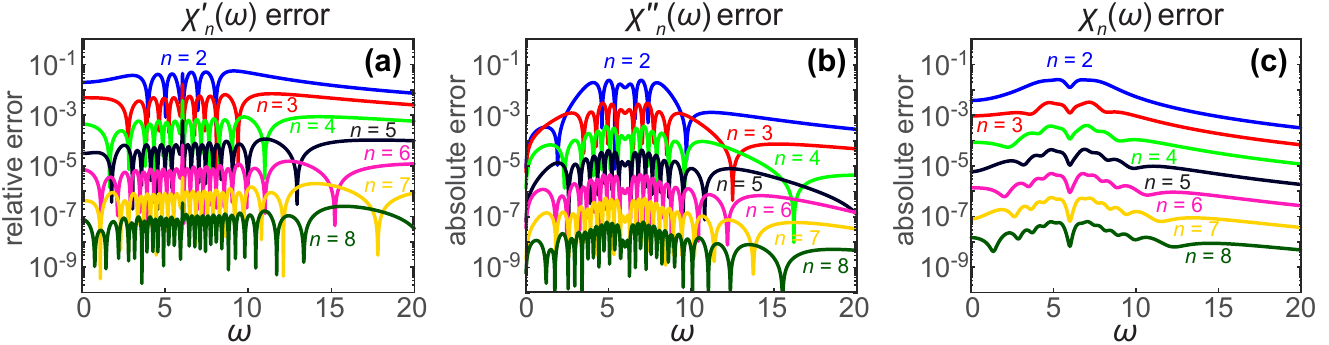}
    \caption{Susceptibility function approximation errors for orders $2\le n \le 8$. Parameters $A=0.5$, $\Omega=6$ and $\sigma=1$ are used for the plots.   }
    \label{fig:GDMerros_n}
\end{figure}

    In Figure~\ref{fig:GDMerros_n} we perform an approximation error check for the complex susceptibility function obtained in the Lemma~1. As expected, the error converges exponentially with approximation order $n$. 
        
    Lemma 1 together with Table~\ref{tab:Approx_Dawson} of coefficients $[a_i, b_i]$ gives a GDM representation~\eqref{eq:GDMfinal} of the Gaussian dispersion. More specifically, the GDM parameters $[a_{0,i}, a_{1,i}, b_{0,i}, b_{1,i}]$ can be found as simple expressions of the oscillator parameters $[a_\text{c}^i, \Gamma_\text{c}^i, \Omega_\text{c}^i, \varphi_\text{c}^i]$ using formulas~\eqref{eq:GDM_TD2FD}. This GDM model can be seamlessly implemented in TD solvers using either Auxiliary Differential Equation (ADE) or Recursive Convolution (RC) implementations for each oscillator. Numerical implementation of arbitrary dispersion in GDM form is presented in detail in Section~\ref{sec:FDTD_GDM} for the FDTD method. Obtained polarization recursions can be also coupled to other TD Maxwell solvers.
    
    It is important that the parameters of the approximation model are derived analytically. In contrast to models generated via mathematical fitting of multivariate curves containing many Gaussian terms, this analytical approach does not require a new fitting procedure for every set of parameters. This allows us to retain connections with the original physics of the Gaussian profile as well as use Gaussian models already available from VASE measurements in the time domain. 

    The lowest order approximations, $n=2$ and $n=3$, in particular provide an efficient, stable, and accurate technique for simulating Gaussian dispersion in the time domain with just 2 and 3 oscillators respectively. For the reader's convenience, these cases are formulated as separate corollaries of Lemma~1 with included constants.
    
    \begin{corollary}[A][Two-oscillator model for Gaussian absorption]
    The Gaussian absorption model in 
    Eqs.~(\ref{eq:CDG_im},\ref{eq:CDG_re},\ref{eq:chi_LGt}) can be approximated within a few percent error with two symmetrically detuned coupled oscillators in the time domain
    \begin{align}\label{eq:corA1}\nonumber
         \chi_\subG(t; {\color{red}a},{\color{red}\Omega},{\color{red}\sigma}) \approx
         \alpha_1  {\color{red}a}
            \Exp^{-\gamma_1 {\color{red}\sigma} t}
            \Bigg[ 
            & \sin(({\color{red}\Omega} + \omega_1{\color{red}\sigma})t - \varphi_1) + \\
            &  \sin(({\color{red}\Omega} - \omega_1{\color{red}\sigma})t +\varphi_1)
            \Bigg]
             \theta(t) = \chi_\subG^2(t; {\color{red}a},{\color{red}\Omega},{\color{red}\sigma}).
    \end{align}
    The latter can be equivalently written in the frequency domain
    \begin{align}\label{eq:corA2}\nonumber
         \chi_\subG(\omega; {\color{red}a},{\color{red}\Omega},{\color{red}\sigma}) \approx 
          \alpha_1 {\color{red}a}
          \Bigg[ 
          \frac
          {({\color{red}\Omega} + \omega_1{\color{red}\sigma})\cos\varphi_1 + \iota(\omega+\iota
          \gamma_1 {\color{red}\sigma})\sin\varphi_1}
          {({\color{red}\Omega} + \omega_1{\color{red}\sigma})^2-(\omega+\iota\gamma_1 {\color{red}\sigma})^2} +\\ \nonumber
          \frac
          {({\color{red}\Omega} - \omega_1{\color{red}\sigma})\cos\varphi_1 - \iota(\omega+\iota
          \Gamma_1 {\color{red}\sigma})\sin\varphi_1}
          {({\color{red}\Omega} - \omega_1{\color{red}\sigma})^2-(\omega+\iota\gamma_1 {\color{red}\sigma})^2}
          \Bigg] =\\ \nonumber
         \alpha_1 \frac{\color{red}a}{2}
          \Bigg[
             \frac{\Exp^{-\iota\varphi_1}}{\omega+\iota\gamma_1 {\color{red}\sigma}+({\color{red}\Omega} + \omega_1{\color{red}\sigma})}
            -\frac{\Exp^{ \iota\varphi_1}}{\omega+\iota\gamma_1 {\color{red}\sigma}-({\color{red}\Omega} + \omega_1{\color{red}\sigma})} + \\ \nonumber
             \frac{\Exp^{ \iota\varphi_1}}{\omega+\iota\gamma_1 {\color{red}\sigma}+({\color{red}\Omega} - \omega_1{\color{red}\sigma})}
            -\frac{\Exp^{-\iota\varphi_1}}{\omega+\iota\gamma_1 {\color{red}\sigma}-({\color{red}\Omega} - \omega_1{\color{red}\sigma})}
          \Bigg]\\
        = \chi_\subG^2(\omega; {\color{red}a},{\color{red}\Omega},{\color{red}\sigma}).
    \end{align}
    \noindent
    Here the approximation constants are 
    $\alpha_1  = 2|b_1| = 1.046764$, 
    $\gamma_1  = \Re[\sqrt{-a_1}] = 0.861192$,
    $\Omega_1  = \Im[\sqrt{-a_1}]= 0.550392$, 
    $\varphi_1 = \arg[b_1] = 1.072804$.  
    \end{corollary}
    \noindent
    Note that the initial condition of the envelope $\left.\chi_\subG^2(t)\sin^{-1}(\Omega t)\right|_{t\rightarrow 0^+} = a$ is satisfied since $4\Re[b_1]=1$. 
    
    \begin{corollary}[B][Three-oscillator model for Gaussian absorption]
    The Gaussian absorption model in 
    Eqs.~(\ref{eq:CDG_im},\ref{eq:CDG_re},\ref{eq:chi_LGt})
    can be approximated with  $<$1\% error with a central Lorentz oscillator and two coupled oscillators in the time domain
    \begin{align}\label{eq:corB1}\nonumber
         \chi_\subG(t; {\color{red}a},{\color{red}\Omega},{\color{red}\sigma}) \approx 
          {\color{red}a} 
            \Bigg[ 
            & \alpha_0\Exp^{-\gamma_0 {\color{red}\sigma} t} \sin({\color{red}\Omega}t) + \\ \nonumber
            & \alpha_1\Exp^{-\gamma_1 {\color{red}\sigma} t} \sin(({\color{red}\Omega} + \omega_1{\color{red}\sigma})t - \varphi_1) + \\
            &  \alpha_1\Exp^{-\gamma_1 {\color{red}\sigma} t} \sin(({\color{red}\Omega} - \omega_1{\color{red}\sigma})t +\varphi_1)
            \Bigg]
             \theta(t) = \chi_\subG^3(t; {\color{red}a},{\color{red}\Omega},{\color{red}\sigma}).
    \end{align}
    The latter can be equivalently written in the frequency domain
    \begin{align}\label{eq:corB2}\nonumber
         \chi_\subG(\omega; {\color{red}a},{\color{red}\Omega},{\color{red}\sigma}) \approx 
           {\color{red}a}
          \Bigg[ 
          &\alpha_0\frac
          {{\color{red}\Omega}}
          {{\color{red}\Omega}^2-(\omega + \iota\gamma_0{\color{red}\sigma})^2} +\\ \nonumber
          &\alpha_1\frac
          {({\color{red}\Omega} + \omega_1{\color{red}\sigma})\cos\varphi_1 + \iota(\omega+\iota
          \gamma_1 {\color{red}\sigma})\sin\varphi_1}
          {({\color{red}\Omega} + \omega_1{\color{red}\sigma})^2-(\omega+\iota\gamma_1 {\color{red}\sigma})^2} +\\ \nonumber
          &\alpha_1\frac
          {({\color{red}\Omega} - \omega_1{\color{red}\sigma})\cos\varphi_1 - \iota(\omega+\iota
          \gamma_1 {\color{red}\sigma})\sin\varphi_1}
          {({\color{red}\Omega} - \omega_1{\color{red}\sigma})^2-(\omega+\iota\gamma_1 {\color{red}\sigma})^2}
          \Bigg] =\\ \nonumber
        = \chi_\subG^3(\omega; {\color{red}a},{\color{red}\Omega},{\color{red}\sigma}).
    \end{align}
    \noindent
    Here the approximation constants are 
    $\alpha_0  = 1.7253301$, 
    $\gamma_0  = 1.0778291$,
    $\alpha_1  = 0.6444771$, 
    $\gamma_1  = 1.0161543$,
    $\omega_1  = 1.02283141$, 
    $\varphi_1 = 2.16847802$. 
    \end{corollary}
    \noindent
    As before, the initial condition of the envelope $\left.\chi_\subG^2(t)/\sin(\Omega t)\right|_{t\rightarrow 0^+} = a$ is satisfied since $2\sum_{i=1}^3 b_i=1$.  

    Lemma 1 (for arbitrary $n$) gives a powerful universal algorithm to adapt any inhomogeneous broadening profile, described by a Lorentz model convoluted with an arbitrary symmetric (with zero skewness) probability density function (PDF) with a given variance $\sigma$, for time-domain simulations. Oscillator parameters~\eqref{eq:lemma1C} in Lemma 1 can be further analyzed depending if the pole is real or complex. 
    
    \textbf{Lorentz oscillator}. In case of a real pole, which is present only for odd $n$ and has the 0-th index by notation, see Eq.~\eqref{eq:Dawson_poles}, $a_0<0$, $b_0>0$, we obtain a classic damped Lorentz oscillator with zero phase $\varphi_\text{c}^0=0$ and aligned to the Gaussian center $\Omega_\text{c}^0 = \Omega$. 
    
    \textbf{Pair of Coupled Oscillators (CO)}. In case of a complex pole, there exist a matching conjugate pair, (indexed by $i$, $i+\floor*{n/2}$, as in Eq.~\eqref{eq:Dawson_poles}): $a_{i+\floor*{n/2}} = \conjugatet{a_i}$, $b_{i+\floor*{n/2}} = \conjugatet{b_i}$. 
    The parameters of the two coupled oscillators, $i$ and $i+\floor*{n/2}$, are not independent, both oscillators share:
    \begin{itemize}
    \item identical amplitude, $a_\text{c}^i=a_\text{c}^{i+\floor*{n/2}}=2 |b_i|a$, defined by the residue constant $|b_i|$ and scaled by the Gaussian amplitude parameter $a$;
    \item identical broadening $\Gamma_\text{c}^i=\Gamma_\text{c}^{i+\floor*{n/2}}=
    \Re[\sqrt{-a_i}]\sigma$, defined by the constant real part of the pole and scaled by the Gaussian broadening $\sigma$;
    \item symmetric oscillation frequencies, 
    $\Omega_\text{c}^{i,i+\floor*{n/2}}=
    \Omega \pm \Im[\sqrt{-a_i}]\sigma$,
    equally offset to the right and to the left from the Gaussian center $\Omega$ and spread apart with Gaussian broadening $\sigma$ normalized by the constant imaginary part of the pole;
    \item reversed phase, $\varphi_\text{c}^i=-\varphi_\text{c}^{i+\floor*{n/2}}=\arg[b_i]$, defined by the constant residue phase that does not depend on the Gaussian parameters [$a$, $\Omega$, $\sigma$].
    \end{itemize}
    From now on, the above system of two oscillators arising from conjugate poles of the Dawson approximation and formulas~\eqref{eq:lemma1C} is called a \textit{Coupled Oscillator} (CO) pair. We also use the notation CO$^-$ and CO$^+$ to indicate the left-shifted and right-shifted oscillator in the pair. 
    
    Using the sum-to-product trigonometric identity, the two coupled oscillators can be combined into one oscillator with a slow varying cosine envelope (typically $|\Im[\sqrt{-a_i}]|\sigma \approx \sigma << \Omega $) 
    \begin{equation}
        \sum\limits_{i\in\{j,j+\floor*{n/2}\}} 
        a_\subc^i\Exp^{-\Gamma_\subc^i t} \sin(\Omega^i_\subc t - \varphi_\subc^i) = 
        2 a_\subc^i\Exp^{-\Gamma_\subc^i t}
        {\color{red} \cos\left(\Im[\sqrt{-a_i}]\sigma t -\arg[b_i]\right)}
        \sin\left(\Omega t \right).
    \label{eq:COpair}
    \end{equation}
    The difference between the two oscillators, the classical Lorentz and CO pair~\eqref{eq:COpair}, is a slowly-varying cosine modulation factor (shown in red) that is capable of efficiently approximating the Gaussian lineshape even for the smallest order $n=2$. We can also convert the oscillator~\eqref{eq:COpair} into exponential notation. The result of this analysis is formulated as Lemma 2 below. 
    \begin{lemma}[Single oscillator approximation for Gaussian absorption]\label{lemma2}
    Approximation of the Gaussian absorption model from Lemma 1 can be written in the single oscillator form
    \begin{align}\label{eq:lemma2}
    \begin{split}
        \chi_\subG(t; a,\Omega,\sigma) = \left[\Exp^{-t^2{\color{red}\sigma}^2/4}\right] a\sin(\Omega t) \theta(t) & \\
        \approx 
            \left[
            2 \sum\limits_{i=1}^n b_i\Exp^{-{\color{red}\sigma}t\sqrt{-a_i}} 
            \right]
            a \sin(\Omega t) \theta(t) &
        = \chi_\subG^n(t; a,\Omega,\sigma),
    \end{split}
    \end{align}
    where constant coefficients $a_i, b_i, i\in\overline{1,n}$ are listed in Table~\ref{tab:Approx_Dawson}.
    \end{lemma}
    \noindent
    Here $a_i,b_i$ are either real with $a_i<0$ or have a conjugate pair and thus the expression in square brackets is always real. Also, because of the minimax constraint  $F_n(x)|_{x\rightarrow\infty}\approx 0.5x^{-1}$, we have $2\sum\limits_{i=1}^n b_i = 1$, and thus the TD initial condition $\left.\chi_\subG^n(t)\sin^{-1}(\Omega t)\right|_{t\rightarrow 0^+} = a$ is always satisfied.
    
    Lemma 2 provides a simple pathway to deriving the time domain compatible approximation that can be applied to any model with inhomogeneous broadening. For example, Gaussian broadening in the time domain is simply substituted with complex exponentials each corresponding to a pole in the minimax approximation of the Dawson function (the Hilbert transform of the Gaussian function). 
    
    This relation also proves that the proposed approximation does not generate random, non-physical oscillations in the time domain, even for large approximation orders $n$. In fact, for any approximation order, the model represents a single oscillation with a slow-varying envelope that depends solely on parameter $\sigma$ and approximates the quadratic Gaussian lineshape with a complex exponential series. This is possible because of the correctly derived analytical form of the susceptibility function (as opposed to fitting). Every oscillator in the approximation is either (i) Lorentz at frequency $\Omega$, or (ii) a coupled oscillator pair that effectively behaves as Lorentz at frequency $\Omega$ modulated with a slow cosine envelope~\eqref{eq:COpair}.
    
\begin{figure}[!ht]
    \center
    \includegraphics[width=0.99\textwidth]{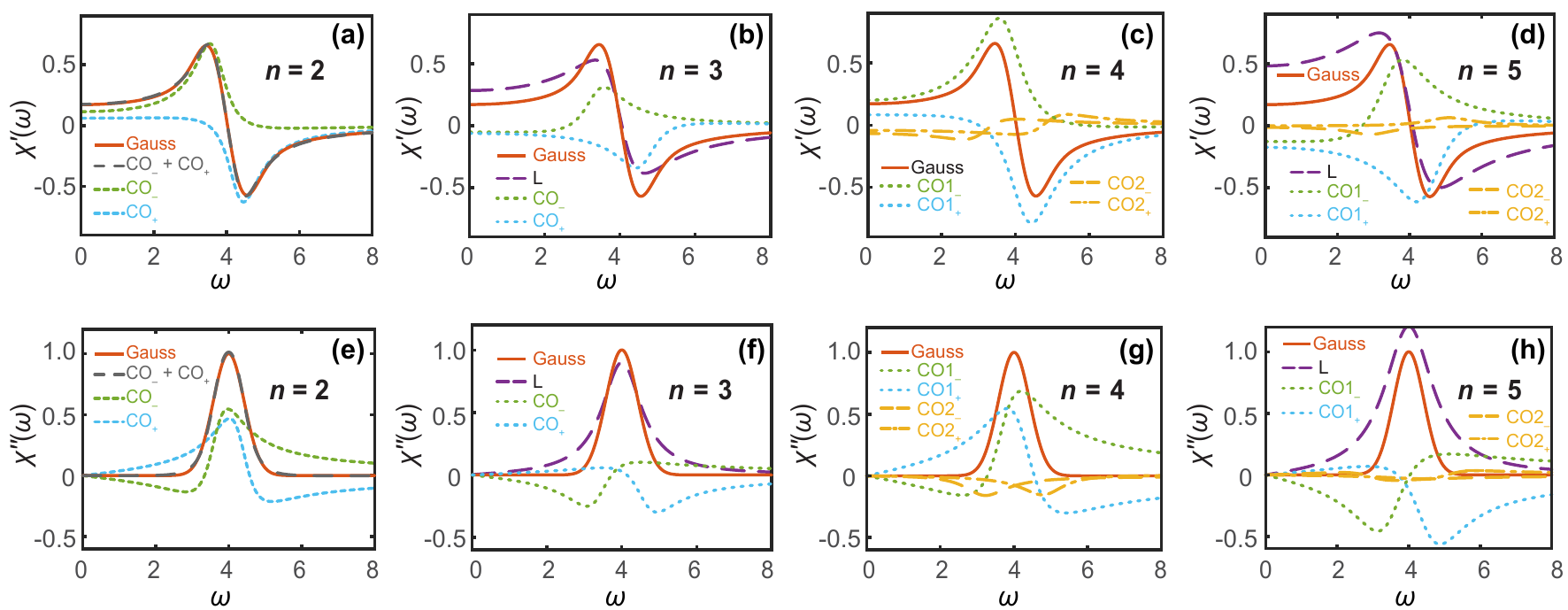}
    \caption{Decomposition of the Gaussian oscillator into: 
    (a,e) 2~oscillators --- a single CO pair ($\text{CO}_{\pm}$);
    (b,f) 3~oscillators --- Lorentz and a CO pair ($\text{L}$, $\text{CO}_{\pm}$);
    (c,g) 4~oscillators --- two CO pairs ($\text{CO1}_{\pm}$, $\text{CO2}_{\pm}$);
    (d,h) 5~oscillators --- Lorentz and two CO pairs ($\text{L}$, $\text{CO1}_{\pm}$, $\text{CO2}_{\pm}$).
    Parameters $A=1$, $\Omega=4$ and $\sigma=0.6$ are used for the plots.}
    \label{fig:GDMapprox_n}
\end{figure}

    Finally, in Figure~\ref{fig:GDMapprox_n} we plot the contribution of each individual oscillator to the approximation vs. the exact Gaussian susceptibility, and show the changes in the oscillator ensemble vs. increasing approximation order (for $n=2,3,4,5$).
    When $n=2$, the right- and left-shifted oscillators seem to be uniquely defined by the negative and positive peaks (respectively) in the real part, without cancelling each other, and thus the model is unlikely to have overfitting problems. The two asymmetric absorption terms sum up to produce a symmetric profile approximating a nearly-Gaussian absorption distribution. 
    For $n=3$, the addition of the Lorentz oscillator, exactly aligned to the Gaussian center, tremendously improves the accuracy from slightly noticeable deviation (few percent error) to indiscernible curves ($<$1\% error). In this case, the central Lorentz oscillator becomes the leading contribution, while the CO pair add missing correction terms for the curve in the peak as well as in the tails. 
    At $n=4$, the two right-shifted oscillators (as well as the two left-shifted ones) have opposite signs and thus partially balance each other to achieve the required curve as a sum. 
    This overcompensation trend becomes even more pronounced when $n=5$ --- the central Lorentz absorption peak increases in the positive direction while the CO pair grows in the negative direction. For $n>5$ the amplitude of the largest term grows with the approximation order even further, thus requiring significant compensation from the additional terms. 
    
    As we show in the simulation section, the higher-order approximations provide mathematical representations of the Gaussian absorption lines in the time domain with machine precision and are stable in TD simulations. Nonetheless, these approximations are unnecessary complicated, computationally expensive, and do not correspond to any simple physical models, compared to the simpler $n=2$ and $n=3$ cases. From both the physical and computational point of view, 2- and 3- oscillator models are the most appealing and likely to be used for engineering and characterization of realistic macroscopic optical responses in media with disorder.

    For an arbitrary, $N$-term Gaussian absorption model with parameters $A_j$, $\Omega_j$, $\sigma_j$, ($j=1,N$), the GDM coefficients, $a_{0,i}$,$a_{1,i}$,$b_{0,i}$,$b_{1,i}$, ($i=1,nN$) and the resulting dielectric function $\veps(\omega) = \sum_{i=1}^{nN} (a_{0,i} -\iota\omega a_{1,i})/(b_{0,i} -\iota\omega b_{1,i} - \omega^2)$ for a given approximation order $n$ can be calculated using the MATLAB function 
    \texttt{[Eps,a0,a1,b0,b1] = Gauss\_n(n,w,A,W,sig)} 
    included in our software package MADIS.

\subsection{FDTD-GDM method for light propagation in dispersive media}\label{sec:FDTD_GDM}
\noindent
In this section, we extend Yee's classical FDTD scheme \cite{yee1966numerical} for solving Maxwell equations \eqref{eq:Maxwell_TD} to the general dispersive case, where an arbitrary dielectric function is given in the GDM form \eqref{eq:GDMfinal}. 
These dispersive FDTD-GDM schemes were originally published in \cite{prokopeva2011optical} and \cite{prokopeva2020time}. In this section we present a brief recap with detailed reference to the codes.

We begin by constructing explicit numerical schemes that recursively calculate the local response $\Pv_i = \veps_0\Ev(a_{0,i} -\iota\omega a_{1,i})/(b_{0,i} -\iota\omega b_{1,i}-\omega^2)$ in the time domain
from the known electric field $\Ev(t)$. Different schemes can be derived with either 
(a) finite-difference approximation of the differential equation on $\Pv_i$ -- this method is called the Auxiliary Differential Equation (ADE) method, see Section~\ref{sec:ADE}, or
(b) numerical integration of the convolution integral -- this method is called Recursive Convolution (RC) method, see Section~\ref{sec:RC}. Here both ADE and RC schemes are implemented with a \textit{universal scheme}, where a table of coefficients is built for 6 different second-order RC and ADE schemes and can be extended to other schemes. The last step in Section~\ref{sec:universal} couples the obtained universal recursion on $\Pv_i$ to a numerical solution of Maxwell equations through the full polarization vector 
$\Pv = \sum_i \Pv_i$ with a \textit{compact universal scheme}.

In sections \ref{sec:ADE}-\ref{sec:RC}, the goal is to write an explicit scheme for the recursive calculation of each $\Pv_i$ in the GDM sum \eqref{eq:GDMfinal} independently. For simplicity of notation the index $i$ is omitted in these sections and only one GDM term is assumed, 
\begin{equation}\label{eq:GDM_1}
    \Pv = \veps_0 \Ev \frac{a_0 -\iota\omega a_1}{b_0 -\iota\omega b_1-\omega^2}.
\end{equation}
In section \ref{sec:universal}, where the full system of numerical equations is derived, we return to the full multi-term notation.

\subsubsection{Auxiliary Differential Equation (ADE) schemes}\label{sec:ADE}
\noindent
The ADE method of incorporating dispersion can be used when the dielectric function gives an immediate ordinary differential equation (ODE) for the polarization terms. In the case of the single-term GDM model, the polarization equation~\eqref{eq:GDM_1}, after substituting $-\iota\omega$ with time derivative $\dot f(t,\xv) = \partial f /\partial t$, gives a second-order ODE
\begin{equation}\label{eq:Pi_ODE}
    \ddot\Pv + b_1\dot\Pv + b_0\Pv = \veps_0 \left( a_1\dot\Ev + a_0\Ev \right).
\end{equation}
To solve \eqref{eq:Pi_ODE} numerically we introduce the finite time step $\tau$ and use the discrete function notation $f^n \approx f(n\tau,\xv)$ that approximates the solution. We seek a second-order accurate scheme, since $\Pv^n$ will be coupled to the second-order Yee's scheme for Maxwell equations. 

\textbf{ADE scheme}. The most common second-order ADE scheme  for dispersion models (called ``ADE'' in this paper) is achieved with centered derivative approximations  
$\dot{f}(n\tau) \approx (f^{n+1}-f^{n-1})/(2\tau)$ and
$\ddot{f}(n\tau) \approx (f^{n+1}-2f^n+f^{n-1})/\tau^2$ and 
$f(n\tau)\approx f^n$ on both sides of Eq.~\eqref{eq:Pi_ODE}. 

\textbf{ADE2 scheme}. An alternative second-order scheme (called ``ADE2'' in this paper) can be formed by replacing $f(n\tau)\approx f^n$
with $f(n\tau) \approx (f^{n+1}+2f^n+f^{n-1})/4$. This second scheme is known as the \textit{bilinear scheme} in the literature, and it corresponds to the consistent approximation of all $j$-th derivatives in the GDM model with powers of the same $z$-transform $(-\iota\omega)^j = \left(2\tau^{-1}(z-1)/(z+1)\right)^j$. This way the internal states of the continuous model are followed more closely by the discrete model. Also, 
the bilinear scheme guarantees stability preservation. For details on the comparison of different ADE approximations and stability analysis see \cite{hulse1994dispersive}.

Both schemes can be written using centered averaging and difference operators, $\mu_t f^n = (f^{n+1/2}+f^{n-1/2})/2$ and $\delta_t f^n = (f^{n+1/2}-f^{n-1/2})/\tau$, 
\begin{align}
    \big(\delta_t^2 + b_1 \delta_t\mu_t + b_0        \big)&\ \Pv^n = 
    \veps_0\big(             a_1 \delta_t\mu_t + a_0        \big)&\!\!\!\!\!\!\!\!\!\!\Ev^n, \qquad &(\text{ADE})&& \label{eq:ADE_oper}\\
    \big(\delta_t^2 + b_1 \delta_t\mu_t + b_0\mu_t^2 \big)&\ \Pv^n = 
    \veps_0\big(             a_1 \delta_t\mu_t + a_0\mu_t^2 \big)&\!\!\!\!\!\!\!\!\!\!\Ev^n.  \qquad &(\text{ADE2})&& \label{eq:ADE2_oper}
\end{align}

\medskip\noindent
\textbf {Single pole,  $\chi = \dfrac{a_1}{b_1 -\iota\omega}$}. 
The subcase of $a_0 = b_0 =0$ is treated separately from the general case, since it allows for the reduction of \eqref{eq:Pi_ODE} to a first-order ODE 
\begin{equation}\label{eq:Pi_ODE1}
    \dot\Pv + b_1\Pv = \veps_0 a_1  \Ev,
\end{equation}
and thus can be implemented with reduced computational cost. Second-order accuracy can be achieved with the Crank-Nicolson scheme
\begin{equation}\label{eq:ADE_oper1}
    \left( \delta_t^- + b_1\mu_t^- \right) \Pv^n = \veps_0a_1\mu_t^-\Ev^n .
\end{equation}
where $\mu_t^- f^n = (f^n+f^{n-1})/2$ and $\delta_t^- f^n = (f^{n}-f^{n-1})/\tau$ are backward averaging and difference operators.

After substituting operators and factoring similar terms, all three schemes
(\ref{eq:ADE_oper},\ref{eq:ADE2_oper},\ref{eq:ADE_oper1}) can be written in the same form, as  an explicit universal recursion
\begin{equation}\label{eq:ADE_scheme1}
    \Pv^{n+1} =                          \beta_1\Pv^n + \beta_2\Pv^{n-1} 
                 + \veps_0\left(
                 \alpha_0\Ev^{n+1} +\alpha_1\Ev  ^n + \alpha_2\Ev^{n-1}
                 \right),
\end{equation}
where the scheme coefficients $\alpha_j, \beta_j$ are given in Table~\ref{tab:universal}.

\subsubsection{Recursive Convolution (RC) schemes}\label{sec:RC}

\noindent
RC schemes are derived from the convolution integral, if the chosen  numerical integration quadrature can be written in a recursive way. For a single-term GDM dispersion model~\eqref{eq:GDM_1}, the convolution integral can be written using TD representations \eqref{eq:chi_D} and \eqref{eq:chi_Lp} as
\begin{equation}\label{eq:convolution}
    \Pv(t) = \veps_0 \int\limits_{-\infty}^{\infty}
    \chi(\tilde t)
    \Ev(t - \tilde{t}) {\mathrm d}\tilde{t},
\end{equation}
with
\begin{equation}\label{eq:chit_cases}
\chi(t) =
\begin{cases}
    a_1\Exp^{-b_1 t}\theta(t),& \text{if } a_0=b_0=0, \\
    a\Exp^{-\Gamma t} \sin({\Omega t - \varphi})\theta(t),& \text{otherwise}.
\end{cases}
\end{equation}
Parameters $a, \Gamma, \Omega, \varphi$ are obtained from $a_0,a_1,b_0,b_1$ using \eqref{eq:GDM_FD2TD}. The susceptibility $\chi(t)$ is a real function that is either one real exponent if $a_0=b_0=0$, or is a sum of two exponents (possibly complex). 

Traditionally, formulation of RC methods (e.g., for Lorentz media) requires that the two exponents are complex conjugate. Here the derivation formulas do not use complex conjugation and work equally for the damped oscillator case ($b_0>b_1^2/4$) which gives two complex conjugate exponents and for the overdamped case ($b_0<b_1^2/4$) which gives two real decaying exponents. Recursive convolution (RC)  schemes are derived regardless of $b_0 \gtrless b_1^2/4$ for any GDM dispersion for variety of second-order quadratures, including Trapezoidal (TRC), quasi-Trapezoidal (TRC2), Piecewise Constant with second order modification (PCRC2), and Piecewise-Linear (PLRC).

\medskip\noindent
\textbf {Single pole,  $\chi = \dfrac{a_1}{b_1 -\iota\omega}$}. The subcase $a_0=b_0=0$ is treated separately since it gives only one exponential term in the susceptibility and can be implemented with less computational cost than the general case.
\begin{align}\label{eq:RC_exp1}
\begin{split}
    \chi(t) &= A \Exp^{B t}\theta(t),\\
    B &= -b_1,\\
    A &= a_1. \\
\end{split}
\end{align}
The parameters $A,B$ are introduced for a unified notation with the general case. For a single exponent~\eqref{eq:RC_exp1}, the recursion for the convolution integral \eqref{eq:convolution} is based on the identity
\begin{equation}\label{eq:RCidentity}
\Pv(t+\tau) = \Exp^{B \tau} \Pv(t)  + A\veps_0  \int_0^{\tau} \Exp^{B(\tau-\tilde{t})} \Ev(t+\tilde{t}) \, \dd\tilde{t}  .
\end{equation}
The derivation can be found in the Appendix. Note that the identity~\eqref{eq:RCidentity} is derived for continuous fields (prior to any approximation). It indicates that $\Pv(t+\tau )$ is determined from $\Pv(t)$ and a finite time history of $\Ev(t)$ for $t\in[t,t+\tau]$.

Next, we assume that the integral in~\eqref{eq:RCidentity} is approximated with a two-point quadrature, leading to a recursive update formula on $\Pv^{n+1}$
\ba
\Pv^{n+1} = \Exp^{B\tau} \Pv^n +
        A\tau\veps_0 \, 
        \left[ 
        \wrc_0 \, \Ev^{n+1} +  \wrc_1 \Exp^{B\tau}  \Ev^{n}
        \right],
  \label{eq:PRCquadrature}
\ea
where the weights $\wrc_0$ and $\wrc_1$ are to be determined by the quadrature rule and may depend on $B$ and $\tau$.
For example, three trivial RC schemes can be built immediately using $\wrc_0=0,\wrc_1=1$ (left Riemann rule), $\wrc_0=1,\wrc_1=0$ (right Riemann rule), and $\wrc_0=\wrc_1=1/2$ (trapezoidal rule, or TRC). Out of three, only the TRC scheme is second-order accurate. TRC was used in one of the first works on dispersive Yee's scheme by Bui et al. in 1991, and even earlier in Bui's master's thesis from 1990 \cite{bui1991propagation}. This seems to be independent from the first Leubbers et al. work the same year \cite{luebbers1990frequency}. The left Riemann rule was used in another classic work on dispersive FDTD by Hawkins\&Kallman in 1993 \cite{hawkins1993linear}, where it was referred to as ``rectangular quadrature'', but it is only first-order accurate. 

Perhaps the most famous works on dispersive FDTD are by Luebbers et al. \cite{luebbers1990frequency,luebbers1991frequency,luebbers1992fdtd}, frequently called ``the original RC method'', also referred to as ``(FD)$^2$TD'' by the authors (Frequency Dependent FDTD). The method name was later established as PCRC (Piecewise Constant RC) or sometimes simply CRC. In PCRC, the electric field in the integral~\eqref{eq:RCidentity} is taken as constant $\Ev(t+\tilde{t}) \approx \Ev(t+\tau)$, while the exponential susceptibility is integrated exactly
\begin{equation}\label{eq:PCRC}
    \int_0^{\tau} \Exp^{B(\tau-\tilde{t})} \Ev(t+\tilde{t}) \, \dd\tilde{t}  
    \approx
    \Ev(t+\tau)\frac{\Exp^{B\tau}-1}{B}.
\end{equation}
This approximation corresponds to $\theta_0=(B\tau)^{-1}(\Exp^{B\tau}-1), \theta_1=0$ and is also only first-order accurate. 

In 1995 Siuchansian\&LoVetri followed Lueberrs approach and increased the accuracy to second order by using the trapezoidal rule for the electric field $\Ev(t+\tilde{t}) \approx \Ev(t+\tau)/2 + \Ev(t)/2$ in the integral~\eqref{eq:RCidentity}  (while still integrating the exponential susceptibility exactly) 
\cite{siushansian1995efficient,siushansian1995comparison,siushansian1997efficient}
\begin{equation}\label{eq:TRC2}
    \int_0^{\tau} \Exp^{B(\tau-\tilde{t})} 
    \Ev(t+\tilde{t})
    \, \dd\tilde{t}  
    \approx
    \frac{\Exp^{B\tau}-1}{2B}\  \Big[ \Ev(t+\tau) + \Ev(t)\Big].
\end{equation}
To avoid confusion with the \textit{pure} trapezoidal (TRC) rule ($\theta_0=\theta_1=1/2$), the approximation~\eqref{eq:TRC2} is sometimes called the \textit{quasi-trapezoidal} (quasi-TRC) and corresponds to ($\theta_0=(B\tau)^{-1}(\Exp^{B\tau}-1)/2, \theta_1=\theta_0\Exp^{-B\tau}$). 

In 1996 and 1998 two second-order RC schemes were published by Luebbers \& co-authors based on his original first-order RC scheme. In \cite{kelley1996piecewise} they used a piecewise linear approximation of the electric field $\Ev(t+\tilde{t}) \approx \Ev(t) + \tilde{t}\left( \Ev(t+\tau)-\Ev(t) \right)/\tau$ in the integral~\eqref{eq:RCidentity}. Integration gives
\begin{equation}\label{eq:PLRC}
    \int_0^{\tau} \Exp^{B(\tau-\tilde{t})} 
    \Ev(t+\tilde{t})
    \, \dd\tilde{t}  
    \approx
    \underbrace{\frac{\Exp^{B\tau}-1-B\tau}{B^2\tau}}_{\tau\theta_0} \Ev(t+\tau) 
    + 
    \Exp^{B\tau}\underbrace{\frac{\Exp^{-B\tau}-1+B\tau}{B^2\tau}}_{\tau\theta_1} \Ev(t).
\end{equation}
This RC approximation is called Piecewise Linear (PLRC). In the second scheme \cite{schuster1998accurate} authors used the same piecewise constant approximation of the electric field as in original PCRC, but shifted the quadrature points to half steps, which corresponds to $\theta_{0,1}=(B\tau)^{-1}(\pm\Exp^{\pm B\tau/2}\mp 1)$. This recursive approximation is called PCRC2.

Table~\ref{tab:RCapproximations} provides a universal list of weights calculated for all the mentioned quadrature rules conventionally used in the literature for RC-FDTD schemes, mainly for Debye and Lorentz media. Using such parameterization through coefficients $\theta_0,\theta_1$, we will implement all the different second-order RC schemes for a general GDM model with the same algorithm and computational cost. Despite the existing discussions in the FDTD literature on the differences in  computational cost of different RC and ADE schemes (e.g. \cite{siushansian1997efficient}) we will show that all these approximations can be implemented in the exact same scheme with different coefficients and thus are computationally equal. Note that in the case of a zero pole, e.g. Drude model ($b_0=a_1=0$) the limiting value of the weights~$\theta_{0,1}$ with $\beta\rightarrow 0$ should be used for proper numerical integration.  

\begin{table}[!ht]
\caption{Weights for first and second order RC schemes~\eqref{eq:PRCquadrature}. Note that if $B=0$ then the expressions should be treated as limit $B\rightarrow 0$.} \label{tab:RCapproximations}
\footnotesize{
\renewcommand{\arraystretch}{2.5}
\begin{tabular*}{\textwidth}{@{\extracolsep{\fill}}cccl@{}}
\whl
$\wrc_0$  &  $\wrc_1$  &  Order  &  Method  \\
\whl
1                               &  0                                              &  1  & Right Riemann   \\
0                               &  1                                              &  1  & \parbox[c]{5cm}{Left Riemann \\ (or ``Rectangular'' \cite{hawkins1993linear})} \\
$\dfrac{\Exp^{B\tau}-1}{B\tau}\approx1$  &  0   &  1  & 
\parbox[c]{5cm}{PCRC \\ (Piecewise Constant \cite{luebbers1990frequency,luebbers1991frequency,luebbers1992fdtd})}   \\
\hline
$\dfrac{1}{2}$                  &  $\dfrac{1}{2}$                                 &  2  & 
\parbox[c]{5cm}{TRC \\ (Trapezoidal \cite{bui1991propagation})}   \\
$\dfrac{\Exp^{B\tau}-1}{2B\tau}  \approx \dfrac{1}{2}$ &  
$\dfrac{1-\Exp^{-B\tau}}{2B\tau} \approx \dfrac{1}{2}$                                 &  2  & 
\parbox[c]{5cm}{quasi-TRC \\ (quasi-Trapezoidal 
\cite{siushansian1995efficient,siushansian1995comparison,siushansian1997efficient})}   \\
$\dfrac{ \Exp^{B\tau/2}-1}{B\tau} \approx\dfrac{1}{2}$  &
$\dfrac{1-\Exp^{-B\tau/2}}{B\tau} \approx\dfrac{1}{2}$  &  2  & 
\parbox[c]{5cm}{PCRC2 \\ (Piecewise Constant \cite{schuster1998accurate})}   \\
$\dfrac{\Exp^{ B\tau}-1-B\tau}{(B\tau)^2} \approx\dfrac{1}{2}$   &
$\dfrac{\Exp^{-B\tau}-1+B\tau}{(B\tau)^2} \approx\dfrac{1}{2}$   &  2  & 
\parbox[c]{5cm}{PLRC \\ (Piecewise Linear \cite{kelley1996piecewise})}   \\
\whl
\end{tabular*}}
\end{table}

In this manuscript, the polarization equation is coupled to the {\it second-order} Yee's FDTD scheme. To maintain the order of accuracy it is best to use second-order RC schemes, for example, TRC, TRC2, PCRC2, and PLRC. This enables global second-order accuracy in the coupled RC-FDTD Yee's scheme. 

\bigskip\noindent
\textbf{General case, $\chi = \dfrac{a_0 - \iota\omega a_1}{b_0 - \iota\omega b_1 -\omega^2}$.}
In this case, the susceptibility is a sum of two exponents, see \eqref{eq:chit_cases}
\begin{align}\label{eq:RC_exp}
\begin{split}
    \chi(t) &= A^+\Exp^{B^+t}\theta(t) + A^-\Exp^{B^-t}\theta(t),\\
    B^{\pm} &= -b_1/2 \pm \iota\sqrt{b_0-b_1^2/4},\\
    A^{\pm} &= \pm \frac{a_0+a_1 B^\pm}{2\iota \sqrt{b_0-b_1^2/4}}. \\
\end{split}
\end{align}
where the two terms are not necessarily complex conjugate to each other and could be both real. This generality allows for overdamped cases ($b_1^2/4>b_0$), including the Drude term ($a_1=b_0=0$).

Applying the recurrence~\eqref{eq:RCidentity} for each complex exponential term in~\eqref{eq:RC_exp}, and then combining them into a single real-valued recurrence gives the same form of the numerical scheme as ADE methods ~\eqref{eq:ADE_scheme1}. Full derivation of this RC recurrence is presented in the Appendix.
The resulting $\theta_{0,1}$-dependent real coefficients $\alpha_i,\beta_i$ for all RC schemes are summarized in Table~\ref{tab:universal}. Here the weights $\theta_{0,1}^+$ and $\theta_{0,1}^-$ are taken according to the chosen 
RC quadrature using Table~\ref{tab:RCapproximations} where 
$\theta_{0,1}^+ = \theta(B = B^+)$ and $\theta_{0,1}^- = \theta(B = B^-)$ respectively. Note that all the coefficients $\alpha_j,\beta_j$ in the table are real since $A^\pm,B^\pm$ are either real or complex conjugate $A^+= \conjugatet{A^-}$, $B^+= \conjugatet{B^-}$.

Thus we unified the ADE and RC schemes into a universal stencil with real recursive accumulators and coefficients, so that switching 
between schemes in the program does not change data types or memory requirements.

In general, an RC scheme can be constructed if the convolution kernel is expressed as a sum of exponents with a {\em linear} argument. The non-linear (quadratic) argument in the Gaussian model \eqref{eq:chi_LGt} significantly complicates its RC (and ADE) realization. Instead of finding a way of direct implementation of \eqref{eq:chi_LGt} 
we use our oscillator approximation model in Section~\ref{sec:GDMapprox} that is accurate and computationally cheap.


\subsubsection{Compact universal GDM-FDTD scheme} 
\label{sec:universal}

\noindent
The universal ADE and RC formulation~\eqref{eq:ADE_scheme1} with coefficients in Table~\ref{tab:universal} can now be coupled to the classic second-order Yee's scheme for Maxwell equations (e.g. \cite{taflove2005computational}) through the polarization vector $\Pv = \sum_{i=1}^N\Pv_i$. Thus we obtain the following universal scheme for solving Maxwell's equation with GDM dispersion (\ref{eq:Maxwell_TD},\ref{eq:GDMfinal})
\begin{equation}\label{eq:YeeUniversal}
\left\{
\begin{aligned}
  \mu_0\,  \delta_t \Hv^{n+1}  &= - \nabla_{h}\times \Ev^{n+1},                         \\
  \veps_0\veps_\infty\, \delta_t \Ev^{n+1/2}  &= \nabla_{h}\times \Hv^{n+1/2}  - \sigma \mu_t \Ev^{n+1/2} -  \delta_t \sum_{i=1}^N\Pv_i^{n+1/2} , \\
\Pv_i^{n+1} &= \beta_{1,i}\,  \Pv_i^n + \beta_{2,i}\, \Pv_i^{n-1} \\
&\qquad\qquad + \veps_0\left(\alpha_{0,i}\, \Ev^{n+1} + \alpha_{1,i} \, \Ev^{n} +  \alpha_{2,i} \, \Ev^{n-1}\right),
                     \quad i\in \overline{1,N}.  \\
\end{aligned}
\right.
\end{equation}
Here the numerical operators as before are $\delta_t f^n = (f^{n+1/2}-f^{n-1/2})/\tau$ and 
$\mu_t f^n = (f^{n+1/2}+f^{n-1/2})/2$, and coefficients $\alpha_{j,i}, \beta_{j,i} \in \Real$ depend on the particular choice of ADE or RC approximation, and are shown in Table~\ref{tab:universal}. 

We choose to keep the contracted notation for the approximation operator $\nabla_h$ of the gradient ($h$ --- spatial step) without explicitly writing the spatial indices. We assume the classic leap-frog Yee's scheme is used for space. The formulas~\eqref{eq:YeeUniversal} only show the modification of the time stencil required for the dispersion implementation. The full spatio-temporal implementation in 1D can be found in the MADIS codes and can easily be extended to the full 3D case.

In contrast with conventional approaches, where each scheme requires a separate implementation, the generalized form~\eqref{eq:YeeUniversal} offers a convenient route to implement various ADE and RC schemes within the same core code with a conditional operator that sets the values of $\alpha_{j,i}$ and $\beta_{j,i}$ for the specific scheme desired. 

Also, while conventional RC implementation 
relies on complex conjugation and does not support the over-damped case with real poles ($b_1 \leq b_0^2/4$),
the universal implementation~\eqref{eq:YeeUniversal} operates with real coefficients and functions and thus works for any generic second-order GDM term without this restriction. In both cases, ($b_1 > b_0^2/4$) and 
($b_1 < b_0^2/4$), Table~\ref{tab:universal} produces correct real coefficients for a universal scheme without restrictions on $b_0, b_1$.

\begin{table}[!ht]
\caption{Coefficients $\alpha_j$ and $\beta_j$ for the universal implementation of ADE and RC schemes     
$\Pv^{n+1} = \beta_1\Pv^n + \beta_2\Pv^{n-1} 
+ \veps_0\left(\alpha_0\Ev^{n+1} +\alpha_1\Ev  ^n + \alpha_2\Ev^{n-1}\right)$}
\label{tab:universal}
\setlength{\tabcolsep}{5pt}
\footnotesize{
\renewcommand{\arraystretch}{1.5}
\begin{tabular*}{\textwidth}{@{\extracolsep{\fill}}c|cc|ccc@{}}
\whl
& \multicolumn{2}{c|}{$ \chi(\omega) = \dfrac{a_1}{b_1-\iota\omega} $}
& \multicolumn{3}{c} {$ \chi(\omega) = \dfrac{a_0-\iota\omega a_1}{b_0-\iota\omega b_1-\omega^2}, \quad \left( \Gamma = b_1/2, \, \Omega = \sqrt{b_0-\Gamma^2} \right)$} \\
& \multicolumn{2}{c|}{ $ A = a_1$ }
& \multicolumn{3}{c} {$A^\pm = \pm\dfrac{a_0+a_1B^\pm}{2\iota\Omega}$} \\
& \multicolumn{2}{c|}{$B = -b_1$}
& \multicolumn{3}{c} {$B^\pm = -\Gamma \pm\iota\Omega$, \quad $\theta^\pm_{0,1} = \theta_{0,1}(B=B^\pm)$  from 
Table~\ref{tab:RCapproximations}} \\
& ADE  &  RC  &  ADE & ADE2 &  RC \\
\whl
$\beta_0$
& $2 + b_1\tau$
& -
& $2 + b_1\tau$
& $4 + 2b_1\tau + b_0\tau^2$
& - \\  [3pt]
$\beta_1$
& $\dfrac{2 - b_1\tau}{\beta_0}$
& $\Exp^{B\tau}$
& $\dfrac{4 - 2b_0\tau^2}{\beta_0}$
& $\dfrac{8 - 2b_0\tau^2}{\beta_0}$
& $2\Exp^{-\Gamma\tau}\cos{\Omega\tau}$ \\ [6pt]
$\beta_2$
& $0$
& $0$
& $\dfrac{-2 + b_1\tau}{\beta_0}$
& $\dfrac{-4 + 2b_1\tau - b_0\tau^2}{\beta_0}$
& $-\Exp^{-2\Gamma\tau}$ \\ [6pt]
$\alpha_0$
& $\dfrac{a_1 \tau}{\beta_0}$
& $A\tau\theta_0$
& $\dfrac{a_1 \tau}{\beta_0}$
& $\dfrac{2a_1\tau + a_0\tau^2}{\beta_0}$
& $\tau\left[ A^+\theta_0^+ + A^-\theta_0^- \right]$ \\ [6pt]
$\alpha_1$
& $\dfrac{a_1 \tau}{\beta_0}$
& $A\tau\theta_1\Exp^{B \tau}$
& $\dfrac{2a_0\tau^2}{\beta_0}$
& $\dfrac{2a_0\tau^2}{\beta_0}$
& 
\begin{tabular}{@{}c@{}} 
$-\alpha_0\beta_1 + \tau\left[ A^+\Exp^{B^+\tau}(\theta_0^++\theta_1^+) \right. $ \\ 
$ + \left. A^-\Exp^{B^-\tau}(\theta_0^-+\theta_1^-)  \right] $
\end{tabular} \\
%
%
$\alpha_2$
& $0$
& $0$
& $\dfrac{-a_1\tau}{\beta_0}$
& $\dfrac{-2a_1\tau + a_0\tau^2}{\beta_0}$
& $\beta_2\tau\left[  A^+\theta_1^+ + A^-\theta_1^- \right] $ \\
\whl
\end{tabular*}}
\end{table} 

Furthermore, as shown in \cite{prokopeva2011optical}, the universal implementation~(\ref{eq:YeeUniversal})
can be written in a \textit{compact form} to calculate $\Ev^n$ with two real recursive accumulators $\Psi_i$ and $\Phi_i$ and therefore requires fewer floating-point operations (FLOPs). The resulting \textit{compact universal} scheme is shown below  
\begin{equation}\label{eq:YeeUniversalCompact}
\left\{
\begin{aligned}
{\bf E}^{n+1}  &=
\xi_0^{-1}\left[
    \xi_1{\bf E}^n + \veps_0^{-1}\tau\nabla_h\times{\bf H}^{n+1/2} - \sum_{i=1}^N{{\bf \Psi}^n_i} 
\right], \\
\Psiv^{n+1}_i &= 
\beta_{1,i}\Psiv^n_i+\Phiv^{n}_i
+\{\eta_{2,i}+(\beta_{1,i}-1)\eta_{1,i}\}{\Ev}^n, \qquad  i\in \overline{1,N}, \\
\Phiv^{n+1}_i &=\beta_{2,i}\Psiv^n_i -\{\eta_{2,i}-\beta_{2,i}\eta_{1,i}\}\Ev^n, \qquad\qquad\qquad\quad  i\in\overline{1,N},
\end{aligned}
\right.
\end{equation}
with 
$\xi_0=\varepsilon_{\infty}+\frac{\sigma\tau}{2\varepsilon_0}+\sum\limits_{i=1}^N\alpha_{0,i}$,\, $\xi_1=\xi_0-\frac{\sigma\tau}{\varepsilon_0}-\sum\limits_{i=1}^N\eta_{1,i}$,\,
$\eta_{k,i}=\alpha_{k,i}+\alpha_{0,i}\beta_{k,i}$, $k=1,2$. Note that for the single pole case $a_0=b_0=0$ we have $\alpha_{2,i}=\beta_{2,i}=\eta_{2,i}={\bf\Phi}_i=0$, and only one recursive accumulator ${\bf\Psi}_i$ can be used for calculations.
The compact scheme \eqref{eq:YeeUniversalCompact} is used in MADIS codes to demonstrate an efficient universal dispersion implementation in a variety of time-domain solvers.

Comparison of the three implementations --- standard complex RC, 
universal~(\ref{eq:YeeUniversal}), and compact universal~\cite{prokopeva2011optical}
--- is summarized in Table~\ref{tab:SchemesComparison} for one GDM term. First, we count the number of elementary operations, assuming that complex multiplication and addition take 6 and 2 FLOPs, respectively, with the universal scheme taking 30\% more FLOPs and the complex RC formulation taking twice as many FLOPs as the compact version.
In terms of storage, all three implementations require storing two numerical arrays, polarization terms
(either the real and imaginary parts of $\Qv_i^n$, or two time levels of the real-valued polarization $\Pv_i^n$ and $\Pv_i^{n-1}$, or two real-valued accumulators $\Psi_i^n$ and $\Phi_i^n$). But organizing computation in the compact universal formulation without additional storage is more convenient since the update formulas involve only one time-step of the electric field, $\Ev^n$.

\begin{table}[!ht]
\caption{Comparison of the GDM implementations (complex numbers are in bold face).}
\label{tab:SchemesComparison}
\footnotesize{
\renewcommand{\arraystretch}{2}
\begin{tabular*}{\textwidth}{@{\extracolsep{\fill}}cccl@{}}
\whl
Scheme & FLOPs & Storage & Notes \\
\whl
\parbox[t]{4.5cm}{
	{\bf Complex RC}: \\
	$\Qv^{n+1} = \av \Qv^n + \bv E^{n+1} + \cv E^n$
} 
&
14 &
\parbox[t]{2.5cm}{
$\Re(\Qv^n)$, $\Im(\Qv^n)$ \\ 
$E^{n+1}$, $E^n$ 
} 
&
\parbox[t]{3.7cm}{
	RC methods only \\
	must be $b_0 > b_1^2/4$
} \\
\parbox[t]{4.5cm}{
	{\bf Universal}, Eq.~\eqref{eq:YeeUniversal}: \\
	$P^{n+1} = a P^n + b P^{n-1} +$ \\
	$c E^{n+1} + d E^n + e E^{n-1}$
} &
9 &
\parbox[t]{2.5cm}{ 
$P^n$, $P^{n-1}$ \\ 
$E^{n+1}$, $E^n$, $E^{n-1}$ } 
&
\parbox[t]{3.7cm}{
	ADE and RC \\
	any $b_0, b_1$
} \\
\parbox[t]{4.5cm}{
	{\bf Compact Universal}, Eq.~\eqref{eq:YeeUniversalCompact} \\
	$\Psi^{n+1} = a \Psi^n + b E^n + \Phi^n$ \\
	$\Phi^{n+1} = c \Psi^n + d E^n$
} &
7 &
\parbox[t]{2.5cm}{ 
$\Psi^n$, $\Phi^n$ \\ 
$E^n$  
} &
\parbox[t]{3.7cm}{
	ADE and RC \\
	any $b_0, b_1$
} \\
\whl
\end{tabular*}}
\end{table} 
\vline
\section{Simulation examples}\label{sec:simul}
    \noindent
    Besides amorphous and polycrystalline materials such as glasses \cite{macdonald2000dispersion, keefe2001curvefitting}, semiconductors, oxides \cite{synowicki2004optical,may2007optical,uprety2017spectroscopic,schoche2017optical}, and polymers \cite{pallapapavlu2011characterization, rauch2012temperature, naqavi2018optical, hilfiker2018dielectric, patel2020diphenylsiloxane} the effective media approximations built on Gaussian models retrieved from VASE are also broadly used to characterize materials with structural and phase disorder, induced either during fabrication or phase transition. Examples of such media with structural and phase disorder include irradiated glasses, electroplated metals, island metal films (see, e.g. \cite{fernandezrodriguez2006modeling, synowicki2017optical,lonvcaric2011optical}), and phase change materials \cite{orava2008optical,abdel2018optical,ramirez2018thermal}. The Gaussian time-domain approximation technique outlined here has been verified on different classes of optical materials including glasses, polymers, and semi-continuous metal films. 
    
    Here, we demonstrate examples for three representative cases, silica, gold island films, and polymers. The first example can be found in the included code package. We show using our GDM approximation that broadband Gaussian models retrieved with VASE can be directly implemented in a time-domain solver. 
    The approximation accuracy and convergence of the dispersive FDTD solver are verified for all cases with 1D film simulation tests.
    
    \subsection{Glasses --- $\text{SiO}_2$}
    \label{sec:silica}
\begin{figure}[!ht]
    \center
    \includegraphics[width=0.99\textwidth]{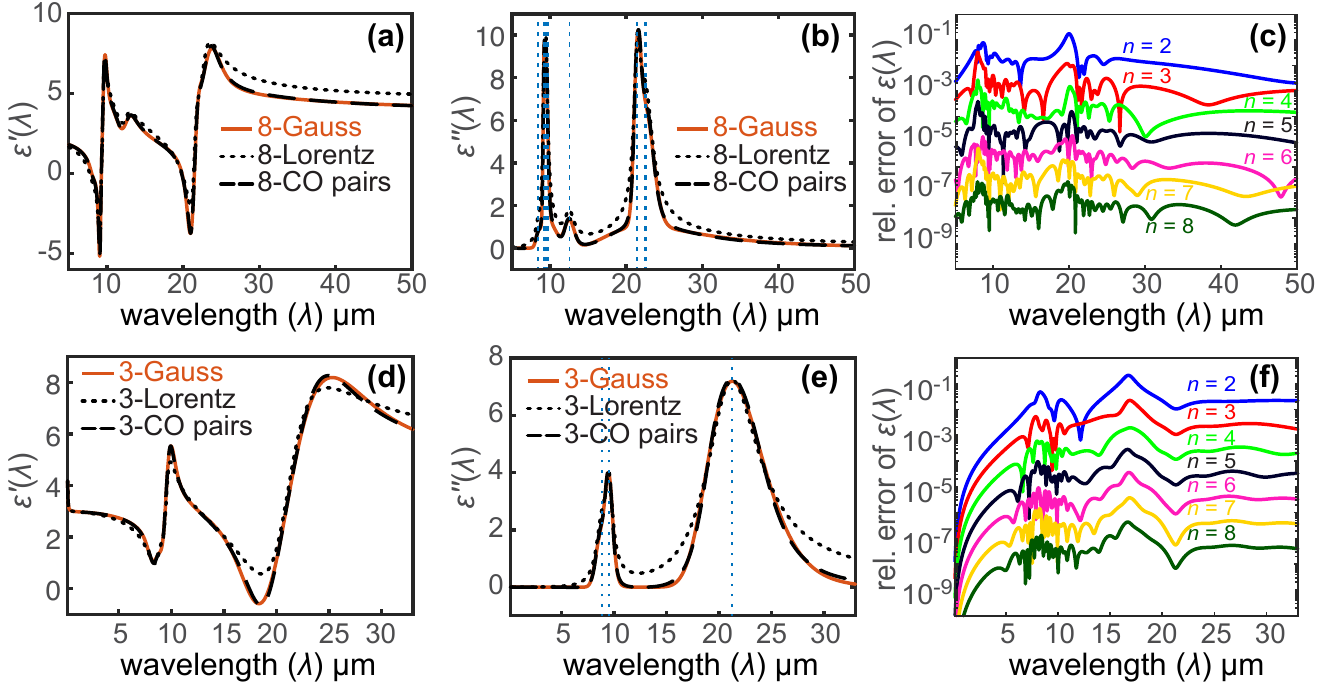}
    \caption{Broadband Gaussian dispersion models for silica and their Lorentz and CO time-domain approximations ($n=2$): (a-b) 8-Gaussian oscillator (8-Gauss) model \cite{kitamura2007optical} for the experimental data for fused silica \cite{popova1972optical}, (d-e) 1-Sellmeier and 3-Gaussian (3-Gauss) model for the multi-sample native $\textrm{SiO}_2$. Vertical lines in (b,e) indicate locations of the Gaussian absorption peaks. (c,f) Approximation errors for CO models of different orders $n=2,...,8$.}
    \label{fig:SilicaModels}
\end{figure}

    \noindent
    Gaussian dispersion analysis was primarily developed for modeling the broadband dispersion of glasses, e.g. \cite{macdonald2000dispersion}. Native silicon oxide ($\textrm{SiO}_2$) films are among the optical material elements that require accurate broadband dispersion characterization for many photonic applications.
    The actual values of the refractive index can vary due to the manufacturing process and presence of defects in samples (e.g., see the 3-Gaussian and 8-Gaussian models in Fig.~\ref{fig:SilicaModels} fitted to different data sets).
    The general absorption bands are caused by OH groups and are relatively consistent at: 2.8, 3.5, and 4.3 $\mu$m. The 9, 12.5 and 22 $\mu$m absorption bands are due to vibration modes. An impressive literature review of silica experimental data and models for wavelengths spanning from 30 nm to 1000 $\mu$m is done in \cite{kitamura2007optical}. As a result of this extensive review, the authors developed a dispersion model for silica with 8 Gaussian oscillators that covers extremely broadband range from 5 to 50 $\mu$m. 

    In some photonic applications, it is possible to neglect minor spectral features in favour of more computationally-efficient models. These models include fewer additional oscillators, but cover more general cases or the average response of multiple samples. Many practical multi-sample dispersion models that contain Gaussian terms are obtained using VASE characterization techniques. For example, a compact model was extracted by J. A. Woollam Co. from multi-sample VASE analysis of native $\textrm{SiO}_2$ as discussed in \cite{johs1999overview}. This example covers a broad spectral range (0.19 -- 33 $\mu$m), contains a Sellmeier pole in the UV (at $\sim0.11~\mu$m), and a set of three Gaussian terms in the IR (at $\sim8.86~\mu$m, $\sim9.54~\mu$m, and $\sim21.3~\mu$m, respectively). Thus, the lowest-order CO approximation ($n=2$) for this model consists of only 7 oscillators total --- 1 Sellmeier and 6 coupled oscillators (2 per Gaussian). 
     
    To demonstrate the accuracy and efficacy of our CO approximations on real Gaussian data we used an extreme wideband 8-Gaussian model (8-Gauss) of fused silica. In Fig.~\ref{fig:SilicaModels} we compare our Gaussian approximation to a simplified Lorentzian fit (with full-width-half-maximum matched to the Gaussian peaks). The absorption centers are shown with vertical lines. The figure indicates that CO approximation catches the finest features of the spectrum even with the lowest approximation order $n=2$, however, the simple Lorentz substitution of the Gaussian terms with the same FWHM is far off. We also tested our model against a more compact 1-Sellmeier 3-Gaussian model (3-Gauss) of native, thermally grown $\text{SiO}_2$  within a narrower range. We once again found that the CO approximation performs very well in this case while a simple Lorentz substitution of Gaussian terms performs quite poorly. The lowest order CO approximation parameters for both the full case and the compact model case can be found in Table \ref{tab:SilicaParam}).
    
     The question remains whether the higher-order approximations give valid FDTD-compatible models that could potentially be used if higher accuracy Gaussian absorption simulations in the time domain are required. In figure~\ref{fig:SilicaModels}(ef) we plotted the approximation errors for all approximations ($n=2,...,8$). These plots demonstrate that the error scales nicely on real Gaussian data with approximation order as $10^{-n}$. This scaling is in good agreement with the approximation errors of the Dawson function analyzed earlier (Fig.~\ref{fig:DawsonConverge}). The maximum error starts at roughly a few percent for $n=2$, and almost reaches single floating point precision 1e-7 for $n=8$. In practical cases, measurement deviations for material characterization is often several percent (comparable to $n=2$ approximation), and thus usage of orders $n=4$ and above CO approximations that double ($n=4$) and triple ($n=6$) the number of additional oscillators may be computationally impractical and unneeded for time-domain implementation.  
    
    To implement the native $\textrm{SiO}_2$ material model discussed previously (1 Sellmeier, 3 Gaussian Terms) in a time-domain solver we again use the $n = 2,...,8$ order CO approximations for each of the Gaussian terms. These approximations will require 7, up to 25 oscillators, respectively. The higher-order approximations are used to demonstrate accuracy convergence and may not be practical for actual simulations due to additional computational costs. However, they provide the possibility of extremely precise simulation if needed. The Sellmeier pole in the model is yet another special case of a second-order GDM term~\eqref{eq:chi_Lp} with $b_1^i=0$ \cite{prokopeva2020time}. All the approximation parameters can be generated using the included MADIS codes and imported into external time-domain Maxwell solvers.
    \begin{figure}[!ht]
    \center
    \includegraphics[width=0.75\textwidth]{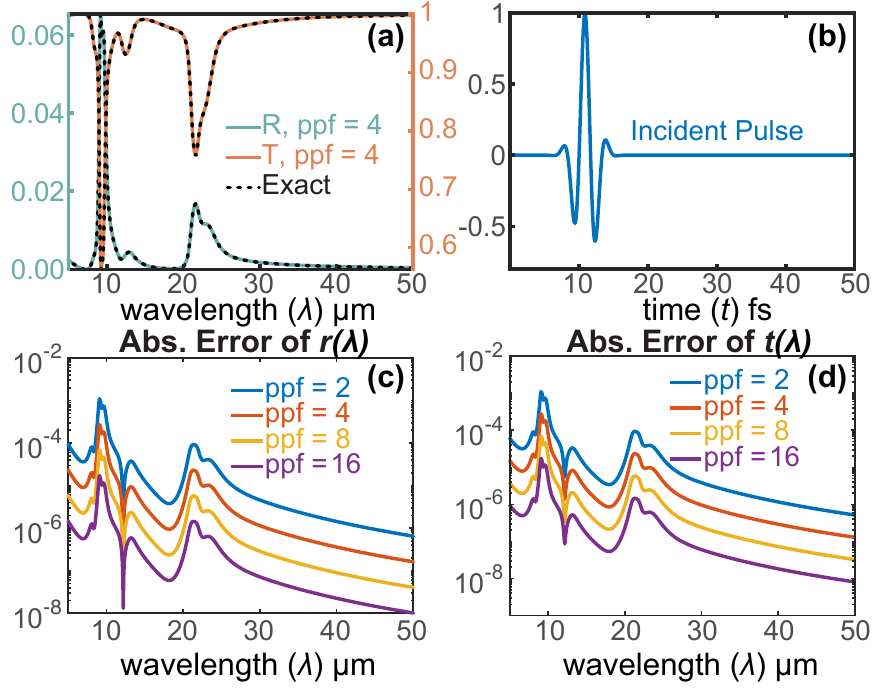}
    \caption{FDTD simulation of a 100-nm silica film using CO approximations of an 8-Gauss model. Parameter {\em points per film} (ppf) indicates the mesh size, $\Delta x =$ 100nm/2, 100nm/4, 100nm/8, and 100nm/16. (a) Comparison of simulated reflectance (R) and transmittance (T) to the exact formulas for a thin film, (b) a short incident Gaussian pulse is used to probe the broadband dispersive response, (c,d) the numerical error of the complex reflection and transmission coefficients $(r,t)$. }
    \label{fig:SilicaFDTD}
    \end{figure}
    
    The approximation models are tested with our FDTD-GDM solver to verify the convergence and stability of the dispersive FDTD simulation. We use a simple one-dimensional formulation of wave propagation through a 100-nm film of silica. To probe the dispersive response of the material numerically we assume an ultra short pulsed plane-wave incident field with a Gaussian TD profile. This test pulse allows us to numerically retrieve a broadband range from 5 to 50 microns with a single time domain simulation,
    \begin{equation}
    E_\text{inc}(t) = \exp \left[ -\dfrac{(t-t_0)^2}{\sigma_0^2} \right] \sin{\dfrac{2\pi c_0}{\lambda_0} t},
    \label{eq:Einc}
\end{equation}
    with a center offset of ${t_0=11}$ fs, width $\sigma_0 = 2$ fs,
    and a carrier wavelength $\lambda_0 = 1~\mathrm{\mu m}$  ($c_0$ is the speed of light in vacuum). 
    
    To compute the numerical error of light propagating through a single film in vacuum,
    we use analytical expressions for reflection and transmission coefficients programmed in the \texttt{RT.m} function supplied with MADIS. For normal incidence, the complex coefficients $r, t$ are given by the classical Drude equation (Eq. (44), \cite{drude1889ueber})
\begin{equation}\label{eq:AnalyticRT}
\left\{
\begin{aligned}
    t(\omega) &=\frac{2}{2\cos{\beta} + \iota(n + n^{-1})\sin{\beta}}, \\
    r(\omega) &= \frac{(n - n^{-1}) \sin{\beta}}{2\iota}\, t(\omega) 
\end{aligned}
\right.
\end{equation}
    where $n=\sqrt{\veps}$ is the film's refractive index, 
    $\beta = -\omega h n/c_0$, $h$ is the thickness, and $\omega$ is the illumination frequency in rad/s. 
    
    The numerical computation of $r$ and $t$ from the FDTD simulation field probe data is done in \texttt{RTnum.m}. To convert the probe data to $r,t$ we first perform the Fourier transform to FD
    \begin{equation}
        \tilde{E}_{\rm{r}} = \operatorname{FFT}(E_{\rm{r}}), 
        \tilde{E}_{\rm{i}} = \operatorname{FFT(E_{\rm{i}}}), 
        \tilde{E}_{\rm{t}} = \operatorname{FFT(E_{\rm{t}}}),
    \end{equation}
    and then perform phase correction according to the exact location of the probes
\begin{equation}\label{eq:NumericalRT}
\left\{
\begin{aligned}
        r_\text{num}(\omega) &= 
        \frac
        { \tilde{E}_{\rm{r}} \exp{(\iota \omega (x_{\rm{a}} - x_{\rm{r}})/c_0}) }
        { \tilde{E}_{\rm{i}} \exp{(\iota \omega (x_{\rm{i}} - x_{\rm{a}})/c_0}) }, \\
        t_\text{num}(\omega) &=
        \frac
        { \tilde{E}_{\rm{t}} \exp{(\iota \omega (x_{\rm{t}} - x_{\rm{b}})/c_0)} }
        { \tilde{E}_{\rm{i}} \exp{(\iota \omega (x_{\rm{i}} - x_{\rm{a}})/c_0}) },
\end{aligned}
\right.
\end{equation}
  where $E_{\rm{r}}$, $E_{\rm{i}}$, and $E_{\rm{t}}$ are the FDTD computed electric fields, $\omega$ is the frequency in rad/s, and $x_{\rm{i}}$, $x_{\rm{a}}$, $x_{\rm{b}}$, $x_{\rm{t}}$ are the positions of the incident/reflected electrical measurement probe, the front and back of the film, and the transmission electrical measurement probe, respectively.
  
  The results of the simulations, shown in Fig.~\ref{fig:SilicaFDTD}, demonstrate that all of the proposed approximation models coupled with the time-domain solver, using any of RC/ADE schemes derived in Section~\ref{sec:FDTD_GDM}, converge with second order. In the figure the numerical error of the bilinear ADE (``ADE2'') scheme is shown as an example. The demonstration codes are available for download and can be used to generate and export the approximation coefficients to other time-domain solvers.

\subsection{Metals --- Gold Island Films} 
    \noindent
    In plasmonics, the dispersion of metals is traditionally modeled in the time domain with a Drude term along with several Lorentz terms (DL-model) \cite{hao2007efficient}.
    Recent works demonstrated that using a combination of a Drude term and critical points (DCP-model) instead of DL-models provides an improvement in the fitting accuracy and computational efficiency due to a reduction in the number of oscillators required \cite{etchegoin2006analytic, etchegoin2007erratum, vial2008comparison}. These recent reports argue that classical Drude–Lorentz theory is incapable of reproducing in detail the profiles empirically observed for many real metallic materials without the addition of nonphysical oscillators. This effect is especially evident when multiple Lorentz oscillators are used to mimic a sharp Gaussian absorption band \cite{orosco2018causal}. To avoid this issue in our approach we derived a physically meaningful analytic representation of the inhomogeneous broadening, with two coupled critical points oscillators for each Gaussian term, rather than fitting the entire ensemble with multiple separate free terms. 
\begin{figure}[!ht]
    \center
    \includegraphics[width=0.99\textwidth]{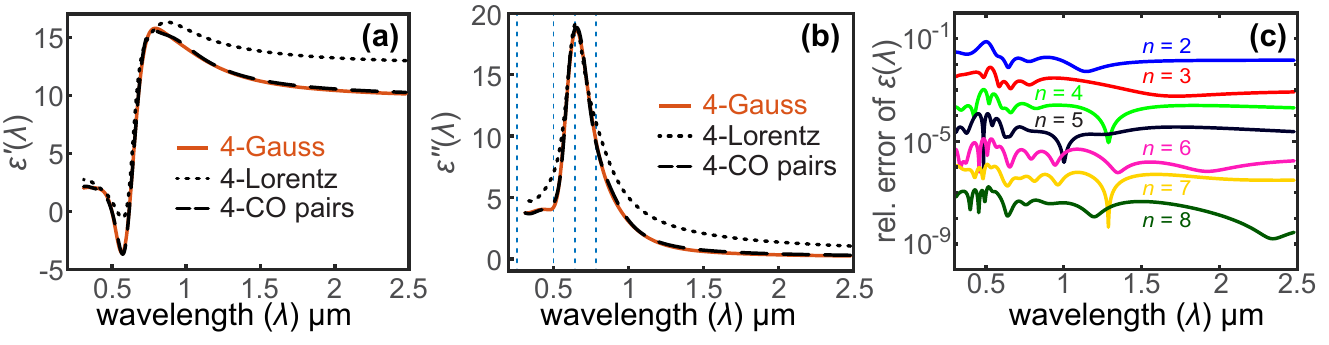}
    \caption{Broadband dispersion models for gold island films: (a-b) 4-Gaussian oscillator (4-Gauss) model \cite{lonvcaric2011optical} along with TD compatible approximations: one Lorentz vs. one CO pair ($n=2$) per Gaussian. (c) Approximation error for CO models of different orders $n=2,...,8$.}
    \label{fig:GoldModels}
\end{figure}    
    
    For demonstration, we tested our approach for a case when metallic material dispersion is combined with geometric structural disorder. Semi-continuous and island metal films are an area of active research. The dispersion characterization of these films is challenging due to the random nature of the material. It is usually performed as a combination of advanced effective medium theory and fitting to extensive amounts of experimental data. It has been shown that the optical constants of island metal films can be successfully obtained by VASE characterization and fitting with a multipole oscillator model that includes Gaussian-type absorption bands. In such applications, Gaussian-based material models significantly outperform conventional Lorentzian fits in terms of residual error. They are also capable of providing meaningful physical insights into the origin or morphology of disorder \cite{lonvcaric2011optical}. 
    
    As a first step, we build CO approximations of orders $2\le n\le 8$ of the 4-Gauss oscillator model derived for the 7-nm island gold film in \cite{lonvcaric2011optical}. The parameters for the $n=2$ TD approximation are given in Table~\ref{tab:SilicaParam}. To showcase the strength of the CO approximation, we also plot a Lorentz approximation with a matched FWHM as a comparison. As before, the Lorentz approximation does not provide acceptable accuracy. However, even a single CO pair ($n=2$) almost completely overlaps with the initial 4-Gauss model, as can be seen in Fig.~\ref{fig:GoldModels}. The error analysis in Fig.~\ref{fig:GoldModels}(c) is consistent with the previous examples, showing approximation errors ranging from a few percent ($n=2$) down to single precision (1e-7) for $n=8$.

\begin{figure}[!ht]
    \center
    \includegraphics[width=0.75\textwidth]{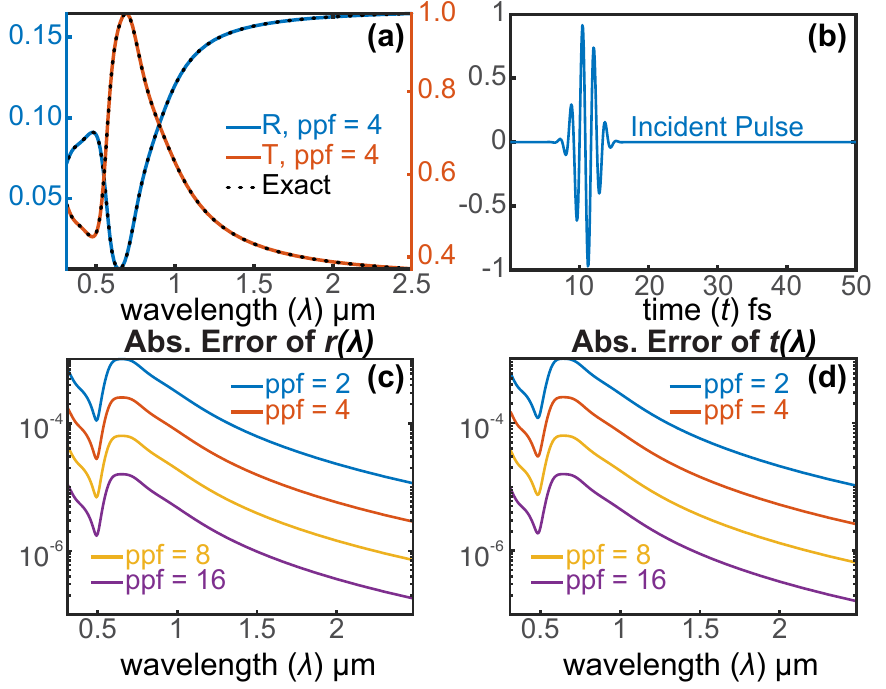}
    \center
    \caption{FDTD simulation of a 7-nm gold island film using the CO approximation of a 4-Gauss model. Parameter {\em points per film} (ppf) indicates the mesh sizes, $\Delta x =$ (7 nm)/2, (7 nm)/4, (7 nm)/8, and (7 nm)/16. (a) Comparison of the simulated reflectance (R) and transmittance (T) vs. the exact reflectance and transmittance calculated analytically. (b) Incident Gaussian pulse used to probe the broadband dispersive response. (c,d) The numerical error of the complex reflection and transmission coefficients $(r, t)$. }
    \label{fig:GoldFDTD}
    \end{figure}

    Next, we verified the CO models in the time-domain for all orders $2\le n\le 8$. FDTD-GDM simulations of a thin 7-nm semi-continuous film illuminated with an incident Gaussian pulse were used to test the model. The pulse parameters are the same as in 
    Section~\ref{sec:silica}, except that the carrier is now $\lambda_0=500$ nm. All the CO approximations showed ideal 2\textsuperscript{nd} order numerical convergence across the entire spectrum for all ADE/RC schemes, as shown in Fig.~\ref{fig:GoldFDTD}.
    
\subsection{Polymers --- DPS-DMS}
\noindent
Polymers are yet another important class of materials with disorder. Gaussian dispersion analysis is widely used for characterizing their optical constants. Here we explore the modeling of diphenylsiloxane-dimethylsiloxane (DPS-DMS) copolymers, which are commonly used as lubricants and adhesives, along with uses in the semiconductor industry and chromatography. The DPS-DMS sample was characterized with 2 Sellmeier and 6 Gaussian poles in \cite{patel2020diphenylsiloxane}. 

\begin{figure}[!ht]
    \center
    \includegraphics[width=0.99\textwidth]{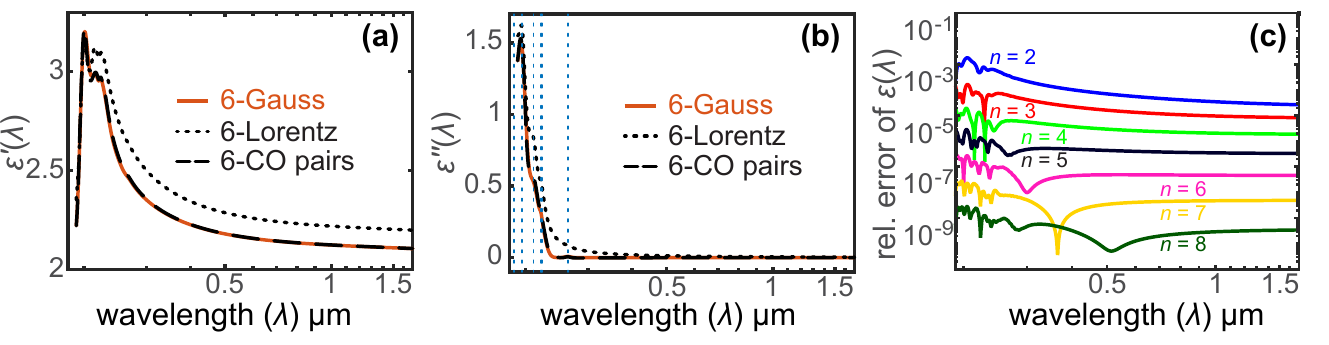}
    \caption{Broadband dispersion models for a DPS-DMS copolymer: (a-b) a 2-Sellmeir, 6-Gaussian oscillator (6-Gauss) experimental data fit \cite{patel2020diphenylsiloxane} and two TD compatible  approximations: Lorentz and CO of order $n=1$. (c,f) Approximation error for CO models of different orders $n=1$, $n=3$, and $n=5$.}
    \label{fig:PolymerModels}
\end{figure}

As before, we first build the CO approximations of all orders $n$, 2 to 8, and check its accuracy vs the exact Gaussian dielectric function, see Fig.~\ref{fig:PolymerModels}. The vertical lines show the locations of the Gaussian peaks. One can clearly see that the Lorentz approximation performs particularly poorly. The CO approximations on the other hand have similarly low error values and trends as the previous examples, along with fast exponential error convergence. 

\begin{figure}[!ht]
    \center
    \includegraphics[width=0.75\textwidth]{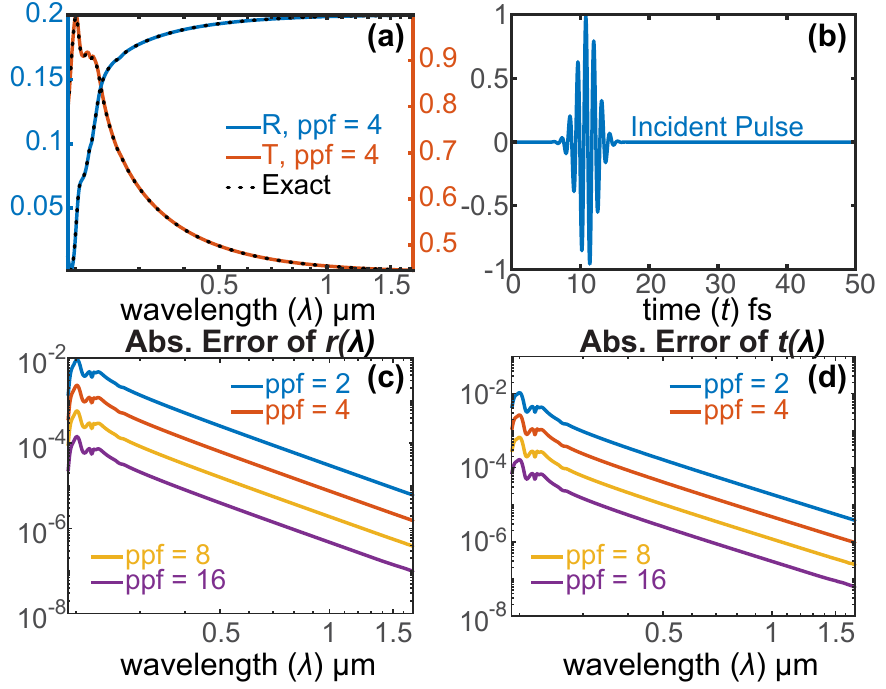}
    \caption{FDTD simulations of a 20-nm DPS-DMS film using CO approximations of a 6-Gauss model. Parameter {\em points per film} (ppf) indicates the mesh sizes, $\Delta x =$ (20 nm)/4, 
    (20 nm)/8, (20 nm)/16, and (20 nm)/32. (a) Comparison of simulated reflectance (R) and transmittance (T),  vs. the analytically computed R, T. (b) The incident Gaussian pulse used to probe the broadband dispersive response. (c,d) The numerical error for the complex reflection and transmission coefficients $(r, t)$. }
    \label{fig:PolymerFDTD}
\end{figure}

All the TD approximation models, $2\le n \le 8$, were tested with one-dimensional Yee-based FDTD codes and follow the theoretical second-order convergence for a light wave propagating through a thin 20-nm film of DPS-DMS. The results can be seen in  Fig.~\ref{fig:PolymerFDTD}. The incident Gaussian probe pulse, depicted in 
Fig.~\ref{fig:PolymerFDTD}(b), is 2-fs long with an offset of 11~fs and a carrier wavelength of 30~nm.

One thing to note is that a full theoretical study regarding the numerical stability of the CO approximations for Gaussian dispersion in FDTD has not been performed in this paper and remains to be explored. Preliminary spectral analysis for the $n = 2$ case revealed two constraints: (i) on phase, $-\pi/2\le \varphi \le 0$, and (ii) on the center/broadening, $\Omega \ge \sigma$. Constraint (i) is satisfied since the phase $\varphi = -1.0728$ is an approximation constant that does not depend on the input Gaussian parameters 
$[A, \Omega, \sigma]$. Constraint (ii) is in agreement with the general limitations of GDA analysis previously discussed in Section 2.1. 

To summarize: we have numerically verified that the proposed Gaussian approximations of order $2\le n \le8$ enable efficient TD simulations with controllable accuracy. This technique works stably and equally well for three different groups of materials: glasses, metals, and polymers, with similar approximation and simulation errors. This method works even for models with large numbers of Gaussian terms across extremely broadband spectral ranges. Furthermore, besides the conventional Yee's FDTD scheme demonstrated here, our approximations can be coupled with other in-house and commercial time-domain solvers to simulate broadband Gaussian material models. This approach enables quick conversion of a significant compendium of the previously characterized and published FD Gaussian-based dielectric functions into a high-performance TD-compatible format. 

\section{Program Description}
\noindent
This section provides a description of the files included in the program package MADIS --- MAterial DIspersion Simulator. Input and output parameters are listed in the tables below for each m-file in the package. For installation unzip the files into your program folder. The package does not have any library dependencies and should work in any version of MATLAB. The package was prepared in version 2020b, but was also tested in older versions including: 2017a, 2018a, and 2019a. A usage example and test run are included in the m-script \texttt{TestRun.m}. This script demonstrates a full simulation workflow --- from the construction of the time-domain Gaussian approximation models to the FDTD simulation and reflection/transmission post-processing.

\subsection{Exact Gaussian Model (Gauss.m)}
\noindent
Function \texttt{Gauss.m} calculates the total permittivity consisting of one or more Gaussian terms (\ref{eq:CDG_im}-\ref{eq:CDG_re}). The output parameter is $\veps_\text{G}(\omega) = \sum_i{ \chi_{\rm{G}}(\omega,i)}$, where
\begin{equation}
    \chi'_\text{G}(\omega,i)
    = \frac{2A_i}{\sqrt{\pi}} \left[ F\left(\frac{\omega+\Omega_i}{\sigma_i} \right)
     - F\left(\frac{\omega-\Omega_i}{\sigma_i} \right)    \right] ,
    \label{eq:Gauss_re2}
\end{equation}
\begin{equation}
    \chi''_\text{G}(\omega,i) = A_i\left[ \Exp^{-(\omega - \Omega_i)^2 / \sigma_i^2}
     - \Exp^{-(\omega + \Omega_i)^2 / \sigma_i^2} \right],
    \label{eq:Gauss_im2}
\end{equation}
and $ \chi_\text{G}(\omega,i) = \chi'_\text{G}(\omega,i)+\iota\chi''_\text{G}(\omega,i)$, see Section~\ref{sec:Gauss} for more details. Exact calculation of the Dawson function is performed using the McCabe algorithm~\cite{mccabe1974continued}. Alternatively, the Dawson function may also be computed with a much slower built-in MATLAB function \texttt{dawson(x)} (introduced in R2014a). 

A summary of the parameters and outputs of the function \texttt{Gauss.m} is given in Table~\ref{tab:gauss}. Parameters [\texttt{w}, \texttt{W}, \texttt{sig}] should be in the same units (e.g. all in eV, all in rad/s, etc).

\begin{table}[!ht]
\caption{Gaussian model (\texttt{Gauss.m})}
\label{tab:gauss}
\scriptsize
\vspace{1mm}

\begin{tabular}{llllll}
\texttt{Gauss.m}     &                         &       &        &                           &          \\
\hline
Parameter          & Type                    & Units & I/O    & Description               & Eqs.    \\
\hline
\texttt{w} ($\omega$)   & double array            & eV    & Input  & Frequency range           & (\ref{eq:CDG_im}-\ref{eq:CDG_re})         \\ 
\texttt{A} ($A$)        & double array            &  -    & Input  & Gaussian amplitudes       & (\ref{eq:CDG_im}-\ref{eq:CDG_re})         \\ 
\texttt{W} ($\Omega$)   & double array            & eV    & Input  & Gaussian offsets           & (\ref{eq:CDG_im}-\ref{eq:CDG_re})         \\ 
\texttt{sig} ($\sigma$) & double array            & eV    & Input  & Gaussian widths            & (\ref{eq:CDG_im}-\ref{eq:CDG_re})         \\ 
\hline
\texttt{Eps} ($\veps_\text {G}(\omega)$) 
    & complex array          & -     & Output & Gaussian relative permittivity & (\ref{eq:GDA_Intro}-\ref{eq:CDG_re}) \\
\hline
\end{tabular}
\end{table}

\subsection{Approximate Gaussian Model (Gauss\_n.m)} 
\noindent
Function \texttt{Gauss\_n.m} is a time-domain compatible $n$-th order approximation of the total Gaussian permittivity: $\veps_{\rm G}^n(\omega) = \sum_i{ \chi^n_{\rm{G}}(\omega,i)} \approx \veps_{\rm G}(\omega)$. For each $i$-th Gaussian term, $\chi^n_{\rm{G}}(\omega,i)$, it uses a constrained minimax rational approximation of order $n$ from 2 to 8 for the Dawson function, Eqns.~(\ref{eq:Dawson_n}, \ref{eq:minimax}), 
rather than the exact Dawson function used in \texttt{Gauss.m}. The full causal approximate complex function $\veps_{\rm G}^n(\omega)$ is restored via Hilbert transform, as described in Section~\ref{sec:GDMapprox}, with the resulting approximation coefficients summarized in Tables~\ref{tab:Approx_Dawson}-\ref{tab:Approx_Faddeeva}.
As shown in Section~\ref{sec:simul}, the lowest order, $n=2$, gives an accuracy around a few percent and is sufficient for most applications. Higher accuracy, $<1\%$ and down to machine single precision (1e-7), can be achieved with $3\le n\le8$ if needed. Along with the approximate relative permittivity $\veps_{\rm G}^n(\omega)\approx \veps_{\rm G}(\omega)$, the function returns the real coefficients $[a_{0,i},a_{1,i},b_{0,i},b_{1,i}]$ representing its GDM representation, 
$\veps_{\rm G}^n(\omega) = 
\sum_i \left( a_{0,i}-\iota\omega a_{1,i} \right)/
\left( b_{0,i}-\iota\omega b_{1,i} -\omega^2\right)$, see Section~\ref{sec:GDM} and Eq.~\eqref{eq:GDMfinal} for more details. 
Using this representation, the multi-term Gaussian model can be efficiently implemented in time domain solvers. 

A summary of the parameters and outputs of the function \texttt{Gauss\_n.m} is given in Table~\ref{tab:gauss_n}. As before, parameters [\texttt{w}, \texttt{W}, \texttt{sig}] should be in the same consistent units (e.g. all in eV, all in rad/s, etc), which will define the units of the output parameters [\texttt{a0}, \texttt{a1}, \texttt{b0}, \texttt{b1}].

\begin{table}[!ht]
\caption{Approximation of Gaussian model (\texttt{Gauss\_n.m})}
\label{tab:gauss_n}
\scriptsize
\vspace{1mm}

\begin{tabular}{llllll}
\texttt{Gauss\_n.m}     &                         &       &        &                           &          \\
\hline
Parameter          & Type                    & Units & I/O    & Description               & Eqs.    \\
\hline
\texttt{n} ($n$)        & integer $\in \overline{2,8}$ & -     & Input  & Model approximation order & (\ref{eq:Dawson_n}, \ref{eq:Dawson_poles})         \\ 
\texttt{w} ($\omega$)   & double array            & eV    & Input  & Frequency range          & (\ref{eq:CDG_im}-\ref{eq:CDG_re})         \\ 
\texttt{A} ($A$)        & double array            & -    & Input  & Gaussian amplitudes       & (\ref{eq:CDG_im}-\ref{eq:CDG_re})         \\ 
\texttt{W} ($\Omega$)   & double array            & eV    & Input  & Gaussian offsets         & (\ref{eq:CDG_im}-\ref{eq:CDG_re})         \\ 
\texttt{sig} ($\sigma$) & double array            & eV    & Input  & Gaussian widths          & (\ref{eq:CDG_im}-\ref{eq:CDG_re})         \\ 
\hline
\texttt{Eps} ($\veps^n_\text {G}(\omega)$) 
    & complex array          & -     & Output & Gaussian relative permittivity & 
    \eqref{eq:lemma1B} \\
\texttt{a0} ($a_{0,i}$)  & double array & eV$^2$ & Output & $a_0$ GDM coefficients             & 
(\ref{eq:lemma1C}, \ref{eq:GDM_TD2FD}) \\
\texttt{a1} ($a_{1,i}$)  & double array & eV     & Output & $a_1$ GDM coefficients                & (\ref{eq:lemma1C}, \ref{eq:GDM_TD2FD}) \\
\texttt{b0} ($b_{0,i}$)  & double array & eV$^2$ & Output & $b_0$ GDM coefficients                & (\ref{eq:lemma1C}, \ref{eq:GDM_TD2FD}) \\
\texttt{b1} ($b_{1,i}$)  & double array & eV     & Output & $b_1$ GDM coefficients                 & (\ref{eq:lemma1C}, \ref{eq:GDM_TD2FD}) \\
\hline
\end{tabular}
\end{table}

\subsection{Prototype of an FDTD Implementation (FDTD1D.m)} 
\noindent
Function \texttt{FDTD1D.m} is an FDTD-GDM Maxwell solver implemented in 1D for simplicity of demonstration. This function is designed as a prototypical example of how our GDM time-domain Gaussian model may be implemented in a TD Maxwell solver. It is also used to test the convergence and stability of the generated material models. These models can consist of a large number of oscillator terms (e.g. an extreme test with a high approximation order $n=8$ and 8 Gauss terms produces 64 oscillators) and requires direct testing in a TD solver. The FDTD-GDM implementation is described in detail in Section~\ref{sec:FDTD_GDM}. 

In particular, this code is built to simulate propagation of a normally incident Gaussian pulse~\eqref{eq:Einc} defined in struct \texttt{S}. The pulse is incident on a film of material with thickness \texttt{Fthick} with  dispersive relative permittivity in GDM format~\eqref{eq:GDMfinal} defined in struct \texttt{D}. The parameter \texttt{ppf} refers to the \textit{points per film} and defines the spatial discretization step as \texttt{Fthick}/\texttt{ppf}. Finally, the \texttt{solver} parameter may be set to ``ADE'', ``ADE2'', ``TRC'', ``TRC2'', ``PCRC2'', or ``PLRC'', as described in Section~\ref{sec:FDTD_GDM}. 

This function returns three simulated electric field probes, \texttt{Eri}, \texttt{Et}, and \texttt{Ei} as arrays over time --- the array index represents the time step from the beginning of the simulation. Probe \texttt{Eri} is the sum of reflected and incident electric field collected in front of the film. Probe \texttt{Et} is the transmitted electric field collected behind the film. Probe \texttt{Ei} is the incident electric field in front of the film. The returned values also include exact locations (adjusted to the numerical mesh) where each probe was recorded and the location of the film --- this is needed for 
the second-order accurate post-processing with phase correction~\eqref{eq:NumericalRT}. A typical mistake
often committed during post-processing is to not take into account half-cell or several-cell phase shifts and thus have only first-order convergence in the resulting reflection/transmission coefficients. Parameter \texttt{Xf = [Xf(1), Xf(2)]} returns the exact location of the beginning and end of the film in the simulation. Parameter \texttt{Xp = [Xp(1), Xp(2), Xp(3)]}
returns the exact locations where probes \texttt{Eri}, \texttt{Et}, and \texttt{Ei} were recorded, respectively. By default, \texttt{Eri} and \texttt{Ei} are set to be 2 cells before the film and \texttt{Et} is 2 cells after the film, while the source is located 5 cells away from the front of the film. Finally, \texttt{dt} refers to the time step used for the simulation. The time step is computed assuming a default maximum Courant number of 1 that can be changed in the function as needed. 

A summary of the parameters and outputs of this function is given in Table~\ref{tab:FDTD1D}.

\begin{table}[!ht]
\caption{GDM-FDTD 1D Implementation (\texttt{FDTD1D.m})}
\label{tab:FDTD1D}
\scriptsize
\vspace{1mm}
\setlength{\tabcolsep}{5pt}

\begin{tabular}{llllll}
\texttt{FDTD1D.m} &              &       &        &                                      & \\
\hline
Parameter & Type & Units & I/O & Description & Eqs. \\
\hline
\texttt{Fthick} & double  & m & Input & Thickness of the material slab       &                                \\
\texttt{ppf}    & integer & - & Input & Points per film                      &                                \\
\texttt{solver} & string  & - & Input & \pbox{20cm}{Solver: ADE, ADE2, TRC,\\ TRC2, PRCR2, PLRC} & \eqref{eq:YeeUniversalCompact} \\
\hline
\texttt{\bf{S}}        & \bf{Struct}       & -   & Input  & Incident pulse parameters struct   &           \\
\hline
\, \texttt{S.lam} ($\lambda_0$)   & double & m     & Input & Carrier wavelength  & \eqref{eq:Einc} \\
\, \texttt{S.sig} ($\sigma_0$)   & double & s   & Input & Pulse width         & \eqref{eq:Einc} \\
\, \texttt{S.t0} ($t_0$)    & double & s   & Input & Pulse center offset & \eqref{eq:Einc} \\
\, \texttt{S.tmax} ($t_\text{max}$)  & double & s   & Input & Simulation runtime  &  \eqref{eq:Einc} \\
\hline
\texttt{\bf{D}}        & \bf{Struct}       & -  & Input  & Film dispersion parameters struct  &  \\
\hline
\, \texttt{D.eps} ($\epsilon_{\infty}$) & double       & -  & Input & High-frequency permittivity     & \eqref{eq:GDMfinal} \\
\, \texttt{D.sig} ($\sigma$) & double       & S/m   & Input & Conductivity   & \eqref{eq:GDMfinal} \\
\, \texttt{D.a0} ($a_{0,i}$) & double array & eV$^2$ & Input & GDM $a_0$ coefficients & \eqref{eq:GDMfinal} \\
\, \texttt{D.a1} ($a_{1,i}$) & double array & eV    & Input & GDM $a_1$ coefficients   & \eqref{eq:GDMfinal} \\
\, \texttt{D.b0} ($b_{0,i}$) & double array & eV$^2$  & Input & GDM $b_0$ coefficients & \eqref{eq:GDMfinal} \\
\, \texttt{D.b1} ($b_{1,i}$) & double array & eV    & Input & GDM $b_1$ coefficients & \eqref{eq:GDMfinal}\\

\hline
\texttt{Eri} ($E_{\rm r}+E_{\rm i}$) & double array & a.u.  & Output & Reflected+incident electric field probe               &  \eqref{eq:NumericalRT} \\
\texttt{Et}  ($E_{\rm t}$)  & double array & a.u.  & Output & Transmitted electric field probe                          &  \eqref{eq:NumericalRT}\\
\texttt{Ei}  ($E_{\rm i}$)  & double array & a.u.  & Output & Incident electric field probe                             &    \eqref{eq:NumericalRT}\\
\texttt{Xf}  ($[x_{\rm a}, x_{\rm b}]$)  & double array(2) & m  & Output & Film location: start and end &   \eqref{eq:NumericalRT}\\
\texttt{Xp}  ($[x_{\rm r}, x_{\rm t}, x_{\rm i}]$)  & double array(3) & m  & Output & Probe locations for \texttt{Eri}, \texttt{Et}, and \texttt{Ei} &     \eqref{eq:NumericalRT}\\
\texttt{dt}  ($\tau$)       & double array & s     & Output & Time step of the simulation                         &         \\          
\hline
\end{tabular}
\end{table}

\subsection{Analytic RT Calculation (RT.m)} 
\noindent
Function \texttt{RT.m} analytically computes the reflection and transmission coefficients of a film at normal incidence~\eqref{eq:AnalyticRT}. 
This function is included for numerical error analysis and to check the convergence order. Numerical error is calculated based on the functions \texttt{RT.m} and \texttt{RTnum.m}. The RT function requires the input frequencies $\omega$ in rad/s, the film thickness $h$ in meters, and the relative permittivity of the film $\veps$ at frequencies $\omega$. This function returns a reflection and transmission coefficient for each $\omega$ inputted. A summary of the parameters and outputs of this function is found in Table~\ref{tab:RT}.

\begin{table}[!ht]
\caption{Analytical calculation of reflection and transmission coefficients (\texttt{RT.m})}
\label{tab:RT}
\scriptsize
\vspace{1mm}

\begin{tabular}{llllll}
\texttt{RT.m} &        &       &        &                                   &       \\
\hline
Parameter & Type  & Units & I/O    & Description                       & Eqs. \\
\hline
\texttt{w} ($\omega$)  & double array  & rad/s & Input  & Frequency range            & \eqref{eq:AnalyticRT}  \\
\texttt{h} ($h$)       & double        & m     & Input  & Film thickness             &  \eqref{eq:AnalyticRT} \\
\texttt{Eps} ($\veps$) & complex array & -     & Input  & Film relative permittivity &  \eqref{eq:AnalyticRT} \\
\hline
\texttt{r}     & complex array      & -  & Output & Reflection coefficient        & \eqref{eq:AnalyticRT} \\
\texttt{t}     & complex array      & -  & Output & Transmission coefficient      & \eqref{eq:AnalyticRT}   \\
\hline
\end{tabular}
\end{table}

\subsection{Numerical (FDTD) RT Calculation (RTnum.m)} 
\noindent
Function \texttt{RTnum.m} computes the reflection and transmission coefficient of a film at normal incidence based on numerical simulation in time-domain, as in Eq.~\eqref{eq:NumericalRT}. As compared with \texttt{RT.m}, this function takes in the electric field probes simulated by FDTD solver, (\texttt{Er}, \texttt{Et}, \texttt{Ei}) and the position of the probes and the film (\texttt{Xp}, \texttt{Xf}) returned by \texttt{FDTD1D.m} along with the simulation time step \texttt{dt}. The returned arrays of reflection and transmission coefficients correspond to the returned $\omega$ frequency array. A summary of the parameters and outputs of this function is found in Table~\ref{tab:RTnum}.

\begin{table}[!ht]
\caption{Numerical reflection and transmission coefficients calculated from field probes returned by a FDTD simulation (\texttt{RTnum.m})}
\label{tab:RTnum}
\scriptsize
\vspace{1mm}

\begin{tabular}{llllll}
\texttt{RTnum.m} &   &       &        &     &       \\
\hline
Parameter   & Type         & Units & I/O    & Description                 & Eqs. \\
\hline
\texttt{Er} ($E_{\rm r}$)      & double array & a.u.   & Input  & Reflected electric field   & \eqref{eq:NumericalRT} \\
\texttt{Et} ($E_{\rm t}$)      & double array & a.u.   & Input  & Transmitted electric field & \eqref{eq:NumericalRT} \\
\texttt{Ei} ($E_{\rm i}$)      & double array & a.u.  & Input  & Incident electric field     & \eqref{eq:NumericalRT} \\
\texttt{Xf}  ($[x_{\rm a}, x_{\rm b}]$)  & double array(2) & m  & Output & Film location: start and end &   \eqref{eq:NumericalRT}\\
\texttt{Xp}  ($[x_{\rm r}, x_{\rm t}, x_{\rm i}]$)  & double array(3) & m  & Output & Probe locations for \texttt{Eri}, \texttt{Et}, and \texttt{Ei} &     \eqref{eq:NumericalRT}\\
\texttt{dt}  ($\tau$)     & double       & s   & Input  & Time step for probe fields  & \eqref{eq:NumericalRT} \\
\hline
\texttt{w} ($\omega$)      & double array & rad/s & Output & Frequency array             & \eqref{eq:NumericalRT} \\
\texttt{r} ($r_\text{num}$)       & complex array       & -  & Output & Reflection coefficient & \eqref{eq:NumericalRT} \\
\texttt{t} ($t_\text{num}$)      & complex array       & -  & Output & Transmission coefficient & \eqref{eq:NumericalRT} \\
\hline
\end{tabular}
\end{table}

\subsection{Example Simulation Script (TestRun.m)} \label{sec: TestRun} 
\begin{figure}[!ht]
    \center
    \includegraphics[width=0.5\textwidth]{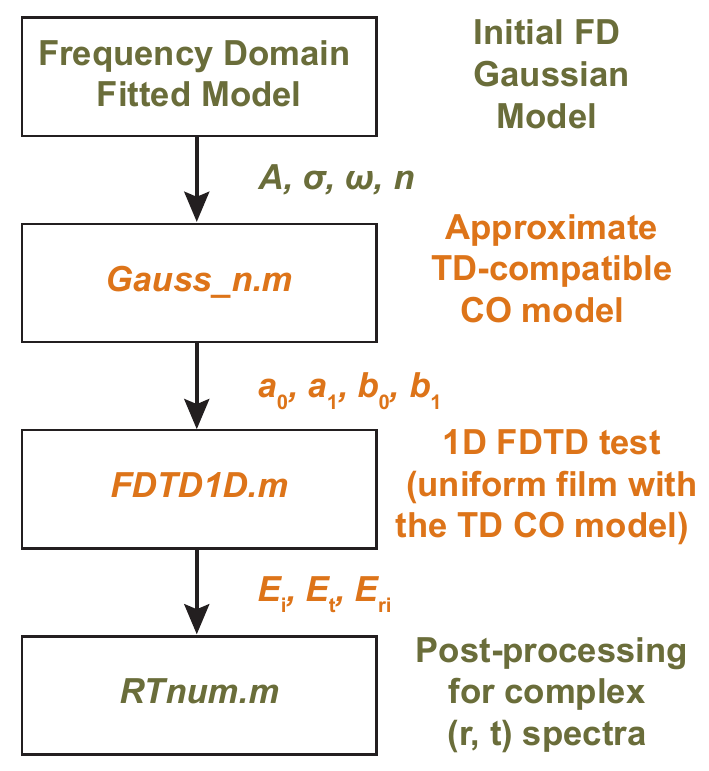}
    \caption{A flow chart illustrating the dataflow through the \texttt{TestRun.m} script (only key variables are indicated). The orange terms correspond to quantities expressed in time domain, 
    and the beige terms are quantities related to the frequency domain.}
    \label{fig:ProgramDescription}
\end{figure}

\noindent
Script \texttt{TestRun.m} is the example program replicating the Silica example from Section~\ref{sec:simul}. A successful run will reproduce Figures~\ref{fig:SilicaModels}, \ref{fig:SilicaFDTD}. The logical program flow of simulation with MADIS is shown
in Figure~\ref{fig:ProgramDescription}. First the frequency domain Gaussian models must be inputted. These are usually obtained from fitting Gaussian profiles to the experimental ellipsometry data. Once the frequency domain model has been constructed the parameters are fed into \texttt{Gauss\_n.m} for conversion of the Gaussian parameters into time-domain solver compatible GDM form defined by parameters ($a_0$,$a_1$,$b_0$,$b_1$). At this point, the GDM model obtained can be adapted for use in other TD solvers, commercial or custom. Once the GDM TD parameters are found, the included FDTD-GDM implementation (\texttt{FDTD1D.m}) is used to simulate a Gaussian pulse passing through a thin film of the modelled material. Function \texttt{FDTD1D.m} returns the simulated electric field probes which are then post processed by \texttt{RTnum.m} into the reflection and transmission spectral curves. 
\section{Conclusions}
\noindent
Over the last several decades, active research has revealed the advantages of employing Gaussian distributions in the approximation of diverse optical material systems with disorder. Recent theoretical and experimental studies have improved the causality foundations and real-life validity of dielectric functions utilizing the difference of two (i) shifted Gaussian absorption peaks and (ii) matching, Hilbert-transformed, Dawson functions. Thus far, the Dawson-Gaussian oscillator model has been obtained solely in the frequency domain. There is no direct analog for time-domain multiphysics computing schemes which severely limits time-domain simulations of many common materials. 

This work builds upon existing work by adapting the frequency-domain Dawson-Gaussian material model to the time-domain. This is done via approximations described by auxiliary differential equations or recursive convolutions. Our time-domain compatible approximations for materials with Gaussian absorption bands is developed within our generalized dispersive material (GDM) model framework. This approach uses constrained minimax rational approximations that minimize the number of oscillators needed for a given accuracy.
This resulting models involve sums of paired phase-relaxed harmonic oscillators with matching amplitudes, phases, and a shared damping parameter with or without a Lorentzian center. The coefficients of the approximation models are explicit functions of the Gaussian parameters --- amplitude, offset, and broadening of the original Gaussian peak. This model's time-domain implementation requires only two second-order auxiliary differential equations per Gaussian peak and is accurate to within several percent error. The accuracy can be increased exponentially by adding more oscillators --- up to 8 per Gaussian peak.


In addition to time-domain applications, the proposed conversion also makes the approximation of the Dawson-Gaussian model extremely efficient in the frequency domain. The approximation of the Dawson function with a rational polynomial, enables ultra-fast optimization of spectral FD approximations. Such optimization is employed in a variety of practical problems of applied spectroscopy and spectroscopic ellipsometry.

Note that this paper considers isotropic nonmagnetic media for explanatory simplicity and its application to optics. However, the proposed concept can be employed to model more general materials with dispersive permittivity and permeability or materials with each component of their anisotropic and bi-anisotropic tensor described as a finite GDM sum (including the GDM representation of Gaussian terms).

The supplied codes named: MADIS (MAterial DIspersion Simulator) contain all necessary MATLAB functions to convert a Gaussian-based model into a time-domain compatible form. These codes also include a one-dimensional Maxwell FDTD solver for the post-conversion time-domain testing of the dispersion model. The package enables seamless implementation of Gaussian-based models normally retrieved from VASE fitting tools into custom and commercial time-domain solvers such as Lumerical FDTD.

Our new TD approximation of Gaussian absorption also provides insight into the physics of Gaussian broadened Lorentzian lineshapes. Equation~\eqref{eq:lemma2} 
explicitly demonstrates that the transition of a classical TD Lorentzian response into an approximate Gaussian profile is achieved as a first-order correction, employing an additional, slowly-varying modulation factor. It is this factor that effectively enables the transition from Gaussian-like inhomogeneously broadened profile to a Lorentzian lineshape homogeneously broadened ($\omega \rightarrow \omega+\iota\Gamma$). In the frequency domain, a different smooth transition between the Gaussian and Lorentzian profiles was implemented by Kim et al. \cite{kim1992modeling} but it has no time-domain conversion. 

Furthermore, the suggested approximation algorithm with minimax optimization and analytically derived modulated oscillators can be applied to other types of inhomogeneous broadening, with other broadening distributions, e.g. Fermi or Boltzmann distributions.

In conclusion, approximating a Gaussian-based dielectric function with pairs of phase-relaxed harmonic oscillators, allows for the implementation of several FDTD-based numerical schemes with controllable accuracy. Experiment-based real-life material examples including glasses, metals, and polymers were presented to demonstrate our techniques efficacy. As such we believe that the proposed method has immediate applicability for broadband time-domain modeling in a variety of photonic systems including conventional and doped glasses, polymers, semiconductors, metals, and composite materials with structural disorder. Essentially, if the model has Gaussian terms caused by molecular, structural, or geometric disorder this method provides a computationally efficient, stable, and physically consistent method of simulation in the time-domain.

\section{Acknowledgements}
\noindent
This work is supported by 
the U.S. Office of Naval Research under award number N00014-21-1-2026,
the DARPA/DSO Extreme Optics and Imaging (EXTREME) Program under Award HR00111720032,
and the Air Force Office of Scientific Research Award FA9550-21-1-0299.
L.J.P. and A.V.K. want to thank their collaborators on recent works \cite{prokopeva2020time, angel2019high} for their contribution to optimization of the GDM model for time domain solvers.
\section{APPENDIX}
\noindent

\begin{enumerate}[leftmargin=*]

\item \textbf{Derivation of integral solution in terms of Faddeeva functions Eq.~\eqref{eq:Faddeeva}}. 

\vspace{2mm}

For this derivation it is convenient to use the convolution integral formulation of Gaussian absorption, \eqref{eq:chi_LG}, with the further assumption of the infinitely narrow band limit $\Gamma\rightarrow +0$. After expanding the Lorentz term into individual poles we obtain two integrals
\begin{align}
\begin{split}    
    \chi_\text{LG}(\omega)
    &= 
    \frac{2A}{\pi}\int\limits_{-\infty}^{\infty}\frac{x\Exp^{-(x-\Omega)^2/\sigma^2}}{x^2-(\omega+\iota\Gamma)^2}\mathrm{d}x \\
    &=
    \frac{A}{\pi}\left(
    \int\limits_{-\infty}^{\infty}\frac{\Exp^{-(x-\Omega)^2/\sigma^2}}{x-\omega-\iota\Gamma}\mathrm{d}x +
    \int\limits_{-\infty}^{\infty}\frac{\Exp^{-(x-\Omega)^2/\sigma^2}}{x+\omega+\iota\Gamma}\mathrm{d}x
    \right)
\end{split}    
\end{align}
Using integral substitutions $y=\pm(x-\Omega)/\sigma$ respectively in both integrals gives
\begin{align}
\begin{split}    
    &=
    \frac{A}{\pi}\left(
    -\int\limits_{-\infty}^{\infty}\frac{\Exp^{-y^2}\mathrm{d}y}{(\omega+\iota\Gamma-\Omega)/\sigma-y} +
    \int\limits_{-\infty}^{\infty}\frac{\Exp^{-y^2}\mathrm{d}y}{(\omega+\iota\Gamma+\Omega)/\sigma-y}
    \right).
\end{split}    
\end{align}
Finally, using integral representation of the Faddeeva function 
$w(z) = \frac{\iota}{\pi}\int\limits_{-\infty}^{\infty}
\frac{\Exp^{-y^2}\mathrm{d}y}{z-y}, (\Im z >0)$
we obtain the solution
\begin{equation}
    = \iota A\left[ 
    w\left( \dfrac{\omega+\iota\Gamma-\Omega}{\sigma} \right)
    -
    w\left( \dfrac{\omega+\iota\Gamma+\Omega}{\sigma} \right)
    \right],
\end{equation}
where taking the infinitely narrow band limit $\Gamma\rightarrow +0$ gives Eq.~\eqref{eq:Faddeeva}.

\item \textbf{Rational approximations of Dawson, Gauss and Faddeeva functions for $2 \le n \le 8$}. 
\begin{table}[!ht]
\caption{Approximations of the Dawson function $F(x)$ for $2\le n \le 8$. \\
         Connection to Tables~\ref{tab:Approx_Gauss},\ref{tab:Approx_Faddeeva} is
         $a_i^F = -(a_i^w)^2 = a_i^G$, $b_i^F = 0.5\sqrt{\pi} b_i^w = 0.5\sqrt{\pi} b_i^G / \sqrt{-a_i^G}$. }
\label{tab:Approx_Dawson}
\centering
\ra{1.3}
\tiny
\addtolength{\tabcolsep}{0pt}
\vspace{2mm}
\begin{tabular}{@{}rr@{}}\toprule
\multicolumn{2}{c}{\scriptsize Dawson function $F(x) \approx x\sum\limits_{i=1}^n \frac{b_i^F}{x^2-a_i^F} = F_n(x)$} \\
\midrule
\multicolumn{1}{c}{\scriptsize $a_i^F$} & \multicolumn{1}{c}{\scriptsize $b_i^F$} \\ [1pt]
\midrule 
\multicolumn{2}{c}{\tiny $n=2$} \\ 
$-0.438720659681 {\pm} \iota0.947986782234$  & $0.250000000000 {\mp} \iota0.459813598153$ \\ 
\midrule 
\multicolumn{2}{c}{\tiny $n=3$} \\ 
$0.013614545833 {\pm} \iota2.078709047946$  & $-0.181332513723 {\mp} \iota0.266376043977$ \\ 
$-1.161715610647$  & $0.862665027447$ \\ 
\midrule 
\multicolumn{2}{c}{\tiny $n=4$} \\ 
$0.635686878634 {\pm} \iota3.281080390081$  & $-0.160133721409 {\pm} \iota0.034566021471$ \\ 
$-1.306199587943 {\pm} \iota1.072802083773$  & $0.410133721409 {\mp} \iota0.834996843474$ \\ 
\midrule 
\multicolumn{2}{c}{\tiny $n=5$} \\ 
$1.348394039593 {\pm} \iota4.521430655277$  & $-0.009436874683 {\pm} \iota0.076036075096$ \\ 
$-1.166813517030 {\pm} \iota2.203849309396$  & $-0.485156399359 {\mp} \iota0.598453435023$ \\ 
$-1.890744864893$  & $1.489186548085$ \\ 
\midrule 
\multicolumn{2}{c}{\tiny $n=6$} \\ 
$2.121195316803 {\pm} \iota5.783242962604$  & $0.030882411871 {\pm} \iota0.014298768740$ \\ 
$-0.856500415978 {\pm} \iota3.372751800665$  & $-0.519723221294 {\pm} \iota0.149428013448$ \\ 
$-2.115483525713 {\pm} \iota1.113518935038$  & $0.738840809423 {\mp} \iota1.584957007127$ \\ 
\midrule 
\multicolumn{2}{c}{\tiny $n=7$} \\ 
$2.937914928185 {\pm} \iota7.059119538495$  & $0.009689437231 {\mp} \iota0.010871384874$ \\ 
$-0.431490105749 {\pm} \iota4.567903347384$  & $-0.032290697828 {\pm} \iota0.338810080854$ \\ 
$-2.118546668713 {\pm} \iota2.256265843118$  & $-1.098611297036 {\mp} \iota1.263262163243$ \\ 
$-2.643176099245$  & $2.742425115265$ \\ 
\midrule 
\multicolumn{2}{c}{\tiny $n=8$} \\ 
$3.788242917171 {\pm} \iota8.345896946524$  & $-0.003216401840 {\mp} \iota0.005130486745$ \\ 
$0.076302555784 {\pm} \iota5.782565813697$  & $0.178494380378 {\pm} \iota0.086186866849$ \\ 
$-1.969817601104 {\pm} \iota3.421842448443$  & $-1.319863092697 {\pm} \iota0.430411295024$ \\ 
$-2.910052425999 {\pm} \iota1.134132616220$  & $1.394585114158 {\mp} \iota3.095210298333$ \\

\bottomrule
\end{tabular}
\end{table} \clearpage
\begin{table}[!ht]
\caption{Approximations of the Gauss function for $2\le n \le 8$.\\
         Connection to Tables~\ref{tab:Approx_Dawson},\ref{tab:Approx_Faddeeva} is
         $a_i^G = -(a_i^w)^2 = a_i^F$, $b_i^G = - a_i^w b_i^w = - 2 \pi^{-1/2} \sqrt{-a_i^F} b_i^F$.}
\label{tab:Approx_Gauss}
\centering
\ra{1.3}
\tiny
\addtolength{\tabcolsep}{0pt}
\vspace{2mm}
\begin{tabular}{@{}rr@{}}\toprule
\multicolumn{2}{c}{\scriptsize Gauss function $G(x) = \Exp^{-x^2} \approx \sum\limits_{i=1}^n \frac{b_i^G}{x^2-a_i^G}=G_n(x)$} \\
\midrule
\multicolumn{1}{c}{\scriptsize $a_i^G$} & \multicolumn{1}{c}{\scriptsize $b_i^G$} \\ [1pt]
\midrule 
\multicolumn{2}{c}{\tiny $n=2$} \\ 
$-0.438720659681 {\pm} \iota0.947986782234$  & $-0.042629823391 {\mp} \iota0.602087244131$ \\ 
\midrule 
\multicolumn{2}{c}{\tiny $n=3$} \\ 
$0.013614545833 {\pm} \iota2.078709047946$  & $-0.515352877308 {\mp} \iota0.096145318257$ \\ 
$-1.161715610647$  & $1.049173140845$ \\ 
\midrule 
\multicolumn{2}{c}{\tiny $n=4$} \\ 
$0.635686878634 {\pm} \iota3.281080390081$  & $-0.155187561979 {\pm} \iota0.300197500306$ \\ 
$-1.306199587943 {\pm} \iota1.072802083773$  & $0.153569026608 {\mp} \iota1.356074937212$ \\ 
\midrule 
\multicolumn{2}{c}{\tiny $n=5$} \\ 
$1.348394039593 {\pm} \iota4.521430655277$  & $0.135606185578 {\pm} \iota0.129914221832$ \\ 
$-1.166813517030 {\pm} \iota2.203849309396$  & $-1.290638448567 {\mp} \iota0.467669057034$ \\ 
$-1.890744864893$  & $2.310577971961$ \\ 
\midrule 
\multicolumn{2}{c}{\tiny $n=6$} \\ 
$2.121195316803 {\pm} \iota5.783242962604$  & $0.082350596335 {\mp} \iota0.047980444387$ \\ 
$-0.856500415978 {\pm} \iota3.372751800665$  & $-0.670413174274 {\pm} \iota0.919914534812$ \\ 
$-2.115483525713 {\pm} \iota1.113518935038$  & $0.588024394058 {\mp} \iota2.993709110109$ \\ 
\midrule 
\multicolumn{2}{c}{\tiny $n=7$} \\ 
$2.937914928185 {\pm} \iota7.059119538495$  & $-0.011444454266 {\mp} \iota0.043972817677$ \\ 
$-0.431490105749 {\pm} \iota4.567903347384$  & $0.493430158336 {\pm} \iota0.658198981159$ \\ 
$-2.118546668713 {\pm} \iota2.256265843118$  & $-2.997471989316 {\mp} \iota1.435261143401$ \\ 
$-2.643176099245$  & $5.030983420934$ \\ 
\midrule 
\multicolumn{2}{c}{\tiny $n=8$} \\ 
$3.788242917171 {\pm} \iota8.345896946524$  & $-0.020684084910 {\mp} \iota0.000255907351$ \\ 
$0.076302555784 {\pm} \iota5.782565813697$  & $0.506678633976 {\mp} \iota0.180461985065$ \\ 
$-1.969817601104 {\pm} \iota3.421842448443$  & $-2.078845443075 {\pm} \iota2.316719699157$ \\ 
$-2.910052425999 {\pm} \iota1.134132616220$  & $1.592850135499 {\mp} \iota6.579845944721$ \\

\bottomrule
\end{tabular}
\end{table} \clearpage
\begin{table}[!ht]
\caption{Approximations of the Faddeeva function for $2\le n \le 8$.\\
         Connection to Tables~\ref{tab:Approx_Dawson},\ref{tab:Approx_Gauss} is
         $a_i^w = -\sqrt{-a_i^F} = -\sqrt{-a_i^G}$, $b_i^w = 2 \pi^{-1/2} b_i^F =  b_i^G / \sqrt{-a_i^G}$.}
\label{tab:Approx_Faddeeva}
\centering
\ra{1.3}
\tiny
\addtolength{\tabcolsep}{0pt}
\vspace{2mm}
\begin{tabular}{@{}rr@{}}\toprule
\multicolumn{2}{c}{\scriptsize Faddeeva function $w(z) \approx \sum\limits_{i=1}^n \frac{b_i^w}{(-\iota z)-a_i^w}=w_n(z)$} \\
\midrule
\multicolumn{1}{c}{\scriptsize $a_i^w$} & \multicolumn{1}{c}{\scriptsize $b_i^w$} \\ [1pt]
\midrule 
\multicolumn{2}{c}{\tiny $n=2$} \\ 
$-0.861192252709 {\pm} \iota0.550392075202$  & $0.282094791774 {\mp} \iota0.518844084903$ \\ 
\midrule 
\multicolumn{2}{c}{\tiny $n=3$} \\ 
$-1.016154290943 {\pm} \iota1.022831407826$  & $-0.204611830803 {\mp} \iota0.300573178637$ \\ 
$-1.077829119410$  & $0.973413245153$ \\ 
\midrule 
\multicolumn{2}{c}{\tiny $n=4$} \\ 
$-1.163272588017 {\pm} \iota1.410280111420$  & $-0.180691555188 {\pm} \iota0.039003578517$ \\ 
$-1.224026977455 {\pm} \iota0.438226486643$  & $0.462786346961 {\mp} \iota0.942193042766$ \\ 
\midrule 
\multicolumn{2}{c}{\tiny $n=5$} \\ 
$-1.298039972818 {\pm} \iota1.741637680641$  & $-0.010648372795 {\pm} \iota0.085797523086$ \\ 
$-1.352864853473 {\pm} \iota0.814511997905$  & $-0.547440373820 {\mp} \iota0.675282388557$ \\ 
$-1.375043586543$  & $1.680367076778$ \\ 
\midrule 
\multicolumn{2}{c}{\tiny $n=6$} \\ 
$-1.421053466936 {\pm} \iota2.034843549931$  & $0.034847070185 {\pm} \iota0.016134432762$ \\ 
$-1.472464958917 {\pm} \iota1.145274045484$  & $-0.586444855563 {\pm} \iota0.168611457355$ \\ 
$-1.501021472283 {\pm} \iota0.370920388415$  & $0.833692577152 {\mp} \iota1.788432467585$ \\ 
\midrule 
\multicolumn{2}{c}{\tiny $n=7$} \\ 
$-1.534301687008 {\pm} \iota2.300434001431$  & $0.010933359112 {\mp} \iota0.012267044209$ \\ 
$-1.584254990907 {\pm} \iota1.441656606292$  & $-0.036436150720 {\pm} \iota0.382306236838$ \\ 
$-1.614549268182 {\pm} \iota0.698729325757$  & $-1.239650100311 {\mp} \iota1.425438707583$ \\ 
$-1.625784764120$  & $3.094495367385$ \\ 
\midrule 
\multicolumn{2}{c}{\tiny $n=8$} \\ 
$-1.639690405623 {\pm} \iota2.544961206672$  & $-0.003629320829 {\mp} \iota0.005789134360$ \\ 
$-1.689196059176 {\pm} \iota1.711632519590$  & $0.201409340263 {\pm} \iota0.097251465030$ \\ 
$-1.720193603530 {\pm} \iota0.994609688532$  & $-1.489306017217 {\pm} \iota0.485667138588$ \\ 
$-1.736850287092 {\pm} \iota0.326491184833$  & $1.573620789558 {\mp} \iota3.492570818419$ \\

\bottomrule
\end{tabular}
\end{table} \clearpage

\item \textbf{Additional pictures}. 
    \begin{figure}[!ht]
    \center
    \includegraphics[width=\textwidth]{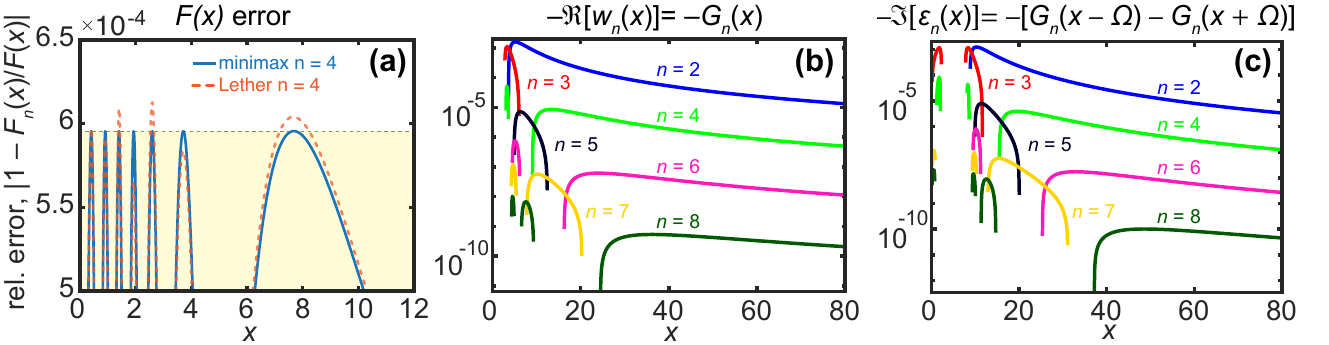}
    \caption{
    (a) Comparison of relative errors of Lether's \textit{near-minimax} approximation \cite{lether1997constrained} and our minimax approximation for the Dawson function $F(x)$. Lether's approximation obtained by the modified Remez algorithm gives a slight deviation from our perfect minimax profile found numerically and results in larger maximum error. (b) In order to build a causal approximation, we take Hilbert transform of the Dawson's minimax approximation $G_n(x) = -\frac{2}{\sqrt{\pi}}\mathcal{H}\left\{ F_n(x)\right\} \approx \Exp^{-x^2}$ which approximates Gaussian function. By plotting the additive inverse of $G_n(x)$ in
    log scale, we analyze the ``criminal'' negative values in the approximation, which are not present in the exact Gaussian function $G(x)=\Exp^{-x^2}>0$. In the picture positive values of $G_n(x)$ go to NaN and cause blank segments in the curves, so only negative values are plotted. Although segments of negative values do exist in the approximation curves $G_n(x)$ (and are infinitely long for even $n$ approximations without Lorentzian center), they go to zero exponentially fast with order $n$. (c) Same behaviour is inherited in the absorption of the approximated dielectric function $\veps''_n(\omega) = G_n(\omega-\Omega) - G_n(\omega+\Omega)$ --- small negative absorption exists but goes to zero exponentially fast.}
    \label{fig:AppendixFig}
    \end{figure}
    
\item \textbf{ Derivation of the RC recursion \eqref{eq:RCidentity} for a single pole,\\* $\chi(t) = A\Exp^{Bt}\theta(t)$ }. 
\vspace{2mm}

\noindent
We start from convolution integral \eqref{eq:convolution} for  $\Pv(t+\tau)$  with $\chi(t) = A\Exp^{B t}\theta(t)$
\bas
   \Pv(t+\tau) &= A\veps_0 \int_0^\infty \Exp^{B\tilde{t}} \Ev(t+\tau-\tilde{t}) \, \dd\tilde{t}, \\
\intertext{replace $\tilde{t}$ by  $\tilde{t}+\tau$,}
   \Pv(t+\tau) &=  A\veps_0 \int_{-\tau}^\infty \Exp^{B(\tilde{t}+\tau)} \Ev(t-\tilde{t}) \, \dd\tilde{t}, \\
\intertext{then split the integral into two parts,}
\Pv(t+\tau) &= \Exp^{B \tau} A\veps_0 \int_0^\infty \Exp^{B \tilde{t}} \Ev(t-\tilde{t}) \, \dd\tilde{t}
       + A\veps_0 \int_{-\tau}^0 \Exp^{B(\tilde{t}+\tau)} \Ev(t-\tilde{t}) \, \dd\tilde{t} .
\eas
Then, replacing $\tilde{t}$ with $-\tilde{t}$ in the second integral, gives the exact two-level recursion \eqref{eq:RCidentity}
\bas
\Pv(t+\tau ) = \Exp^{B \tau} \Pv(t)  + A\veps_0 \int_0^{\tau} \Exp^{B(\tau-\tilde{t})} 
\Ev(t+\tilde{t}) \, \dd\tilde{t} .
\eas

\item \textbf{Derivation of the RC recursion \eqref{eq:PRCquadrature} for a pole pair,\\* $\chi(t) = A^+\Exp^{B^+t}\theta(t) + A^-\Exp^{B^-t}\theta(t)$}. 
\vspace{2mm}

\noindent
We start by writing the recurrence \eqref{eq:PRCquadrature} for each of the two poles, 
$\Pv_{\pm}(t) = A^\pm\Exp^{B^\pm t}\theta(t) * \veps_0\Ev(t) $
\begin{align}\label{eq:Deriv2_1}
\begin{split} 
\Pv_+^{n+1} &= \Exp^{B^+\tau} \Pv_+^n + 
A^+\tau\veps_0 \, \left[ \wrc_0^+ \, \Ev^{n+1} +  \wrc_1^+ \Exp^{B^+\tau}  \Ev^{n} \right],\\
\Pv_-^{n+1} &= \Exp^{B^-\tau} \Pv_-^n + 
A^-\tau\veps_0 \, \left[ \wrc_0^- \, \Ev^{n+1} +  \wrc_1^- \Exp^{B^-\tau}  \Ev^{n} \right].
\end{split}    
\end{align}
Here expressions for $\theta^\pm_{0,1}$ are given in Table~\ref{tab:RCapproximations} as functions of $B$ for different quadratures, e.g. $\theta^+_0 = \theta_0(B=B^+)$.
Now we make linear transform from complex functions $[P_+,P_-]$ to functions $P = P_+ + P_- \in \mathcal{R}$ and $Q = P_+ - P_-$
\begin{align}\label{eq:Deriv2_2}
\begin{split}    
&\Pv_+^{n+1} \pm \Pv_-^{n+1} = 
\left[ \frac{\Exp^{B^+\tau} \pm \Exp^{B^-\tau}}{2} \right] \Pv^n +
\left[ \frac{\Exp^{B^+\tau} \mp \Exp^{B^-\tau}}{2} \right] \Qv^n +\\
&\quad \tau \, \left[ A^+\wrc_0^+ \pm A^-\wrc_0^- \right] \veps_0\Ev^{n+1}+
\tau \, \left[ A^+\Exp^{B^+\tau}\wrc_1^+ \pm A^-\Exp^{B^-\tau}\wrc_1^- \right] \veps_0\Ev^{n}.
\end{split}    
\end{align}
To exclude $\Qv^n$ from the $\Pv$-equation in \eqref{eq:Deriv2_2} we multiply both sides of \eqref{eq:Deriv2_2} by $\left( \Exp^{B^+\tau} \pm \Exp^{B^-\tau} \right)/2$ and then subtract the $\Pv$-equation from the $\Qv$-equation
\begin{align}\label{eq:Deriv2_3}
\begin{split}    
  \left[ \frac{\Exp^{B^+\tau} - \Exp^{B^-\tau}}{2} \right]\Qv^{n+1} 
- \left[ \frac{\Exp^{B^+\tau} + \Exp^{B^-\tau}}{2} \right]\Pv^{n+1} = 
- \left[ \Exp^{B^+\tau}\Exp^{B^-\tau} \right]\Pv^n \\
- \tau\left[ A^+\theta_0^+\Exp^{B^-\tau} + A^-\theta_0^-\Exp^{B^+\tau}  \right]\veps_0\Ev^{n+1}
- \tau\left[ A^+\theta_1^+ + A^-\theta_1^- \right]\left( \Exp^{B^+\tau}\Exp^{B^-\tau} \right)\veps_0\Ev^n,
\end{split}    
\end{align}
Lastly,  in Eq.~\eqref{eq:Deriv2_3}  we reduce index $n$ by one and use it to substitute $\Qv^n$ in the $\Pv$-equation of \eqref{eq:Deriv2_2}
\begin{align}\label{eq:Deriv2_4}
\begin{split}    
  \Pv^{n+1} &=
   \underbrace  {\left[\Exp^{\beta^+\tau} + \Exp^{\beta^-\tau}    \right]}_{2\Exp^{-\Gamma\tau}\cos\Omega\tau=\beta_1} \Pv^n + \\
&+  \underbrace  {\left[-\Exp^{\beta^+\tau} \Exp^{\beta^-\tau}    \right]}_{-\Exp^{-2\Gamma\tau}=\beta_2} \Pv^{n-1}  +\\
&+  \underbrace  {  \tau\left[ A^+\theta_0^+ + A^-\theta_0^-  \right]}_{\alpha_0}\veps_0\Ev^{n+1} + \\
&+  \underbrace  {  \tau\left[ A^+\Exp^{\beta^+\tau}\theta_1^+ + A^-\Exp^{\beta^-\tau}\theta_1^
                             -A^+\Exp^{\beta^-\tau}\theta_0^+ - A^-\Exp^{\beta^+\tau}\theta_0^-    \right]}_
     {A^+\Exp^{\beta^+\tau}(\theta_0^+ + \theta_1^+) + A^-\Exp^{\beta^-\tau}(\theta_0^- + \theta_1^-) -\alpha_0\beta_1}\veps_0\Ev^n +\\
 &+  \tau\left[ A^+\theta_1^+ + A^-\theta_1^-  \right]
 \underbrace  {\left[ -\Exp^{\beta^+\tau} \Exp^{\beta^-\tau} \right]}_{-\Exp^{-2\Gamma\tau}=\beta_2}  \veps_0\Ev^{n-1}.
\end{split}    
\end{align}
Here we used $-2\Gamma = B^+ + B^-$, since $B^\pm=-\Gamma\pm\iota\Omega$.

\item \textbf{Derivation of the \textit{universal compact} FDTD scheme \eqref{eq:YeeUniversalCompact} \\* for RC/ADE methods}. 
\vspace{2mm}

We start from the universal formulation \eqref{eq:YeeUniversal} for which the coefficients $\alpha_{i,j}, \beta_{i,j}$ were derived in Sections~\ref{sec:ADE}-\ref{sec:RC} for different ADE and RC methods and summarized in Table~\ref{tab:universal}. In this formulation we will assume the case of one second-order term $\Pv = \Pv_i$, since the case of single pole ($a_0=b_0=0$) and generalization to the full sum are obvious,
\begin{equation}
\left\{
\begin{aligned}
  \veps_\infty\, \dfrac{\Ev^{n+1}-\Ev^{n}}{\tau}
  &= 
  \veps_0^{-1}\nabla_{h}\times \Hv^{n+1/2}  
  - \dfrac{\sigma}{\veps_0} \dfrac{\Ev^{n+1}+\Ev^{n}}{2}
  - \veps_0^{-1}\dfrac{\Pv^{n+1}-\Pv^{n}}{\tau} , \\
\Pv^{n+1} &= \beta_1\,  \Pv^n + \beta_2\, \Pv^{n-1}
 + \veps_0\left(\alpha_0\, \Ev^{n+1} + \alpha_1 \, \Ev^{n} +  \alpha_2 \, \Ev^{n-1}\right).  \\
\end{aligned}
\right.
\end{equation}

First, we introduce the substitute $\Qv^n = \veps_0^{-1}\left(\Pv^{n+1}-\Pv^{n}\right) -\alpha_2(\Ev^{n+1}-\Ev^n)$, which represents the finite difference of polarization minus the instantaneous response. The curl equation and the $\Qv$-recurrence become
\begin{equation}
\left\{
\begin{aligned}
&\left[\veps_\infty + \dfrac{\sigma\tau}{2\veps_0} + \alpha_0   \right] \Ev^{n+1} = 
\left[\veps_\infty - \dfrac{\sigma\tau}{2\veps_0} + \alpha_0   \right] \Ev^n
-\Qv^n
+ \tau\veps_0^{-1}\nabla_{h}\times \Hv^{n+1/2}, \\
&\Qv^{n+1} = \beta_1\,  \Qv^n + \beta_2\, \Qv^{n-1}
 + \eta_1\, \Ev^{n+1} + (\eta_2-\eta_1) \, \Ev^n - \eta_2 \, \Ev^{n-1}.  \\
\end{aligned}
\right.
\end{equation}
where $\eta_k = \alpha_k + \alpha_0\beta_k, \, k=1,2$. After introducing the final recursive accumulator $\Psiv^n = \Qv^n-\eta_1\Ev^n$ we obtain equations
\begin{equation}
\left\{
\begin{aligned}
&\left[\veps_\infty + \dfrac{\sigma\tau}{2\veps_0} + \alpha_0   \right] \Ev^{n+1} = 
\left[\veps_\infty - \dfrac{\sigma\tau}{2\veps_0} + \alpha_0  -\eta_1 \right] \Ev^n
-\Psiv^n \\
&\qquad\qquad\qquad\qquad\qquad\qquad\qquad\qquad\qquad + \tau\veps_0^{-1}\nabla_{h}\times \Hv^{n+1/2}, \\
&\Psiv^{n+1} = \beta_1\,  \Psiv^n + \beta_2\, \Psiv^{n-1}
 + \left( \eta_2 + (\beta_1-1)\eta_1 \right) \Ev^n 
 - \left( \eta_2 - \beta_2\eta_1 \right) \Ev^{n-1} ,  \\
\end{aligned}
\right.
\end{equation}
in which introduction of an additional recursive accumulator 
$\Phiv^n = \beta_2\Psiv^{n-1} - (\eta_2 - \beta_2\eta_1) \Ev^{n-1}$ leads to \eqref{eq:YeeUniversalCompact}.

\item \textbf{Oscillator parameters obtained for different Gaussian models in Section~\ref{sec:simul}}. 
\begin{table}[!ht]
\caption{CO approximation parameters 
for thermal (th.) silica, fused (f.) silica and gold island films (IF) (in eV).}
\label{tab:SilicaParam}
\footnotesize{
\renewcommand{\arraystretch}{0.8}
\begin{tabular*}{\textwidth}{c|c|c|c|c}
\whl
& & 3-Gauss (Th. Silica) & 8-Gauss (F. Silica) & 4-Gauss (Gold IF)\\
\whl
offset & $\veps_{\infty}$ & 2.321 & 2.1232 & 2.21 \\
\hline 
pole & $A_{\text S}$ & 81.154 & & \\
{$\frac{A_{\text S}}{\Omega_{\text S}^2-\omega^2}$} & $\Omega_{\text S}$ & 11.042 & &\\
\hline 
      & $a_\CP$  & 0.022480 & 0.010511 & 4.776753 \\  
$i=1$ & $\Gamma_\CP$  & 0.005757 & 0.002017 & 0.248772 \\
      & $\Omega_\CP^{\pm}$  & 0.126288 $\pm$ 0.003680 & 0.133817 $\pm$ 0.001289 & 1.762008 $\pm$ 0.158992 \\
\hline 
      & $a_\CP$  & 0.085812 & 0.004072 & 4.319933 \\  
$i=2$ & $\Gamma_\CP$  & 0.008699 & 0.006442 & 1.049910 \\
      & $\Omega_\CP^{\pm}$  & 0.052771 $\pm$ 0.005560 & 0.143139 $\pm$ 0.004117 & 1.808994 $\pm$ 0.671006 \\
\hline 
      & $a_\CP$  & 0.030018 & 0.010086 & 8.489484 \\  
$i=3$ & $\Gamma_\CP$  & 0.011089 & 0.005874 & 1.655031 \\
      & $\Omega_\CP^{\pm}$  & 0.132827 $\pm$ 0.007087 & 0.095158 $\pm$ 0.003754 & 3.772257 $\pm$ 1.057743 \\
\hline 
      & $a_\CP$  & & 0.043404 & 1.832247 \\  
$i=4$ & $\Gamma_\CP$  & & 0.004050 & 0.325834 \\
      & $\Omega_\CP^{\pm}$  & & 0.128612 $\pm$ 0.002588 & 1.371757 $\pm$ 0.208243 \\
\hline 
      & $a_\CP$  &  & 0.024953 & \\  
$i=5$ & $\Gamma_\CP$  &  & 0.017641 & \\
      & $\Omega_\CP^{\pm}$  &  & 0.044038 $\pm$ 0.011275 & \\
\hline 
      & $a_\CP$  &  & 0.021379 & \\  
$i=6$ & $\Gamma_\CP$  &  & 0.002900 & \\
      & $\Omega_\CP^{\pm}$  &  & 0.053072 $\pm$ 0.001853 & \\
\hline 
      & $a_\CP$  &  & 0.012627 & \\  
$i=7$ & $\Gamma_\CP$  &  & 0.001454 & \\
      & $\Omega_\CP^{\pm}$  &  & 0.056822 $\pm$ 0.000929 & \\
\hline 
      & $a_\CP$  &  & 0.026435 & \\  
$i=8$ & $\Gamma_\CP$  &  & 0.014886 & \\
      & $\Omega_\CP^{\pm}$  &  & 0.117781 $\pm$ 0.009514 & \\
\whl 
\end{tabular*}}
\end{table}

\end{enumerate}

\clearpage

\bibliographystyle{vancouver-modified}
\bibliography{references}

\end{document}